\documentclass[twocolumn]{aastex631}
\usepackage{ae,aecompl}
\usepackage{newtxtext,newtxmath}
\usepackage{CJK}
\usepackage{graphicx, url, times}
\usepackage{multirow}
\usepackage{amsmath}
\usepackage{booktabs}
\usepackage{threeparttable}
\usepackage[linesnumbered,ruled,vlined]{algorithm2e}

\graphicspath{{./pics/}}

\newcommand{\hkpc}{{\ifmmode{h^{-1}{\rm kpc}}\else{$h^{-1}$kpc}\fi}}

\newcommand{\Fig}[1]{Figure~\ref{#1}}

\newcommand{\Tbl}[1]{Table~\ref{#1}}
\newcommand{\Sec}[1]{Section~\ref{#1}}

\newcommand{\tabincell}[2]{
    \begin{tabular}{@{}#1@{}}
        #2
    \end{tabular}
}

\newcommand{\rev}[1]{{#1}}

\def\Mpc{{\rm Mpc}}

\def\tng{TNG100-1 }
\def\SFR{{\rm SFR}}

\newcommand{\papertitle}{The multi-component fitting to the star formation histories in the TNG simulation}

\newcommand{\affa}{Peng Cheng Laboratory, No.2 Xingke 1st Street, Nanshan District, Shenzhen 518000, China}
\newcommand{\affb}{CSST Science Center for Guangdong-Hong Kong-Macau Great Bay Area, Zhuhai 519082, China}

\accepted{}
\submitjournal{ApJ}
\shorttitle{Fitting the SFHs}
\shortauthors{Wang et al.}

\newlength{\figwidth}
\newlength{\resplot}
\hypersetup{ pdfauthor={Yang Wang}, pdftitle={\papertitle},
pdfsubject={pdfsub}, urlcolor=blue, }

\begin{document}
\begin{CJK*}{UTF8}{gbsn}

	\title[]{\papertitle}

	\correspondingauthor{Yang Wang}
	\email{wangy18@pcl.ac.cn}

	\author[0000-0002-1512-5653]{Yang Wang (汪洋)}
	\affil{\affa}
	\affil{\affb}

	\author[0000-0002-5823-0349]{Chenxing Dong (董辰兴)}
	\affil{\affa}

	\author{Hengxin Ruan (阮恒心)}
	\affil{\affa}

	\author{Qiufan Lin (林秋帆)}
	\affil{\affa}

	\author[0000-0002-9300-2632]{Yucheng Zhang (张宇澄)}
	\affil{\affa}

	\author{Shupei Chen (陈树沛)}
	\affil{\affa}


	\label{firstpage}
	\begin{abstract}
		The star formation history (SFH) is a key issue in the evolution of galaxies.
		In this work,
		we developed a model based on a Gaussian and gamma function mixture to fit SFHs with varying numbers of components.
		Our primary objective was to use this model to reveal the shape of SFHs and the corresponding physical driving factors.
		Specifically, we applied this model to fit SFHs from the TNG100-1 simulation.
		Our study led to the following findings: 1) Our model fits with TNG star formation histories well, especially for high-mass and red galaxies; 2) A clear relationship exists between the number and shape of fitted components and the mass and color of galaxies, with notable differences observed between central/isolated and satellite galaxies. 3) Our model allowed us to extract different episodes of star formation within star formation histories with ease and analyze the duration and timing of each star formation episode. Our findings indicated a strong relationship between the timing of each star formation episode and galaxy mass and color.
	\end{abstract}


	\section{Introduction}
	\label{sec:intro}

	Star formation is one of the most important processes driving the evolution of galaxies.
	Well measured star formation history (SFH) can help us understand many aspects in galaxy evolution,
	such as the underlying physics of boosting and quenching of star formation,
	and the timing and time scale of these mechanisms
	\citep{Thomas2005, Gallazzi2005, Panter2007, Choi2014, Conroy2014, Pacifici2016, Carnall2018, Schreiber2018}.
	On the other hand, well modeled SFH is important when inferring galaxy properties, such as stellar mass, SFR, metallicity, and dust contents, from the observed spectra or spectral energy distribution (SED)
	\citep{Conroy2013, Kauffmann2014, Janowiecki2017, Leja2017, Telles2018, Zhou2020}.

	Currently, two commonly used methods exist for modeling the SFH: parametric and nonparametric.
	Parametric models employ simple analytical formulas with a small number of parameters to describe the SFH.
	Despite the inherent complexity of SFHs in real galaxies, these simple functions demonstrate their ability to derive the SFHs and constrain galaxy properties \citep{Carnall2019}.
	Researchers explore various forms of parameterized formulas to enhance predictive accuracy, including exponential decline ($\tau$ model) \citep{Reddy2012, Wuyts2011}, delayed exponential decline \citep{Gavazzi2002, Behroozi2010, Lee2010}, rising form \citep{Buat2008, Maraston2010, Papovich2011}, log-normal \citep{Gladders2013, Abramson2015, Diemer2017}, double power laws \citep{Carnall2018, Pacifici2016}, modified exponentially declining \citep{Simha2014, Ciesla2016}, $\Gamma$ function \citep{Lu2015,Zhou2020a, Lu2016}, Gaussian function \citep{Bellstedt2020}, and Bessel function \citep{Iyer2017}.
	Parametric methods offer the advantages of computational efficiency, conceptual simplicity, and strong physical motivation.

	Parametric models, however, cannot capture all SFHs and may introduce bias if inappropriate functions are utilized \citep{Simha2014,Carnall2019}.
	Consequently, researchers have introduced nonparametric models to accommodate the increased complexity and variability of SFH shapes.
	The simplest nonparametric model employs a piecewise constant function to fit the $SFR(t)$ \citep{Ocvirk2006, Kelson2014, Leja2017, Chauke2018}.
	The piecewise function method can be enhanced by incorporating adaptive time binning \citep{Tojeiro2007}.
	Other methods include direct fitting of SFH patterns from theoretical models \citep{Finlator2007, Pacifici2012}, polynomial expansion of the SFH \citep{JimenezLopez2022}, and Gaussian process regression \citep{Iyer2019}.
	Although nonparametric methods offer improved flexibility and accuracy,
	they are computationally demanding and are susceptible to various degeneracies due to the large number of involved free variables \citep{Walcher2011, Conroy2013, Lower2020}.

	Regardless of the methods employed, the shape of SFHs is always a significant consideration.
	There hasn't been universal agreement on the best form(s) of SFH models.
	Furthermore, \cite{Carnall2019} and \cite{Leja2019} have shown that the form of prior significantly impacts the posterior distributions of SFHs.
	Thus, selecting an appropriate prior, which is based on our comprehension of SFH shapes, is crucial in accurately reconstructing SFHs.

	With the emergency of hydrodynamical cosmological simulations (e.g., Illustris\citep{Vogelsberger2014}, IllustirsTNG\citep{Nelson2019,Pillepich2018}, EAGLE\citep{Schaye2015, Crain2015}, FIRE\citep{Wetzel2016, Hopkins2018}), it is possible to get numerous realistic and physics motivated SFHs.
	These simulated SFHs provide  references and training sets for studies aimed at recovering SFHs \citep[e.g. ][]{Iyer2017, Iyer2019}.
	Analyzing simulated SFHs will assist in selecting appropriate priors for observational tasks.
	Moreover, analyzing the SFHs is necessary to infer the underlying physics of galaxy formation.
	The simulations encompass numerous processes that play significant roles in galaxy SFHs. These processes include environment-dependent accretion, stochastic variations, minor and major mergers, gas-to-star conversion dictated by physical conditions in the interstellar medium, ejection of gas through galactic winds, and subsequent recycling of ejected material through accretion.
	Explicit functions can provide insights into the roles and quantities of these physical processes in galaxy formation.
	For these reasons, the parameterization of SFHs derived from simulations is a topic worthy of investigation \citep[e.g. ][]{Simha2014}.

	In this study, our approach involves fitting the SFHs from \tng using a combination of two basis forms: Gaussian distribution and $\Gamma$ distribution.
	Our idea aligns closely with the dense basis approach advocated by \cite{Iyer2017}, with the distinction that we utilize a reduced number of basis forms.
	Our goal is to uncover the underlying patterns of SFHs and offer valuable insights for studies focused on recovering SFHs.

	The paper is organized as following.
	In Section \ref{sec:data}, we present the dataset employed in this study.
	In Section \ref{sec:r2}, we analyze the performance of fitting SFHs.
	In Section \ref{sec:type}, we classify the SFHs based on our fitting results and investigate the relationship between SFH types and galaxy properties.
	In Section \ref{sec:epi}, we examine the specific features of SFHs, including peak positions, peak widths, and peak separations. Additionally, we analyze the relationship between SFH characteristics and galaxy properties.
	Finally, we present our conclusions and engage in additional discussions in Section \ref{sec:con}.

	\section{Data and Methodology}
	\label{sec:data}

	This section provides a brief introduction to the data used in our study.
	We explore the star formation history of simulated galaxies using \tng simulation.

	\subsection{The Simulation and Samples}

	To conduct our analysis, we select galaxies from the \tng simulation and reconstruct their SFHs.
	The \tng simulation is a large-scale, cosmological simulation which uses the AREPO moving mesh code developed by \cite{springel2010} and includes the effects of gravity and magnetohydrodynamics. Its cosmological parameters, in agreement with Planck2015 \citep{Collaboration2016}, are $\Omega_{\Lambda} = 0.6911$, $\Omega_m = 0.3089$, $\Omega_b = 0.0486$, $\sigma_8 = 0.8159$, $n_s = 0.9667$, and $h = 0.6774$. The simulation covers a box size of $110.7^3 \Mpc^3$ with $1820^3$ dark matter particles and $1820^3$ initial hydrodynamic cells. The dark matter particle mass resolution is $7.5 \times 10^6 M_{\odot}$ and the initial mass resolution of baryons is $1.4 \times 10^6 M_{\odot}$. The simulation has 100 output snapshots, covering the redshift range 127 to 0. The TNG100-1 simulation features an updated physical model that encompasses the revised recipes for star formation and evolution, chemical enrichment, cooling, and feedbacks \citep{Weinberger2017,Pillepich2018,Pillepich2018a,Nelson2018}. It also features a modified AGN feedback model to regulate massive galaxies \citep{Weinberger2017} and a galactic winds model to shape low-mass galaxies \citep{Pillepich2018}. More details about the simulation and its data release can be found in the introductory paper series of TNG100-1 \citep{Pillepich2018, Springel2018, Nelson2018, Naiman2018, Marinacci2018} and in the data release itself \citep{Nelson2019}.

	A total of $43,440$ galaxies were identified using the SUBFIND algorithm \citep{Springel2001a}.
	These galaxies satisfy two criteria, with $M_*(z=0)>10^8M_{\odot}/h$ at redshift $0$, and a merger tree length constraint such that it is longer than 10 snapshots.
	We obtained the merger trees of these galaxies from the TNG project supplementary file, which are generated using the SubLink algorithm \citep{Rodriguez-Gomez2015}.

	The SFH of a galaxy is determined based on its stellar population.
	Each stellar particle records its age, measured in terms of lookback time $t_L$ in units of giga-years (Gyr), as well as its initial stellar mass.
	We divide their initial stellar masses into different time bins based on their age, using a time bin width of $\Delta t = 0.1\ {\rm Gyr}$.
	The SFH of each galaxy is then calculated as $\SFR(t_L)=(M_{*,initial}(t_L-\Delta t) - M_{*,initial}(t_L))/\Delta t$.
	This definition of SFH closely resembles those reconstructed through observations, with numerous studies deriving SFR estimates using identical methods \citep{Donnari2019, Hahn2019, Matthee2019, Conroy2013, Johnson2013}.
	The time bin width of $0.1\ {\rm Gyr}$ is widely employed in previous works.
	The chosen of time bin may affect the reconstruction of SFH.
	We represent the results of tests on different time bin widths in Appendix \ref{app:timebin}.
	We found that the time bin width of $0.1\ {\rm Gyr}$ is appropriate for the SFH reconstruction based on the TNG100-1 data.
	\rev{Different from previous studies} \citep{BenitezLlambay2015,Digby2019,Joshi2021}, we employ the initial stellar mass rather than the stellar mass at $z=0$ to reconstruct the SFH.
	In the \tng simulation, the stellar particles return approximately $40\%$ of in their mass to inter stellar medium during their evolution \citep{Pillepich2018a} .
	Hence, utilizing the initial stellar mass allows us to account for stellar mass loss and obtain a more accurate estimation of the SFH.
	Appendix \ref{app:mass} shows the difference between the SFH derived from the initial stellar mass and the SFH derived from the stellar mass at $z=0$.
	\begin{figure*}
		\includegraphics[width=0.5\linewidth]{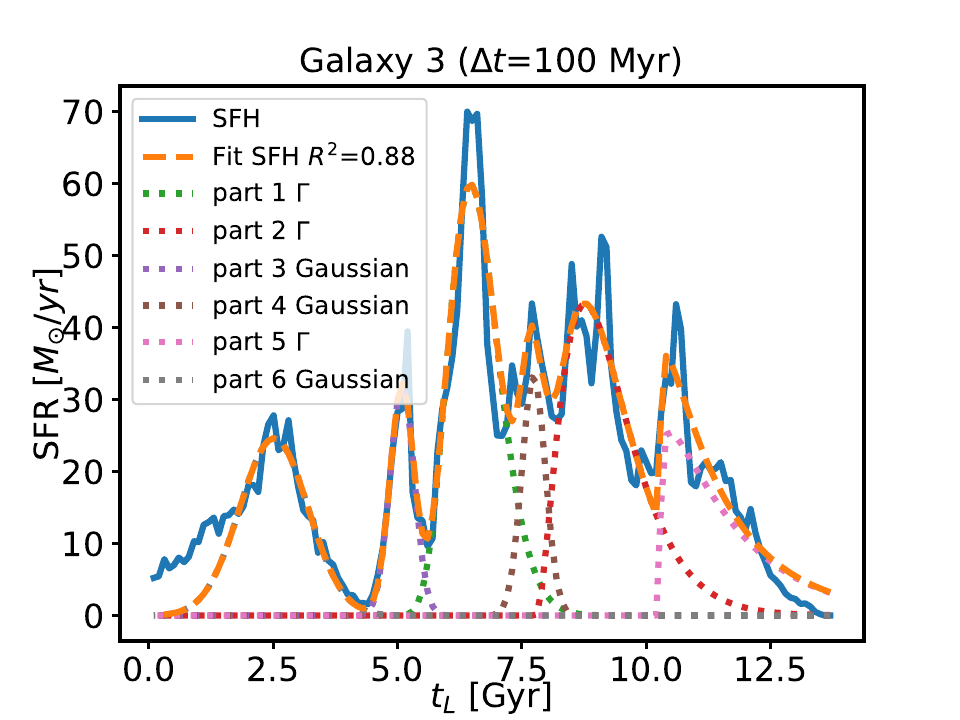}
		\includegraphics[width=0.5\linewidth]{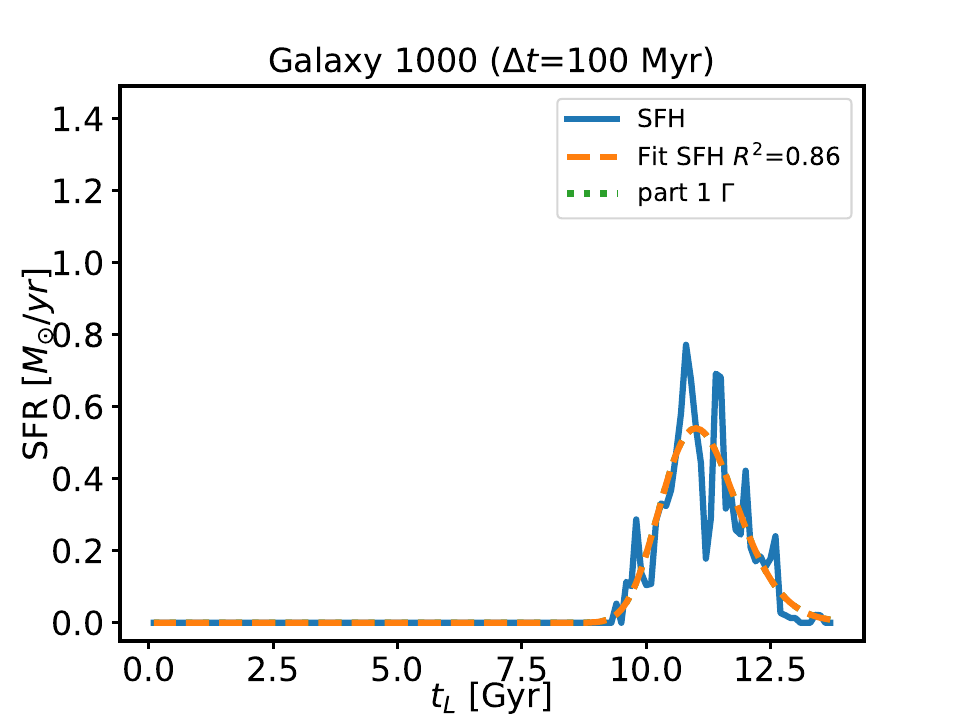} \\
		\includegraphics[width=0.5\linewidth]{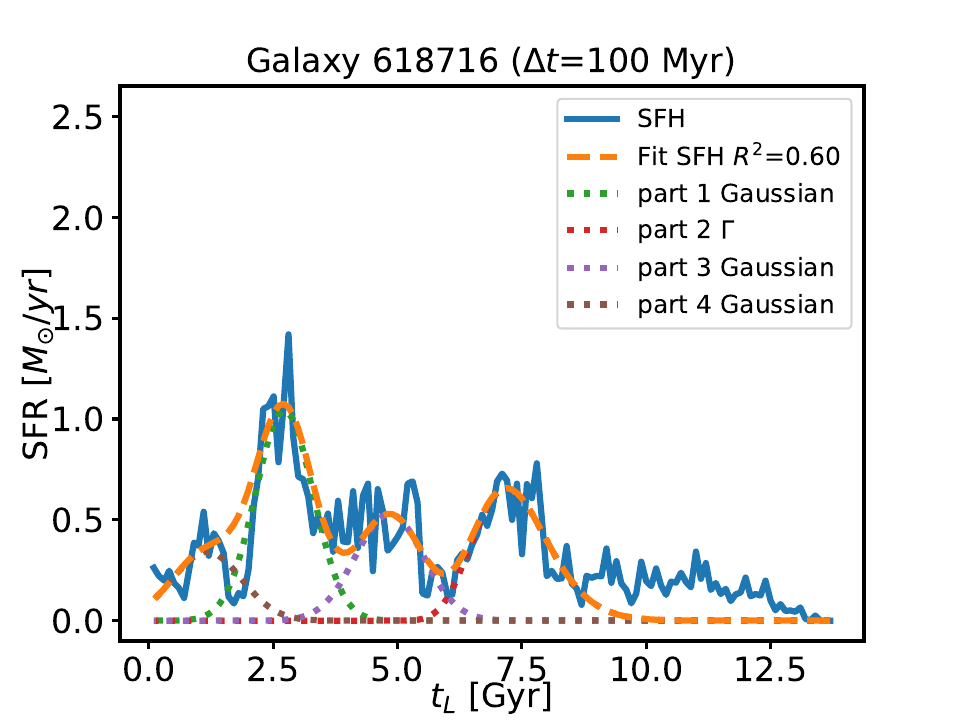}
		\includegraphics[width=0.5\linewidth]{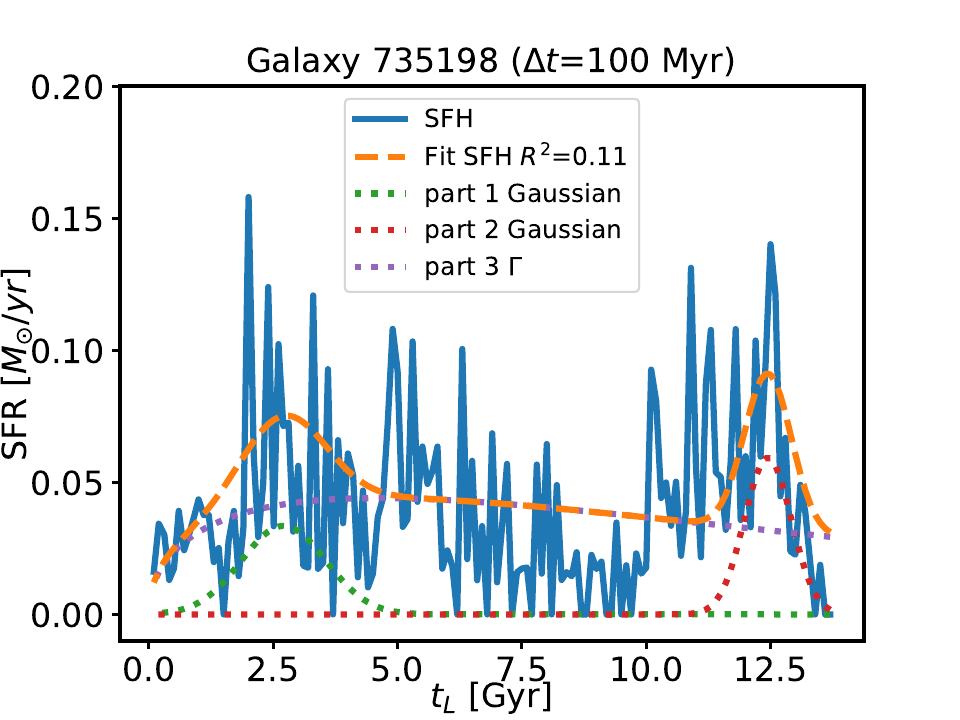} \\
		\caption{Four examples of the fittings to the SFHs.
			The blue solid line represents the actual SFH data, and the orange dashed line represents the corresponding fitted curve.
			The dotted lines show the decomposition of components in the fitted curve.
			The goodness of fit for the overall fitting is displayed in the legend.
			We illustrates the situations for different level of fitting goodness: good fit containing multiple components (top left), good fit containing one component (top right), median fit (bottom left), and bad fit (bottom right).
		}
		\label{FigFitAll}
	\end{figure*}

	\subsection{Fitting Method}
	\label{sec:fit}

	We employed a range of combinations of Gaussian and Gamma distributions to establish the best-fit to the SFH curves.
	The reason we use Gaussian and Gamma distributions as basic functions is because these two patterns have been widely applied in previous works on SFH reconstruction and spectral energy distribution (SED) fitting \citep{Iyer2017,Zhou2020a}.
	Moreover, the shapes of these two distributions can respectively fit symmetric and asymmetric distributions, thereby enabling the overall fitting function to have better universality while limiting the number of basis.
	The general fitting formula was expressed as follows:
	\begin{equation}
		SFR(t_L) = \sum_{i=1}^{N_c} \varphi_i(t_L),
	\end{equation}
	where $\varphi_i(t_L)$ could be either a Gaussian distribution,
	\begin{equation}
		\varphi_i(t_L) = C_i\frac{e^{-(t_L-\mu_i)^2/2\sigma_i^2}}{\sqrt{2\pi}\sigma_i},
	\end{equation}
	or a Gamma distribution,
	\begin{equation}
		\varphi_i(t_L) = C_i\frac{\beta_i^{\alpha_i} (t_L-\nu_i)^{\alpha_i-1} e^{-\beta_i(t_L-\nu_i)}}{\Gamma(\alpha_i)},
	\end{equation}

	Here, $C_i$, $\mu_i$, $\sigma_i$, $\alpha_i$, $\beta_i$, and $\nu_i$ are fitting parameters.

	Also, we utilized the $R^2$ to evaluate the goodness of our fitting through all steps.
	For a sequence $y_i$ and its corresponding fitting values $y_{fit,i}$, the $R^2$ is calculated as:
	\begin{equation}
		R^2=1-\frac{\sum_i(y_{fit,i}-y_i )^2}{n-p-1} / \frac{\sum_i(y_i - \overline{y})^2}{n-1},
	\end{equation}
	where $\overline{y}$ represents the mean of the sequence $y_i$,
	$n$ is the number of data points involved in the fitting,
	and $p$ is the number of variables, which is set to $1$ in this case.
	The $R^2$ ranges between $-\infty$ and $1$, with the results closer to $1$ indicating better agreement between data and model.
	While generally ranging from $0$ to $1$, $R^2$ may be negative if the fitting is severely poor.

	The fitting process is given in pseudo-code Algorithm \ref{alg:fit}.
	The main idea is to obtain the best fit of each component independently first and then combine them together.
	Each component is adjusted to match either the local or global curve of the SFH, and both Gaussian and gamma distribution shape of fittings are tested. The best component fit is determined by the maximum $R^2$ value.
	The overall fit is obtained by summing all the components together.
	Overall fits with different numbers of maximum components $N_c$, ranging from $1$ to $6$, are created.
	Then, the best fit is found based on the $R^2$ values of overall fit candidates.
	We did not test the SFH fittings with $N_c$ greater than $6$ due to the very low fraction of these samples.
	Our fitting results indicate a significant decrease in the number of SFHs with an increasing number of components.
	In current set, fraction of SFH fits with $6$ components is $4\%$.
	The fraction of SFH fits with more than $6$ components should make up less than this value.

	\begin{algorithm*}[!ht]
		\KwIn{$x=t_L$, $y=SFR(t_L)$}
		\KwOut{parameters of fiting $\hat{y}$}
		\For{$N_c \in [1, 6]$ }{
		\tcc{\textit{Try different maximum number of components $N_c$}}
		\For{$i \in [1, N_c]$}{
			\tcc{\textit{Fitting the $i$th component}}
			find the maximum point of $y(x)$ as $(x_i, y_i)$ \;
			select \textbf{local} data $y_{local} = y( x_i - 1 Gyr < x \le x_i + 1 Gyr) $ \;
			fit $y_{local}$ with Gaussian distribution, get $\hat{y}_{local,gau}$ \;
			fit $y_{local}$ with Gamma distribution, get $\hat{y}_{local,gam}$ \;
			select \textbf{global} data $y_{global} = y(all range) $ \;
			fit $y_{global}$ with Gaussian distribution, get $\hat{y}_{global,gau}$ \;
			fit $y_{global}$ with Gamma distribution, get $\hat{y}_{global,gam}$ \;
			\tcc{\textit{the priors for Gaussian fitting are $\mu_i = x_i$, $\sigma_i = 1$}}
			\tcc{\rev{\textit{the ranges of parameters are $\mu_i \in [0, 14]$, $\sigma_i \in (0,10)$ }}}
			\tcc{\textit{the priors for Gamma fitting are $\nu_i = x_i$, $\beta_i = 1$, $\alpha_i=1$}}
			\tcc{\rev{\textit{the ranges of parameters are $\nu_i \in [0, 14]$, $\beta_i \in [0.1,+\infty], \alpha_i \in [0,100]$ }}}
			calculate $R^2$ of $y_{local,gau}$, $y_{local,gam}$,$y_{global,gau}$ and $y_{global,gam}$ \;
			find best component fit $\hat{y}_i = \arg\max\limits_{R^2}(y_{local,gau}, y_{local,gam},y_{global,gau}, y_{global,gam}$) \;
				remove fitted component $y = y - \hat{y}_i$ \;
				change negative point to zero $y[y<0] = 0$ \;
		}
		combine all components  $\hat{y}'_{N_c} = \sum_{i}^{N_c}{\hat{y}_i}$ \;
		use $\hat{y}'_{N_c}$ as prior to redo the fitting to $SFR(t_L)$ to get tuned $\hat{y}''_{N_c}$ \;
		calculate $R^2$ of $\hat{y}''_{N_c}$ \;
		}
		select the best fitting $\hat{y} = \arg \max \limits_{R^2}(\hat{y}''_{1}, \hat{y}''_{2}, ... , \hat{y}''_{6})$
		\caption{Fitting algorithm for one SFH}
		\label{alg:fit}
	\end{algorithm*}

	\Fig{FigFitAll} illustrates the fittings to four galaxies SFHs.
	We select these four SFHs to cover a range of scenarios with regard to different fitting components and goodness.
	As it shows, our method is capable of accurately capturing the trends of SFHs, even when the $R^2$ value is low.
	Furthermore, the bottom right sub-plot reveals that the extremely small $R^2$ value is attributed to the significant fluctuations of the origin SFH.
	By applying the F-test to determine significance, we find that only $281$ SFH fittings fail to reach the $5\%$ significance level (i.e., $p$ value is greater than $0.05$),
	indicating that only these $281$ fittings should be considered unreliable.
	The $R^2$ values of thess unreliable fittings are from $-0.067$ to $0.021$.
	However, in order to improve the credibility of following analysis, we further define a group of well-matched samples by selecting SFH fittings that satisfy an arbitrary criterion of $R^2 \ge 0.5$.
	There are $23,229$ SFHs remained in this group.
	In the following sections, we will specify whether we use all samples or well-matched samples for our investigations.

	The SFHs of galaxies are classified based on the properties of the fittings.
	Six categories labeled as ``C1'' to ``C6'' are assembled based on the number of fitting components.
	The SFHs are also divided into three additional categories based on the major shapes of the fitting components.
	These categories are ``G'' type SFH with more Gaussian components, ``$\Gamma$'' type SFH with more Gamma components, and ``G$\Gamma$'' type SFH with an equal number of both types of components.
	\Tbl{TabFitNumR} presents the number of SFHs of each type.

	As \Tbl{TabFitNumR} shows,
	fittings with fewer components are more prevalent than those with multiple components.
	This is because the number of galaxies increases exponentially as the mass decreases, and smaller mass galaxies dominate in terms of quantity.
	The star formation process in these smaller mass galaxies is usually shorter, therefore it can generally be described with fewer components.
	On the other hand, there are also numerical reasons for this, such as our method tending to achieve fitting with the fewest possible components.
	\Tbl{TabFitNumR} also indicates that SFHs that match our fitting model well contain somewhat more Gaussian components than Gamma components, with the exception of single-component galaxies which exhibit a tendency towards a Gamma-shaped distribution.

	\begin{table*}
		\begin{center}
			\caption{The numbers of SFHs with different fitting components. The three middle columns show the statistics for all SFHs, while the rightmost columns show the results for SFHs that are well fitted with the Gaussian+Gamma format.
				Types ``C1'' to ``C6'' mean the SFH types classified by component number from $1$ to $6$.
				Types ``$\Gamma$'', ``G$\Gamma$'' and ``G'' mean the SFH types classified by the major component shape, corresponding to SFHs with more Gamma components, with an equal number of two components, or with more Gaussian components.}
			\label{TabFitNumR}
			\begin{tabular}{c|llll|llll}
				\toprule
				             & \multicolumn{4}{c|}{all samples} & \multicolumn{4}{c}{$R^2 \ge 0.5$ samples}                                                            \\
				\hline
				types        &                                  & $\Gamma$                                  & G$\Gamma$ & G     &       & $\Gamma$ & G$\Gamma$ & G     \\
				\hline
				C1           & 19462                            & 10290                                     & 0         & 9172  & 8162  & 4733     & 0         & 3429  \\
				C2           & 10897                            & 2933                                      & 5140      & 2824  & 5786  & 1259     & 2626      & 1901  \\
				C3           & 5719                             & 2618                                      & 0         & 3101  & 3680  & 1358     & 0         & 2322  \\
				C4           & 3450                             & 888                                       & 1181      & 1381  & 2446  & 462      & 856       & 1128  \\
				C5           & 2121                             & 822                                       & 0         & 1299  & 1671  & 596      & 0         & 1075  \\
				C6           & 1791                             & 449                                       & 533       & 809   & 1484  & 346      & 443       & 695   \\
				\hline
				total number & 43440                            & 18000                                     & 6854      & 18586 & 23229 & 8754     & 3925      & 10550 \\
				\bottomrule
			\end{tabular}
		\end{center}
	\end{table*}

	\section{Goodness of Fit}
	\label{sec:r2}

	\begin{figure*}
		\includegraphics[width=0.5\linewidth]{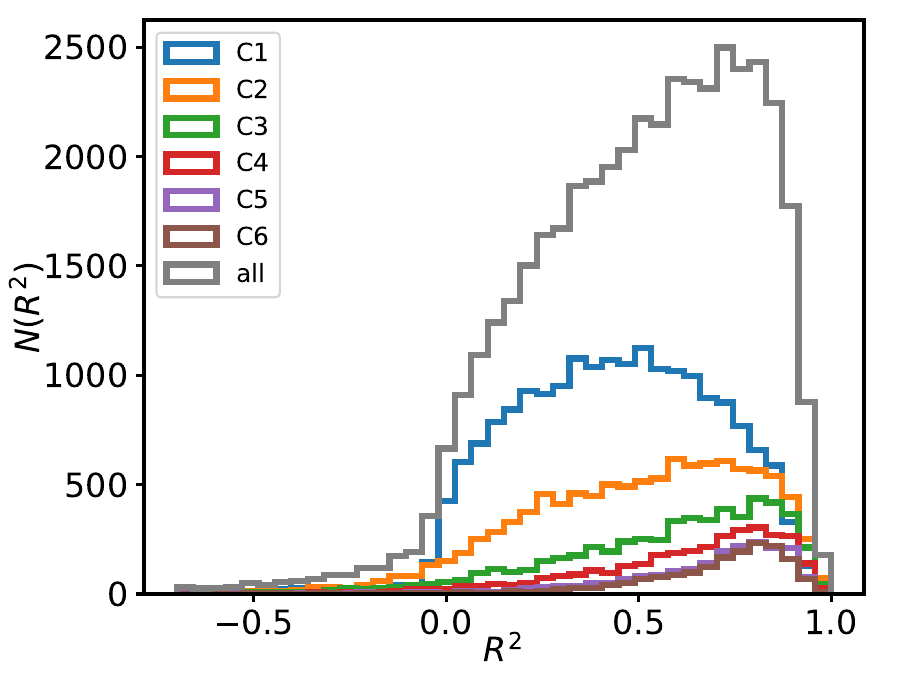}
		\includegraphics[width=0.5\linewidth]{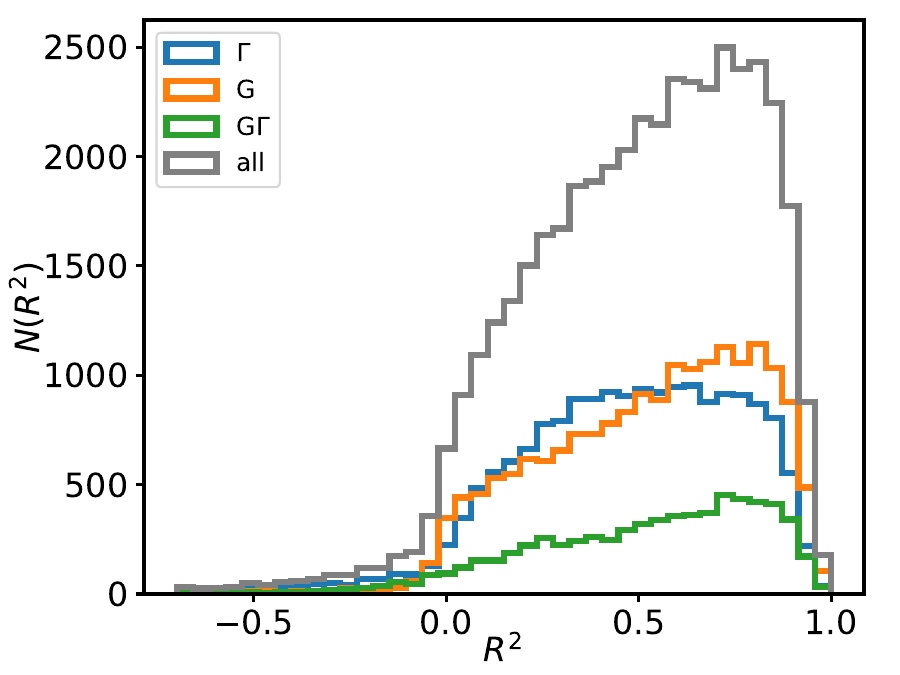}
		\caption{
			Histograms of $R^2$ of SFH fittings.
			The overall distribution is shown as a gray histogram.
			The histograms of $R^2$ values for SFHs of different types of fittings are shown in different colors.
			The SFH fittings are classified by the number of components (left plot) or by the shape (Gamma or Gaussian distribution) of  dominant component(right plot).
		}
		\label{FigR2}
	\end{figure*}

	\begin{figure}
		\includegraphics[width=1\linewidth]{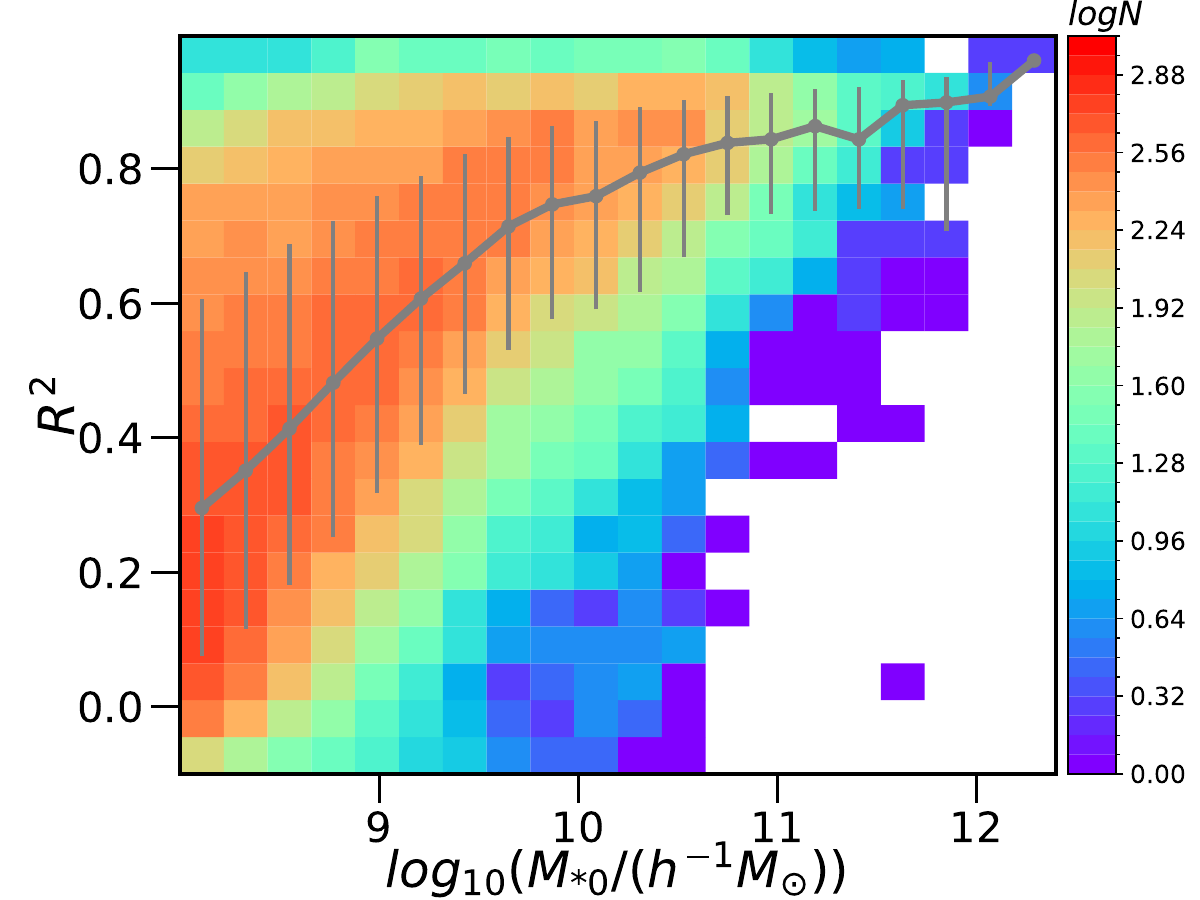}
		\caption{The two-dimensional distribution of $R^2$ of SFH fittings and stellar mass of corresponding galaxies at $z=0$.
			The gray line displays the median value of $R^2$ for galaxies of different masses, while the error bars represent the range of $R^2$ values from $20\%$ to $80\%$ in each mass bin.
		}
		\label{FigR2Mass}
	\end{figure}

	\begin{figure}
		\includegraphics[width=1\linewidth]{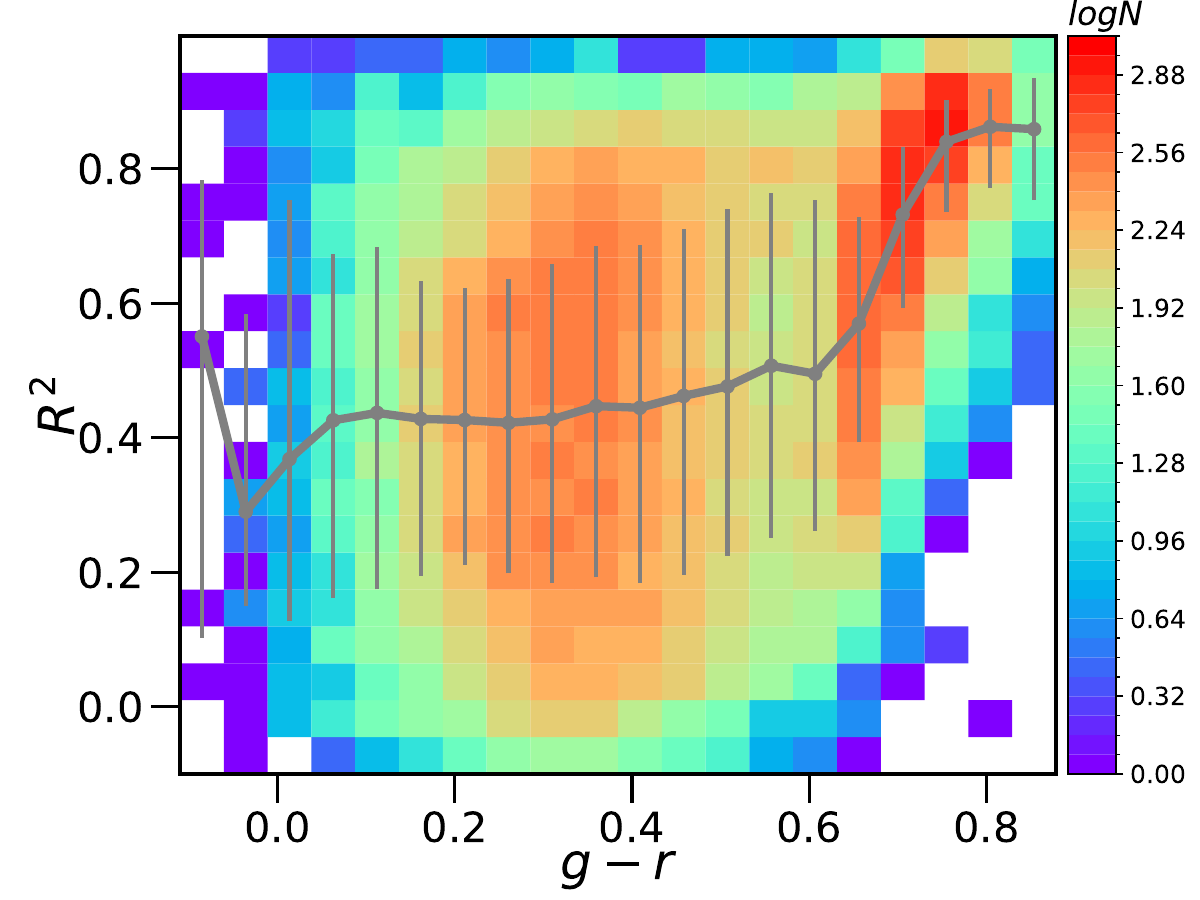}
		\caption{The two-dimensional distribution of $R^2$ of SFH fittings and color of corresponding galaxies at $z=0$.
			The gray line displays the median value of $R^2$ for galaxies of different colors, while the error bars represent the range of $R^2$ values from $20\%$ to $80\%$ in each color bin.
		}
		\label{FigR2Color}
	\end{figure}

	\begin{figure}
		\includegraphics[width=1\linewidth]{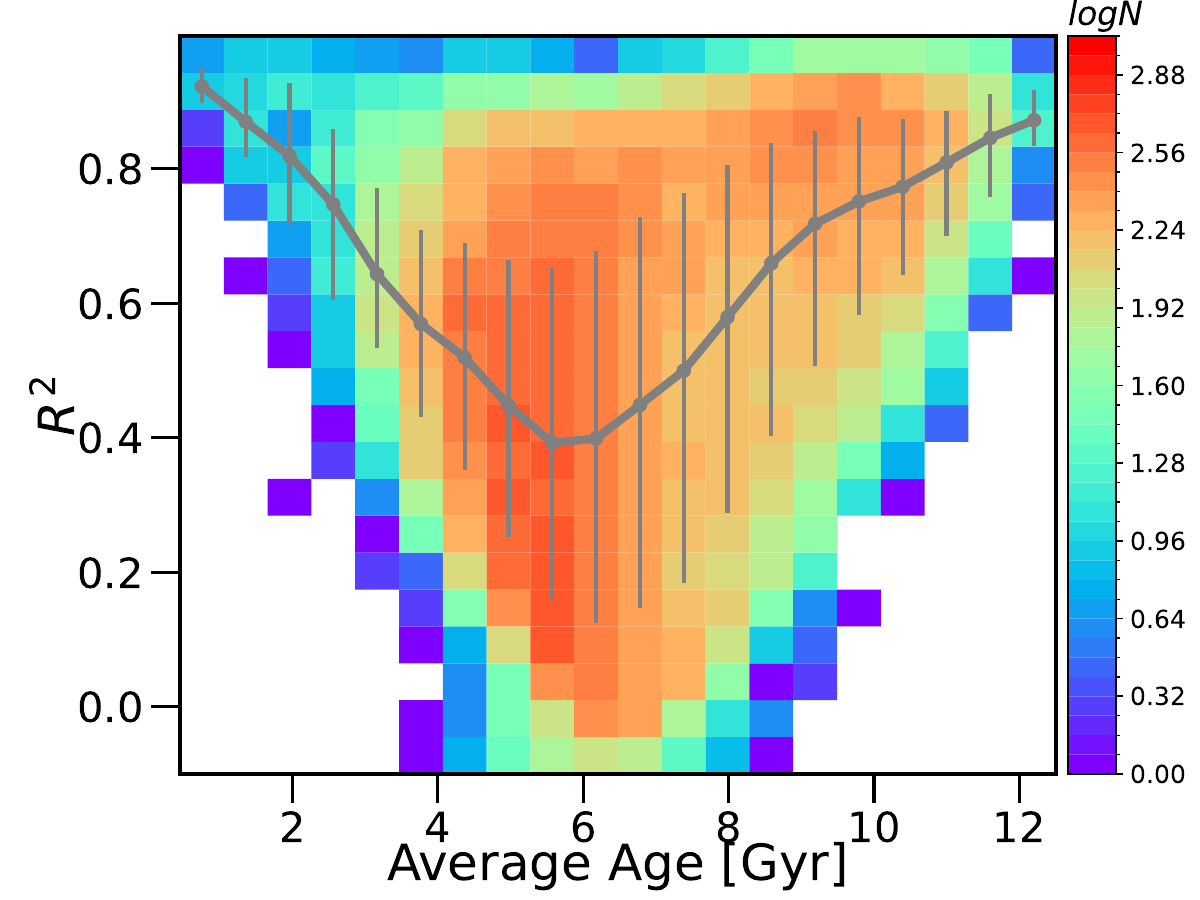}
		\caption{The two-dimensional distribution of $R^2$ of SFH fittings and the aga of them.
			The gray line displays the median value of $R^2$ for galaxies of different SFH lengths, while the error bars represent the range of $R^2$ values from $20\%$ to $80\%$ in each age bin.
		}
		\label{FigR2Len}
	\end{figure}

	\begin{figure}
		\includegraphics[width=1\linewidth]{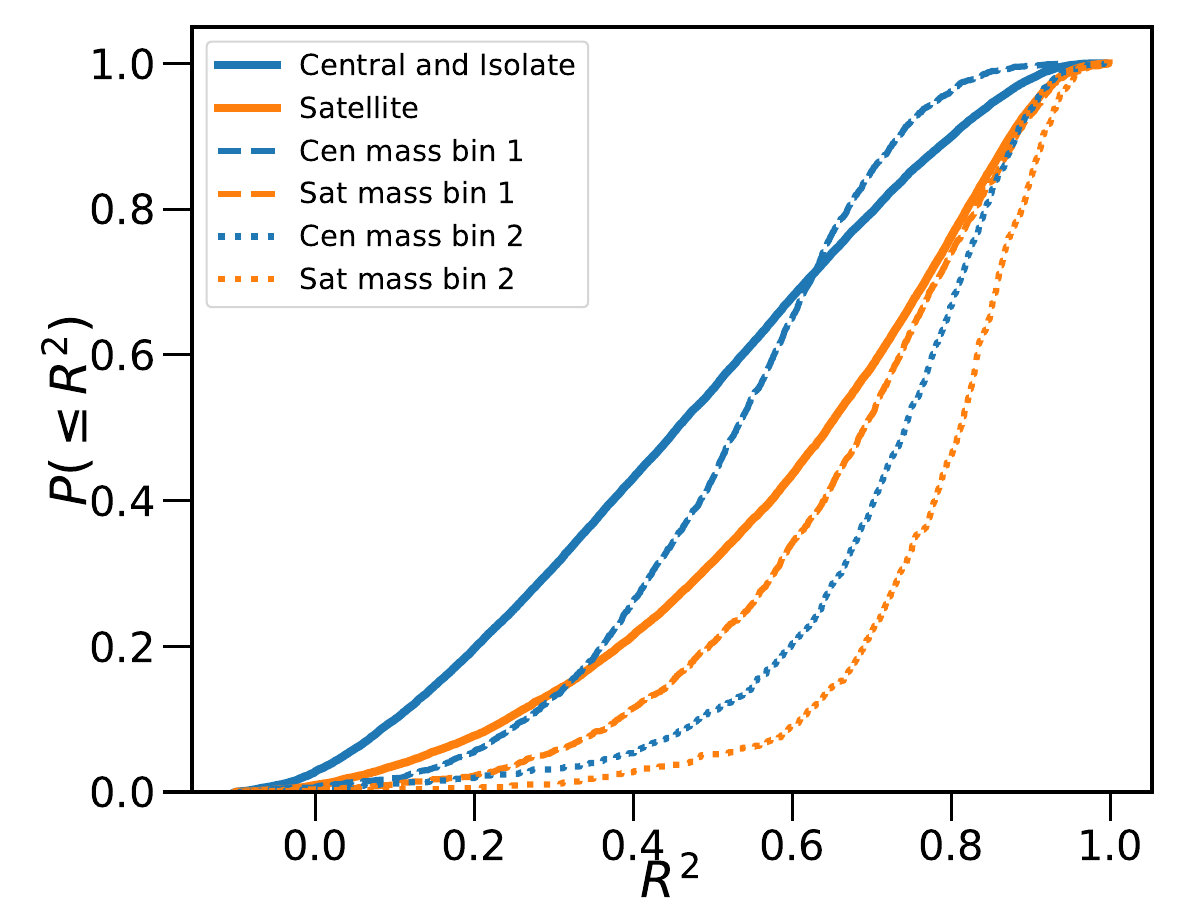}
		\caption{The cumulative density function (CDF) of $R^2$ of SFH fittings for both central and satellite galaxies.
		The blue lines indicate statistics of central and isolated galaxies.
		The oranges lines indicate statistics of satellite galaxies.
		Solid lines denote the statistics of all samples.
		The dashed lines show the CDF of sub-samples with stellar masses between $10^9M_{\odot}h^{-1}$ and $2\times10^9M_{\odot}h^{-1}$.
		The dotted lines show the CDF of sub-samples with stellar masses between $10^{10}M_{\odot}h^{-1}$ and $2\times10^{10}M_{\odot}h^{-1}$.  }
		\label{FigR2Sat}
	\end{figure}

	Before conducting analysis,
	it is important to evaluate the ability of our fitting method to accurately describe the SFHs of galaxies.
	We examined the distribution of the goodness of fit for all samples, which is shown in \Fig{FigR2}.

	For the total trends (grey line in \Fig{FigR2}), the median of $R^2$ value is  $0.53$, the peak of $R^2$ distribution is at $R^2=0.70$.
	The median value here is consistent with \cite{Iyer2017}.
	To quantitatively evaluate the performance at reconstructing the SFHs from mock catalogs, e.g., SAMs \citep{Somerville2015},  hydrodynamic simulation MUFASA \citep{Dave2017} and stochastic SFHs \citep{Kelson2014}, \cite{Iyer2017} used $R^2$, which is in the same definition as this work, to quantify the accuracy of reconstruction.
	The $R^2$ distribution plot indicates that the ability of reconstruction of our model reaches acceptable accuracy, therefore it can be used to fit the majority of SFHs of galaxies.
	Nonetheless, the long tail in the $R^2<0$ region suggests that there is a small number of SFHs that cannot be fitted with our model.
	Generally, fittings with multiple components exhibit good accuracy in fitting, where the fitting quality of Gaussian components is slightly superior to that of Gamma components, as shown consistently in the results presented in \Tbl{TabFitNumR}.

	The shape of the SFH curve of a galaxy is tightly related to its underlying physical processes.
	In our study, the "goodness of fit" value can indicate to some extent whether the shape of the SFH curve of a galaxy approximates a certain type of function/distribution, specifically a combination of Gaussian and Gamma distributions.
	Hence, exploring how the $R^2$ value relates to the properties of galaxies at $z=0$ will be useful in our investigation of star formation history.

	\Fig{FigR2Mass} depicts the relationship between the $R^2$ value of SFH fittings and the stellar mass of their corresponding galaxies at $z=0$.
	It is evident from the figure that higher mass galaxies exhibit better fitting quality.
	This implies that the star formation history of high-mass galaxies conforms more to Gaussian or Gamma distribution shapes.
	This $R^2 - M_*$ relation is partially ascribed to the resolution effect present in the simulation.
	Low-mass galaxies contain a lower number of particles, leading to reduced accuracy and greater noise in reconstructing SFHs.
	Consequently, the overall precision of the fitting is lowered.
	Although, we have already set a minimum mass threshold to ensure sufficient galaxy particles to ease the resolution effect.
	A more comprehensive assessment of this point would necessitate additional simulations with different resolutions, which are beyond the scope of this work.

	\Fig{FigR2Color} depicts the relationship between the $R^2$ value of SFH fittings and the color of their corresponding galaxies at $z=0$.
	It indicates that the correlation between SFH fitting goodness and color is weak for blue galaxies, where fitting goodness is relatively low.
	Conversely, at the red end, our model yields better fitting results, as redder colors correspond better with SFH peak functions that resemble Gaussian or Gamma distributions.
	We attribute this phenomenon to the quenching process that red galaxies underwent during stellar formation history, leading to distinctive SFH with peak-shaped functions that resemble Gaussian or Gamma functions more closely.

	\Fig{FigR2Len} illustrates the relationship between the $R^2$ values and the age of galaxies.
	Following \cite{Nelson2018}, the age of a galaxy is defined as the mass weighted average age of all stellar particles within it.
	It represents the overall distribution of star formation time, as well as a effective length of the SFH
	\Fig{FigR2Len} exhibits that young and old galaxies have better fitting goodness to our model.
	This result is straightforward to comprehend.
	Young galaxies have shorter and simpler shapes, which results from fewer physical mechanisms in the star formation process.
	Old galaxies corresponds to those massive and red galaxies, which can achieve better fitting results to our model, as have been discussed in previous paragraphs.
	\rev{For both old or young galaxies, their SFHs should have relatively significant peaks at early or late stages, which makes the fittings easier. On the other hand, the mass weighted age reaches half of the SFH length, approximately $6\ Gyr$ here, when the SFR distributes evenly accross the whole SFH. In this case, the SFH is hard to be fitted by Gaussian or Gamma distributions, thus results in the extremely low $R^2$ value at $Age\simeq 6 Gyr$. In subsequent content, the \Fig{FigTypeSFHlen} illustrates that the age distribution of SFHs with more components tends to concentrate around $6\ Gyr$, which also verifies the argument here.}

	We evaluated the fitting goodness of different galaxy types, as depicted
	in \Fig{FigR2Sat}.
	The galaxies were categorized into two groups: satellite galaxies and central/isolated galaxies.
	In the simulation, we define the biggest galaxy in a dark matter halo as the central or isolated galaxy, and other galaxies as satellites.
	The results show that satellite galaxies have a relatively better fitting.
	We also plot the cumulative density function (CDF) for sub-samples with a stellar mass around $10^{9} M_{\odot}/h$ (dashed lines) and $10^{10} M_{\odot}/h$ (dotted lines).
	The discrepancy between centrals and satellites still exists, negating the possibility of mass effect.
	A plausible explanation is that satellites are more likely to be quenched in their history, resulting in SFH curves that are similar to Gaussian or Gamma-like peaks.
	This explanation aligns with the color dependence of $R^2$.

	Generally, only half of the galaxies possess SFH shapes that are consistent with the Gaussian-Gamma combined distribution.
	High-mass galaxies, red galaxies, and satellite galaxies generally demonstrate a better fitting for these shapes.
	We hypothesize that these results are mainly due to mechanisms that can shape distinctive peaks in SFHs, such as rapid increases and quenching of star formation.

	\section{Dependency of SFH types}
	\label{sec:type}

	\begin{figure}
		\includegraphics[width=1\linewidth]{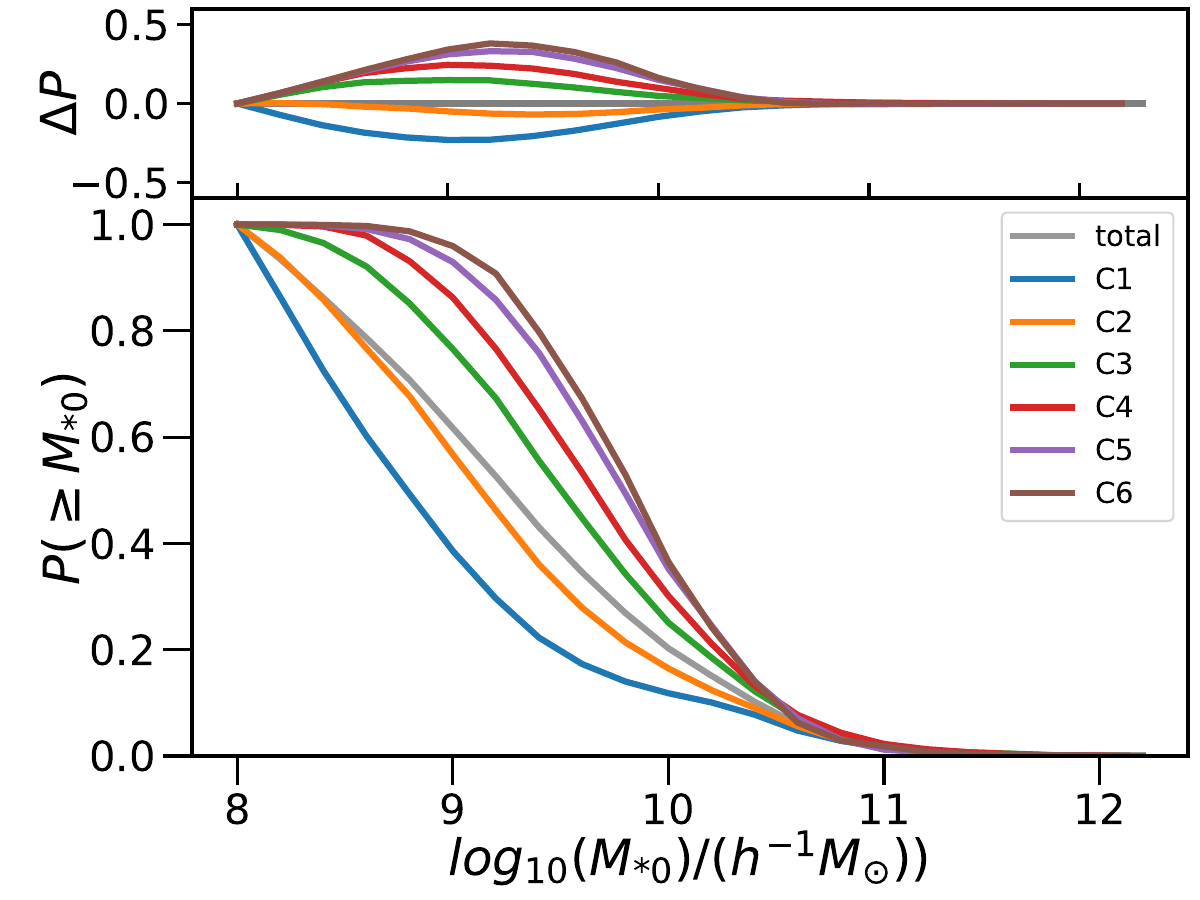}
		\includegraphics[width=1\linewidth]{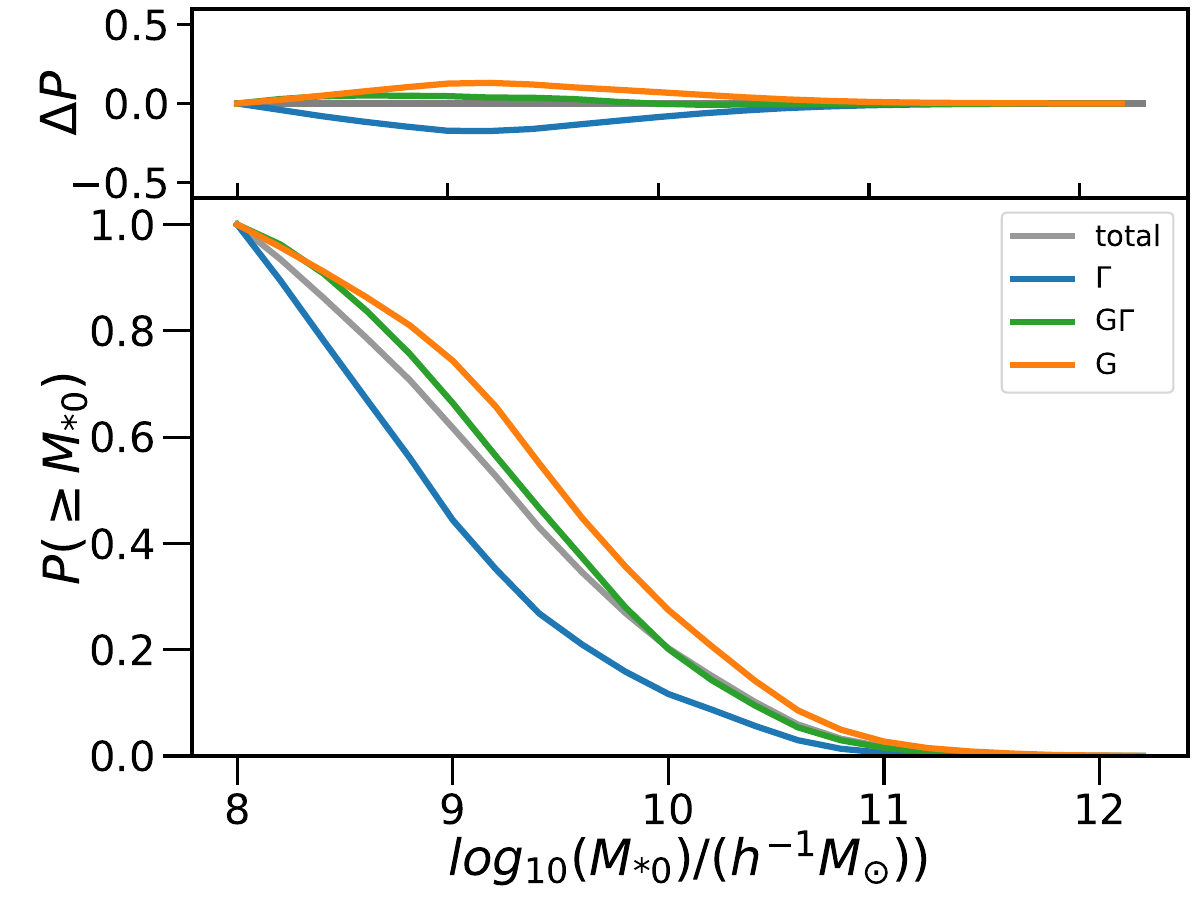}
		\caption{The cumulative stellar mass function of galaxies at $z=0$.
			Different colors represent galaxies with different SFH types.
			The gray line shows the cumulative stellar mass function of all samples.
			The SFHs types are classified by the number of components (left plot) or by the shapes of dominant components (right plot).
			In the top sub-panel of each plot, the residuals to the cumulative mass function of all samples for the samples with different SFH types are given.  }
		\label{FigTypeMass}
	\end{figure}

	\begin{figure}
		\includegraphics[width=1\linewidth]{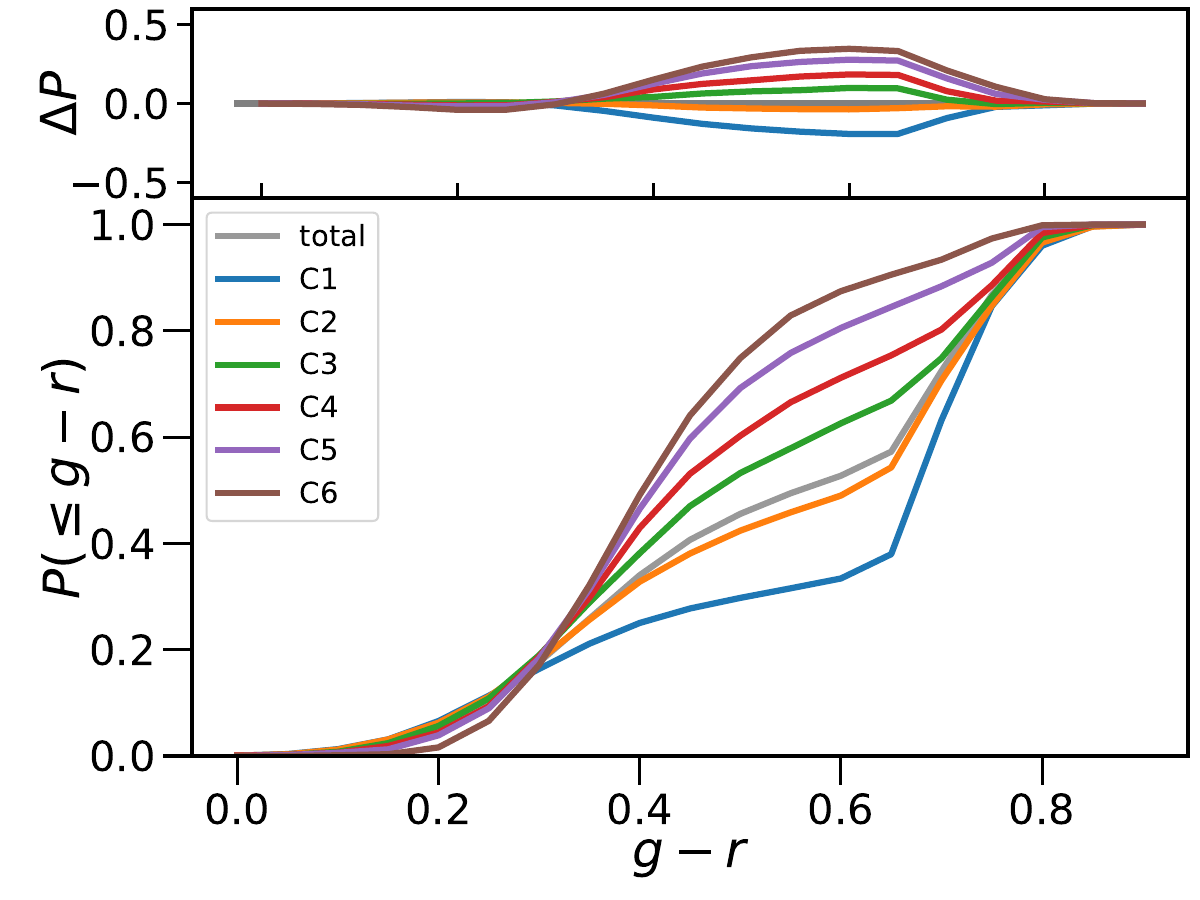}
		\includegraphics[width=1\linewidth]{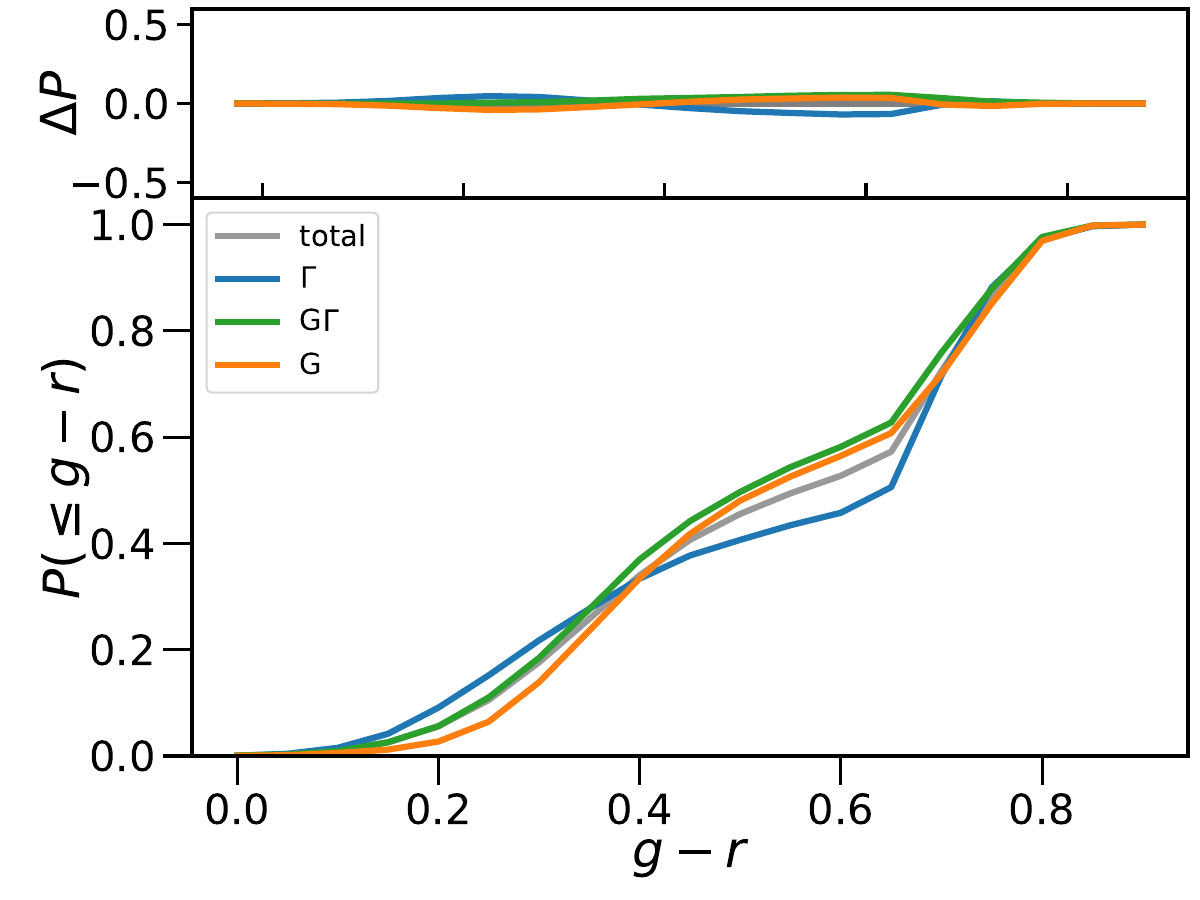}
		\caption{The cumulative density function of galaxies color ($g-r$) at $z=0$. Different colors represent different SFH types. The gray line shows the color distribution of all samples.
			The SFHs types are classified by the number of components (left plot) or by the shapes of dominant components (right plot).
			The top sub-panel of each plot shows the residuals to the color distribution of all samples for samples with different SFH types.}
		\label{FigTypeColor}
	\end{figure}

	\begin{figure}
		\includegraphics[width=1\linewidth]{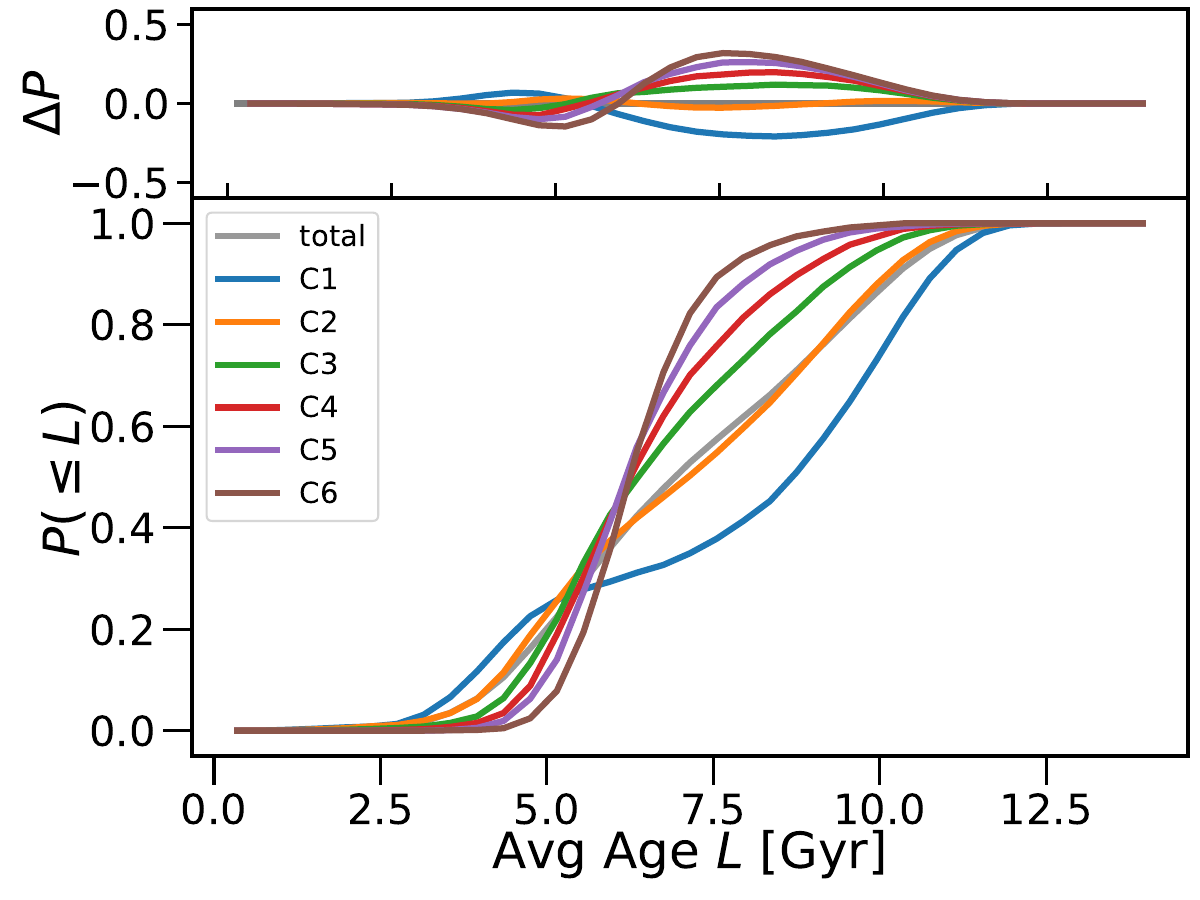}
		\includegraphics[width=1\linewidth]{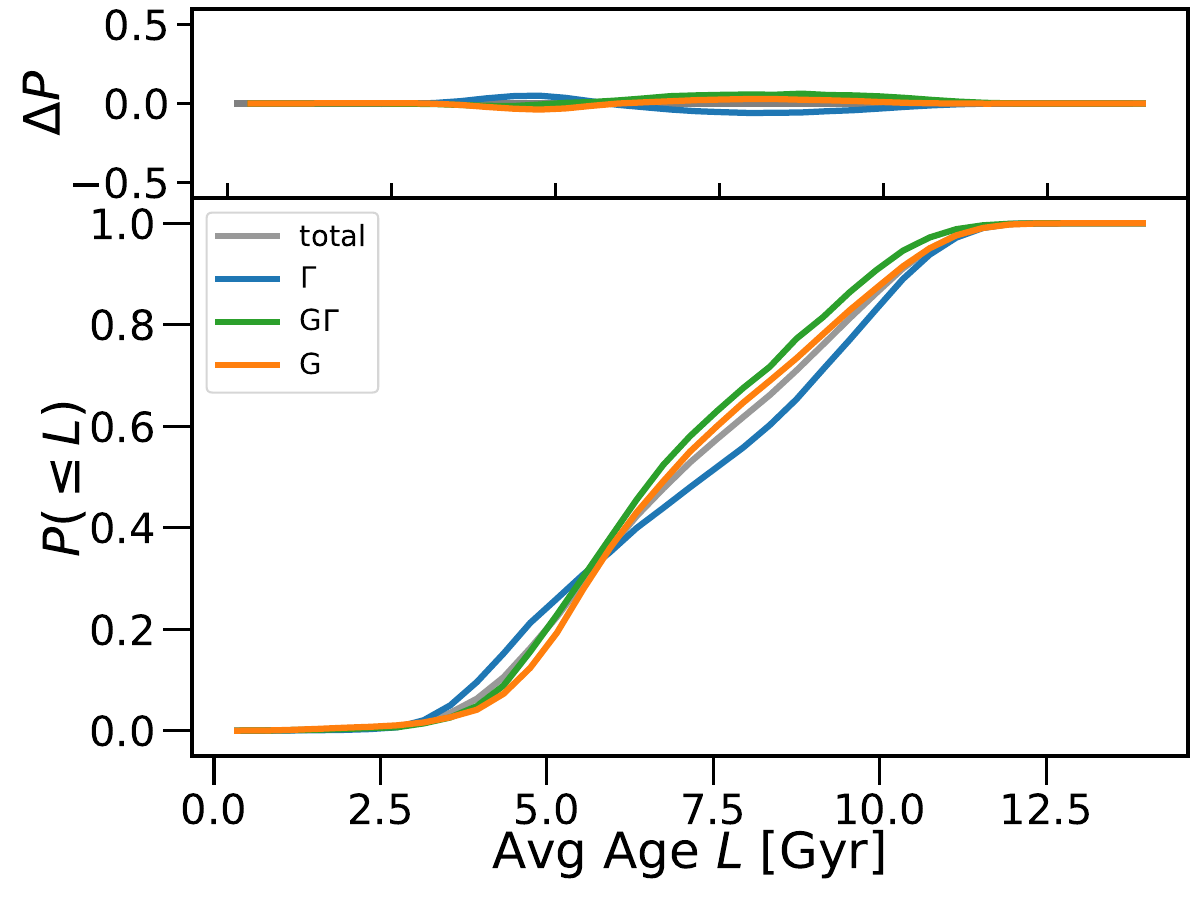}
		\caption{The cumulative density function of average galaxy age with different SFH types. Different colors of histograms represent different SFH types. The gray histogram shows the average galaxy age distribution of all samples.
			The SFHs types are classified by the number of components (left plot) or by the shapes of dominant components (right plot).
			The top sub-panel of each plot shows the residuals to the SFH length distribution of all samples for samples with different SFH types.
		}
		\label{FigTypeSFHlen}
	\end{figure}

	The SFHs that have been fitted are classified into different types based on the number of components and their major shapes, as explained in \Sec{sec:fit}.
	This section aims to investigate how galaxy properties differ depending on the type of SFHs.
	In order to ensure the validity of our study, we will only investigate well-matched samples of SFHs in this section.
	This is because it is not convincing to attribute poorly fitted SFHs to specific types as defined by our classification.

	\Fig{FigTypeMass} shows the stellar mass function of galaxies at $z=0$ corresponding to the different types of SFHs.
	High-mass galaxies tend to have SFHs with more decomposition components, which aligns with our intuition.
	Concerning the component shapes, high-mass galaxies tend to exhibit a prevalence of Gaussian components, whereas low-mass galaxies tend to be dominated by Gamma components.
	The first two rows of \Fig{FigExample1}, \Fig{FigExample2} and \Fig{FigExample3} show SFH examples of low mass and high galaxies.

	\Fig{FigTypeColor}
	shows the distribution of galaxy color at $z=0$ corresponding to different types of SFHs.
	Multiple-component SFHs are predominantly found in blue galaxies, while single-component SFHs are largely associated with red galaxies.
	The color of galaxies shifts towards the blue area as the number of decomposed components in their history increases.
	A simple understanding of this is that galaxies with fewer stellar formation epochs are less likely to experience star formation in their later stages, when compared to galaxies with multiple stellar formation epochs.
	Consequently, the colors of the former tend to be red.
	Galaxies with more Gamma components tend to have a slightly bluer color than others.
	This is primarily because the typical pattern of star formation for many blue galaxies is a slow but continuous rise until recent time.
	The Gamma distribution provides a good fit for this type of curve.
	The third and fourth rows of \Fig{FigExample1}, \Fig{FigExample2} and \Fig{FigExample3} show SFH examples of red and blue galaxies.

	\Fig{FigTypeSFHlen} shows the cumulative density distribution of the age of galaxies.
	The age of a galaxy is defined as the mass weighted age of its all particles.
	Galaxies with a single-component SFH are concentrated around two area, approximately age of 4 Gyr and 10 Gyr.
	This corresponds to two types of galaxies.
	The former represents young galaxies that have just formed and only experienced one period of star formation.
	The latter refers to older quiescent galaxies that have remained in a quiescent state after an initial period of star formation,
	hence having only one component in their early stage.
	As the number of components increases, the distribution of galaxy ages gradually converges towards the central position of 6 Gyr.
	The duration of star formation periods is determined by the underlying physical mechanisms thus will not extend infinitely.
	Therefore, for SFHs with finite lengths (cannot exceed the age of the universe), there must be more effective star formation times to allow for additional star formation cycles and thus more SFH components.
	When a galaxy maintains a constant rate of star formation from the beginning of the universe,
	its average age of stellar particles tends to approach half of the age of the SFH length, approximately 6 Gyr in this case.
	So, more components of SFHs indicates more effective star formation time across a galaxy's life.
	The galaxy age does not show a strong correlation with the shape (Gaussian or Gamma) of the SFH components.
	It implies that different physical drivers of star formation have relatively weak correlations with the galaxy age, indicating that they do not have a particular preference for the timing of occurrence.
	The last two rows of \Fig{FigExample1}, \Fig{FigExample2} and \Fig{FigExample3} show SFH examples of  galaxies with large and small ages.

\begin{figure}
		\includegraphics[width=1\linewidth]{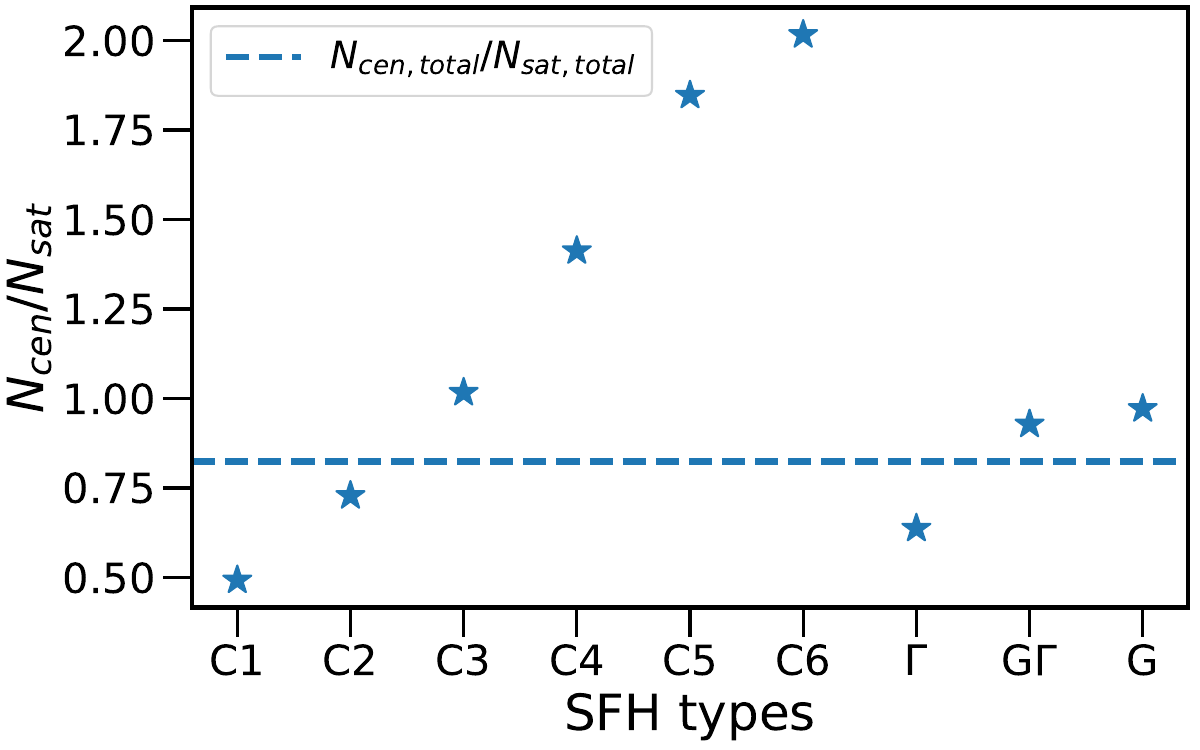}
		\caption{The ratio of central and isolated galaxies over satellite galaxies for different types of SFHs. The horizon line indicate the ratio for the whole sample. }
		\label{FigTypeCS}
	\end{figure}

\begin{figure*}
		\includegraphics[width=1\linewidth]{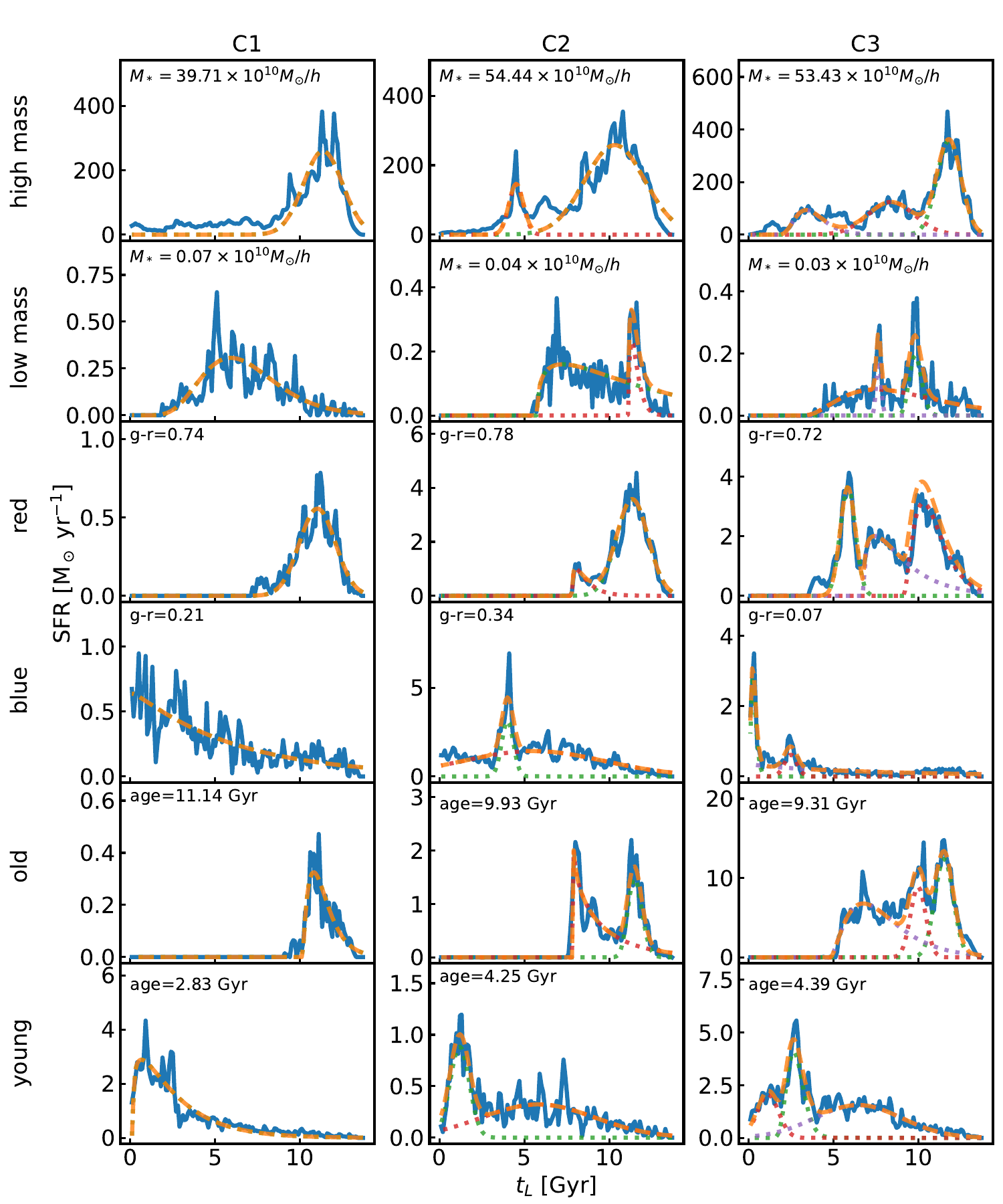}
		\caption{Examples of C1, C2 and C3 type SFHs with different galaxy mass, color and age. Each column show one kind of SFHs as labeled on the top. Each row represent one galaxy property range as labeled on the left. In each sub figure, the solid blue line represents the origin SFH, and the dashed orange line represents the best fit to it. The colored dashed lines represent decomposed components.}
		\label{FigExample1}
	\end{figure*}

\begin{figure*}
		\includegraphics[width=1\linewidth]{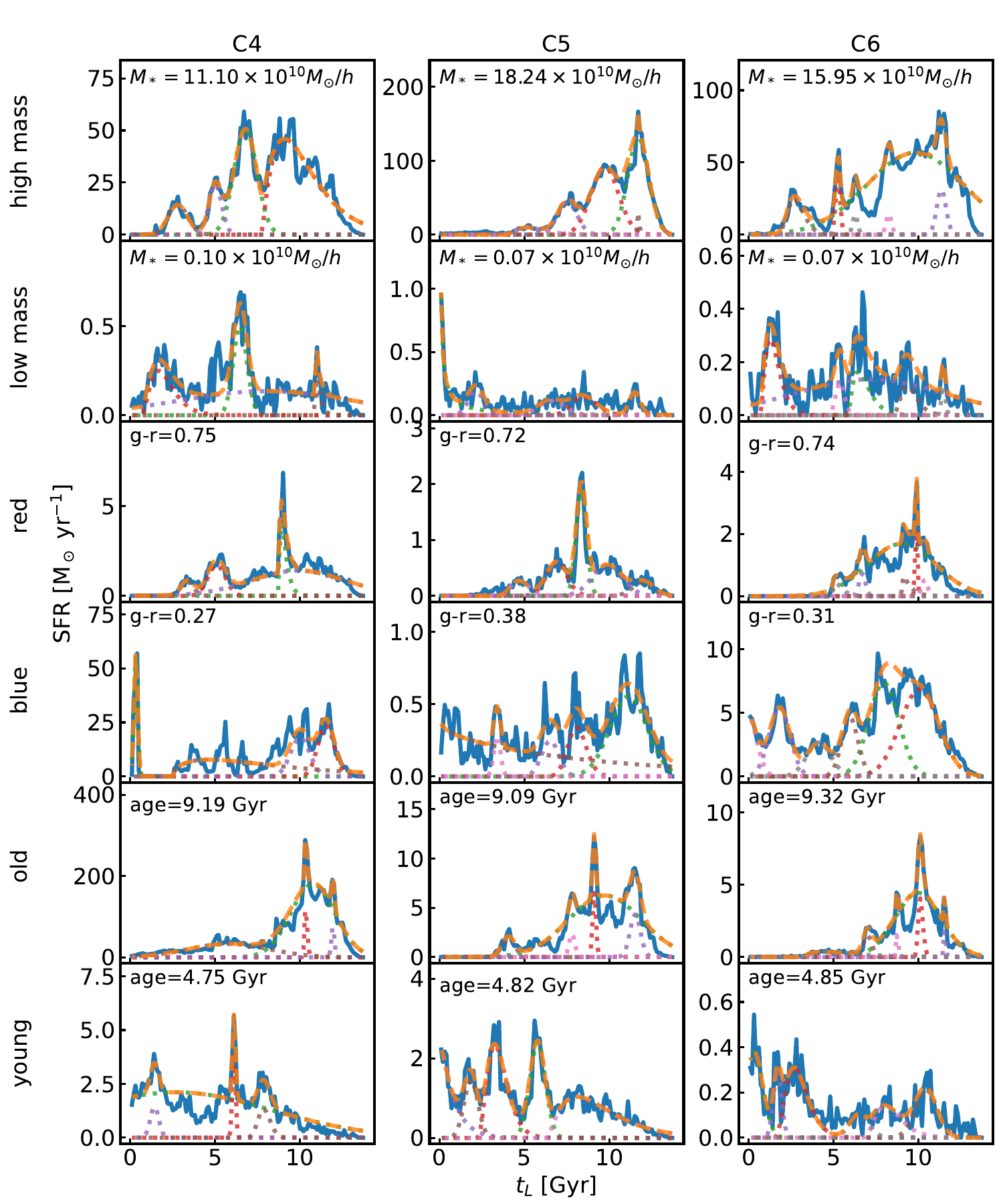}
		\caption{Examples of C4, C5 and C6 type SFHs with different galaxy mass, color and age. Each column show one kind of SFHs as labeled on the top. Each row represent one galaxy property range as labeled on the left.
			The blue solid line represents the actual SFH data, and the orange dashed line represents the corresponding fitted curve.
			The dotted lines show the decomposition of components in the fitted curve.  }
		\label{FigExample2}
	\end{figure*}

\begin{figure*}
		\includegraphics[width=1\linewidth]{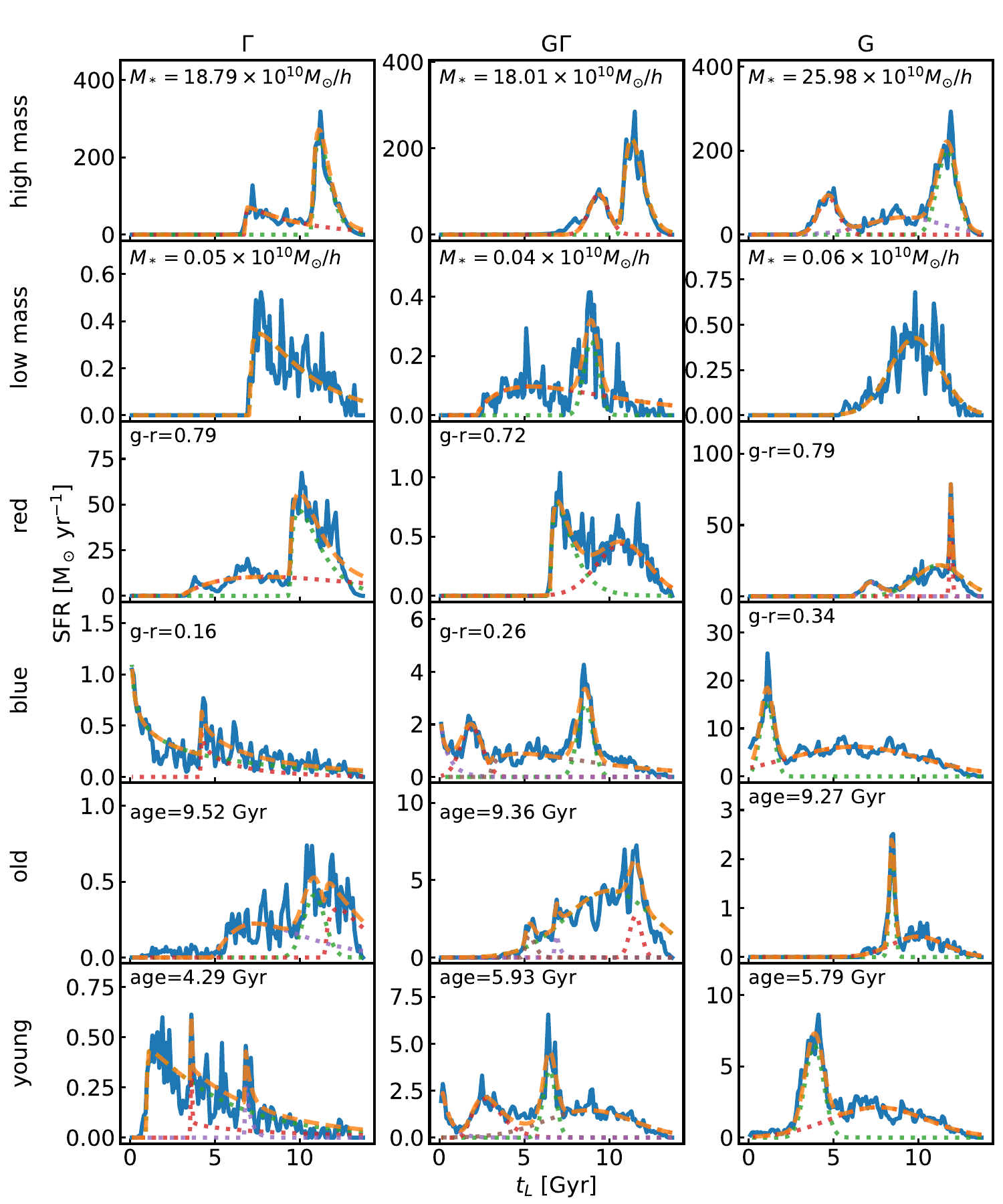}
		\caption{Examples of $\Gamma$, G$\Gamma$ and G type SFHs with different galaxy mass, color and age. Each column show one kind of SFHs as labeled on the top. Each row represent one galaxy property range as labeled on the left. In each sub figure, the solid blue line represents the origin SFH, and the dashed orange line represents the best fit to it. The colored dashed lines represent decomposed components.}
		\label{FigExample3}

	\end{figure*}

	\Fig{FigTypeCS} shows the ratio of the number of central and isolated galaxies to satellite galaxies in galaxies with different SFH types.
	The horizontal dashed line represents the overall sample ratio.
	Data points above this line indicate fewer satellite galaxies in this subsample,
	and vice versa.
	The proportion of central and isolated galaxies increases with increasing decomposed SFH components, indicating more star formation epochs within central and isolated galaxies.
	Because most of the satellite galaxies are quenched in late stages, they naturally have fewer star formation epochs.
	The proportion of satellite galaxies with Gamma SFHs is slightly higher than those with Gaussian SFHs.
	We believe that it is mainly due to the influence of the cluster stripping,  the star formation process in satellite galaxies is quenched more quickly, resulting in the asymmetric gamma distribution of the star formation history curve.

	Overall, high-mass, reddish, long-lived galaxies, and central and isolated galaxies tend to have more SFH components, suggesting more star formation periods.
	Meanwhile, the shapes of the main components of SFHs in these galaxies tend to be slightly Gaussian like.
	Nevertheless, there is considerable overlap in attribute distributions of galaxies with distinct SFH types, make it a challenging to establish a clear standard for strictly constraining the shape and number of SFH components to apply it in reconstructing the SFH of a given galaxy.

	\section{Episodes of SFH}
	\label{sec:epi}

	Both Gaussian and Gamma distributions have well-defined shapes,
	hence, the fitting results can be used to characterize the episodes of each star formation history.
	By exploring the parameters of the fitting function, the peak position and full width at half maximum (FWHM) of each episode can be easily identified and used to further understand the occurrence and time scale of star formation process in galaxies.
	The position and width of SFH peaks are key parameters in this analysis, as they help us gain insight into the star formation history of galaxies.
	We continue to constrain our samples with $R^2 \ge 0.5$ in this section.

	\subsection{Number of episodes}
	\label{sec:episodeunm}

	In this section, we recalculated the number of peaks in the SFH to determine the number of star formation episodes present in a given SFH.
	This is because some fitted components may overlap with others.
	We combine two fitted components into one peak when the distance between their peaks is smaller than the minimum FWHM of them.
	The numbers of SFHs with varying numbers of peaks obtained after applying this procedure are presented in the third column of \Tbl{TabPeakNum}, with the numbers of SFHs having the same number of components shown in the second column.
	The table indicates that for each type of multi-component SFH, there are approximately $250$ SFHs whose components can be combined, which is quite a small fraction to the total number of SFHs.
	This suggests that in most cases, the number of fitted components is consistent with the number of star formation episodes in a SFH.

	\begin{table}
		\caption{{Number of SFHs with different numbers of fitting components (left) and actual peaks (right), limited to the sample with fitting goodness of $R^2>0.5$.
				}}
		\label{TabPeakNum}
		\centering
		\begin{tabular}{c|c | c}
			\toprule
			$N$ & \tabincell{l}{SFH with        \\$N$ components} & \tabincell{l}{SFH with\\ $N$ peaks} \\
			\hline
			1   & 8162                   & 8406 \\
			2   & 5786                   & 5868 \\
			3   & 3680                   & 3626 \\
			4   & 2446                   & 2388 \\
			5   & 1671                   & 1779 \\
			6   & 1484                   & 1162 \\
			\bottomrule
		\end{tabular}
	\end{table}

	\Tbl{TabPeakNum} indicates that $64.9\%$ of the SFHs have a better fit with two or more peaks.
	We examined the intrinsic peak numbers of SFHs in the \tng simulation and confirmed that the reconstructed multi-peaks SFH fraction aligns with the intrinsic fraction.
	However, these results appear to contradict the findings of \cite{Iyer2017} and \cite{Iyer2019}, who reported a fraction of multi-episode SFHs ranging from $10\%$ to $20\%$.
	They validated their method using hydro simulation and SAMs data, which derived a similar fraction from intrinsic peaks.

	The main purpose of \cite{Iyer2017} and \cite{Iyer2019} was to constrain the SFH through SED fitting.
	It is hard to extract information about all stages in a galaxy's life via SED.
	\rev{
		The accuracy of fitting the episodes of SFH is highly sensitive to the timing, duration, and strength of the star formation rate, all of which are challenging to reconstruct solely from SED photometry/spectroscopy.
		Due to the degeneracy of stellar population on spectrum and the noise, the reconstructed SFH from SED is always more smoothed-out compared to a direct fit to the original SFH.
		Another possibility is that the discrepancy arises from the method used to count the intrinsic peaks.
		However we speculate this possibility.
		Considering the intrinsic peak number, it is unlikely that the \tng simulation reproduces SFHs with significantly more peaks than those in the MUFASA simulation and SAMs.
	}
	We conducted tests using different sets for intrinsic peak counting, and the results are listed in Appendix \ref{app:peak}.
	In summary, \cite{Iyer2017} and \cite{Iyer2019} \rev{find less peaks, which is reasonable considering the nature of reconstructing SFHs from SED. Take advantages of direct fitting, }
	our approach attempts to distinguish more small peaks compared with the work in \cite{Iyer2017,Iyer2019}.
	This approach only concerns the performance of direct fitting to SFH.
	Whether the same performance can be achieved in observations and whether such capability can benefit studies are questions necessitate further exploration and analysis.


	\subsection{Timing and time scale of episodes}

\begin{figure}
		\includegraphics[width=1\linewidth]{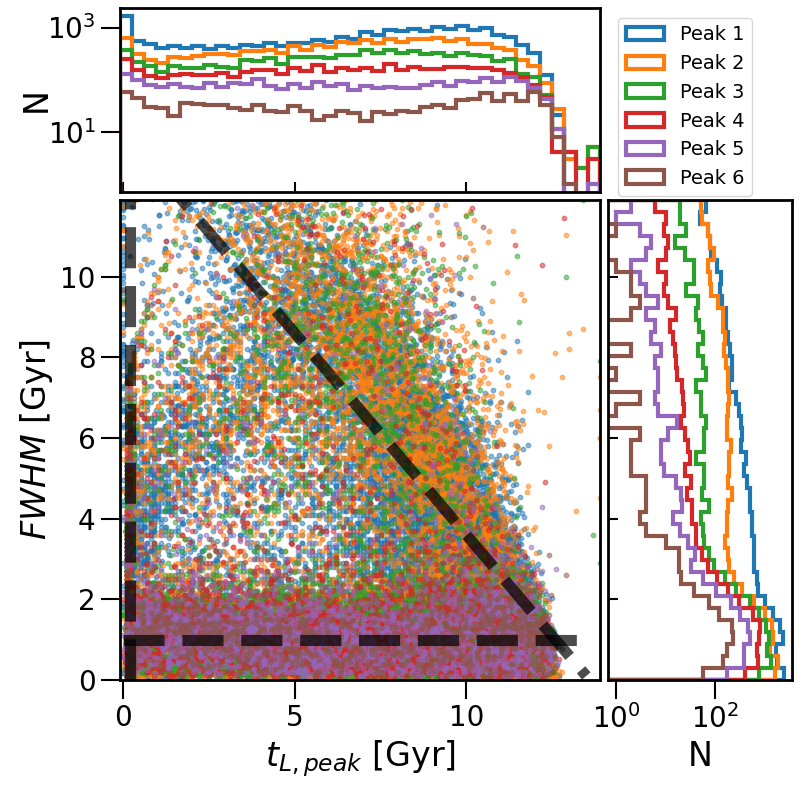}
		\caption{The position and width of peaks of SFHs.
			$t_{L,peak}$ is the lookback time of the peaks.
			$FWHM$ is the full width at half maximum of the peaks.
			In the middle panel, each point represents data of one peak.
			The upper panel shows the histogram of the peak positions,
			and the right panel shows the histogram of the peak widths.
			For each SFH, we sort the peaks by their height and distinguish the distribution of the peaks in different colors according to their order in the SFH.
			'Peak 1' is the strongest peak, 'Peak 2' is the second strongest peak, and so on.
			The three wide dashed lines in the figure correspond to $FWHM=13.6 - t_{L,peak}$, $t_{L, peak}=0$ and $FWHM=1$.
		}
		\label{FigPeaks}
	\end{figure}

	\Fig{FigPeaks} shows the distribution of peak positions and widths of SFHs.
	Each data point in the figure corresponds to one peak.
	In order to provide readers with a better understanding of the distribution of different intensities of peaks in the multiple-component SFH, we ranked the peaks within the same SFH according to their height and group peaks with same index as sub-samples.
	Scatters and histograms with different colors in \Fig{FigPeaks} represent the results of peaks with different index.
	However, we found that the position and width distributions of peaks across different orders are remarkably similar.
	Therefore, in all subsequent discussions, unless otherwise stated, the subject of the discussion is all peaks.

	The distribution of SFH peak position $t_{L,peak}$ reveals two conspicuous peaks. The first occurs at approximately $t_{L,peak} \approx 0$ Gyr, while the second arises at $t_{L,peak} \approx 10$ Gyr ($z\approx 1.8$).
	The peak occurring at $t \approx 10\ Gyr$ is consistent with the general trend of cosmic star formation rate, which found that the star formation rate density peaks at $z \approx 2$ \citep[e.g., ][]{Behroozi2013,Vogelsberger2014,Pillepich2018a}.
	The peak at $t \approx 0\ Gyr$ corresponds to SFH components keep rising.
	\cite{Iyer2017} likewise documented a significant fraction of recent peaks.

	The FWHM of most peaks concentrate in the region of $0\ Gyr < FWHM < 2\ Gyr$, and reach the maximum of distribution at $FWHM \simeq 1.2\ Gyr$.
	Many works emphasis the importance of SFR regulators in this timescale \citep[e.g., ][]{Tacchella2020,Matthee2019,Katsianis2020}, which correspond to various physical process like gas inflow and outflow, AGN feedbacks, galactic wind, giant molecular cloud life cycles, gas recycling, halo dynamical time-scale, etc. \cite[see Table C1 in][for a list of time-scale estimates]{Iyer2020}.
	The peak of FWHM distribution of $1.2\ Gyr$ coincides with the time scale of dark matter halo's dynamical timescale\citep{Mo2010,Lilly2013}.
	This suggests that the dynamical process of dark matter halos significantly affects the star formation process of galaxies.

	The FWHM distribution also includes a part of peaks featuring wider widths that correspond to long-term star formation processes, primarily associated with mergers\citep{Robertson2006, Jiang2008,BoylanKolchin2008,Hani2020}, metallicity evolution\citep{Torrey2018}, and galaxy quenching\citep{Sales2015, Nelson2018a,RodriguezMontero2019, Wright2019}.
	However, the amount of the wide peaks is $1$ to $2$ orders of magnitude smaller than that of the narrow peaks.
	It worth noting that, from the aspect of algorithm, our method is designed to get wider peaks to better represent the whole trend of the SFH of a galaxy.
	But the FWHM distribution in \Fig{FigPeaks} still prefers narrow peaks.
	Therefore, it demonstrates the importance of short time-scale processes in shaping the SFH of the galaxy.

	The joint distribution of SFH width and position shows three major branches, as indicated by the three thick dashed lines in \Fig{FigPeaks}.
	One of these branches is located vertically near $t_{L, peak}=0$, which corresponds to galaxies exhibiting a sustained increase in star formation rate until present time.
	Another branch manifests in the form of a horizontal line with $2\ Gyr$ width near $FWHM \simeq 1.2\ Gyr$.
	This branch suggests that a large number of star formation processes occurred within the timescale of $0$ to $2\ Gyr$.
	These processes do not exhibit any clear preference for specific timing of occurring.
	The third branch occurs like a negative correlation between $FWHM$ and $t_{L,peak}$.
	For these components, the summation of looking back time of peaks position and $FWHM$ of peak width is close to the length of the SFH they belong to, namely $t_{L,peak} + FWHM \approx SFH\ Length$ .
	These components are in line with the process of sustained SFR increase from the birth of the galaxy, succeeded by a decline at a specific point in time.
	In \Fig{FigPeaks}, the line $FWHM=13.6-t_{L,peak}$ is shown as a reference, not meaning the regression for these group of components.

\begin{figure}
		\includegraphics[width=1\linewidth]{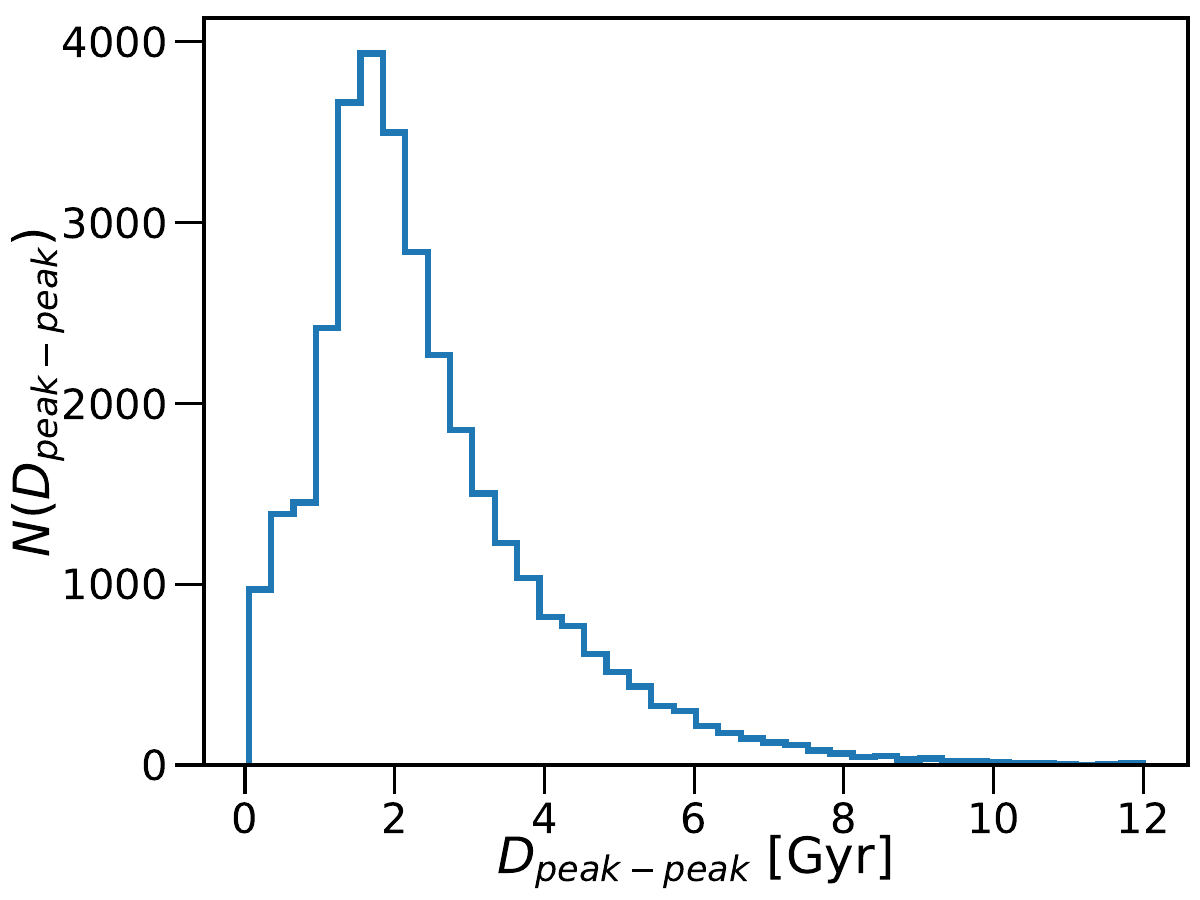}
		\caption{The histogram of separations between neighboring peaks.
		}
		\label{FigPeakSep}
	\end{figure}

	It is important to note that the bimodal distribution of peak positions does not imply the distance between adjacent peaks in a single SFH.
	Within a single SFH, it is found that adjacent peaks still tend to occur at shorter time intervals.
	This can be seen in Figure \ref{FigPeakSep}.
	The median peak separation, accompanied by a $1 \sigma$ scatter, is $2.07^{+1.76}_{-0.92}\ Gyr$.
	This value is significantly smaller than the one measured in the study by \cite{Iyer2019}.
	In their study, the peak spacing was determined to be $0.42^{+0.15}_{-0.10}t_{univ}$, where $t_{univ}$ is the age of the universe at the time of observation.
	Their estimate corresponds to $5.72^{+2.04}_{-1.36}\ Gyr$ in the context of this work.
	However, the SFH reconstruction performed by \cite{Iyer2019} used a Gaussian process, which is different from the methodology used in this work.
	In their study, \cite{Iyer2019} attempted to use the fewest possible number of time points, which resulted in the smoothing of the SFH and the omission of certain short-term fluctuations.
	Additionally, as mentioned in the previous section, the SED fitting process naturally leads to fewer peaks being found when reconstructing the SFH.
	As a consequence, larger intervals between peaks are expected.
	On the other hand, although similar in methodology, the results from \cite{Iyer2017} are not as convincing for comparison.
	According to Figure 15 in \cite{Iyer2017}, the median peak separation is about $0.07 t_{univ}$, which is much smaller than the value found in this work and in \cite{Iyer2019}.
	However, there are only a few tens of SFHs with two episodes in \cite{Iyer2017}, which reduces the statistical significance.
	In summary, comparing peak separation between different studies is challenging due to differences in data and fitting methodologies.
	These differences warrant further investigation.

	\subsection{Mass dependency}

	\begin{figure*}
		\includegraphics[width=1\linewidth]{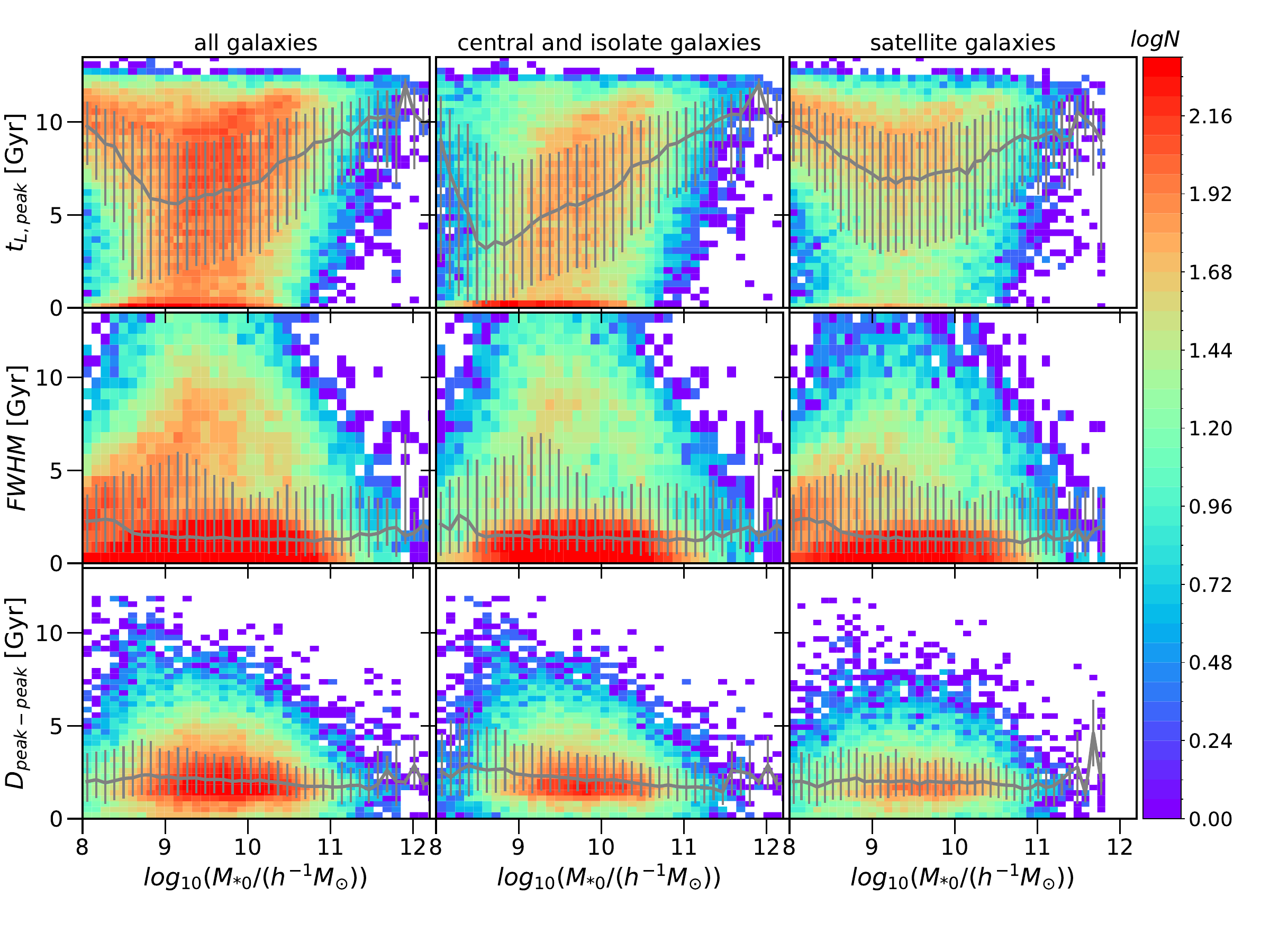}
		\caption{
			The dependence of SFH peak position (top row), FWHM (middle row) and neighboring peak separation on stellar mass at $z=0$.
			The colors correspond to logarithmic galaxy count in the corresponding SFH-properties-stellar-mass bins.
			The median values of the peak position, FWHM, or peak separation in different stellar mass bins are displayed by the gray lines.
			The error bars denote the range of the same value from $20\%$ to $80\%$ within the bin.
			The statistics for all galaxies are presented in the left row, those for central and isolated galaxies are in the middle row, while satellite galaxies are on the right row.
		}
		\label{FigMassDep}
	\end{figure*}

	The SFH of a galaxy is strongly associated with the galaxy properties.
	Consequently, it is reasonable to anticipate the presence of correlations between the extracted SFH features(peak position, $FWHM$ and peaks separation) and the galaxy properties, including stellar mass and color.
	Figure \ref{FigMassDep} illustrates the relationship between the stellar mass at $z=0$ and the SFH features.
	For each SFH peak, we record its position, $FWHM$ and the stellar mass of the galaxy it belongs to.
	Also, we select all neighboring peak pairs from all SFHs and record their separation and the stellar mass of the galaxy they belong to.
	Then we plot the 2D histograms of these SFH features and the stellar mass of corresponding galaxy.

	The peak position of SFH in a galaxy exhibits a U-shaped relationship with the galaxy's mass.
	The peak of SFH occurs earlier in galaxies with high and low masses, while it occurs later in galaxies with moderate masses.
	Generally, the formation of high-mass galaxies requires larger dark matter halos with deeper potential wells to attract more gas.
	The gas accretion into these strong potential wells triggers star formation earlier.
	As the galaxy evolves, mechanisms such as AGN feedback and metal enrichment gradually strengthen to suppress the process of stellar formation.
	This suppression effect is more pronounced in higher mass galaxies, preventing their star formation rate from surpassing the early levels in subsequent times.
	The combination of these mechanisms leads to the earlier appearance of the SFH peaks in higher mass galaxies.
	On the other hand, for galaxies at the lowest mass end, their limited gas reservoirs result in an early cessation of stellar formation, leading to smaller final masses.
	Consequently, the peak position of their SFH remains in the early stages.
	As for satellite galaxies, their stellar formation process stops earlier due to accretion and gas stripping from host galaxies, causing the peak position of SFH in moderate mass galaxies to shift towards earlier times

	The correlation between the width of SFH peak and galaxy mass is relatively weak.
	With the exception of galaxies at low and high mass ends, there is a slight decrease in the $FWHM$ as galaxy mass increases.
	Analyzing the joint distribution of FWHM and mass, we observe a reduction in the number of broader peaks with the  mass range of approximately $9 \lesssim log_{10}(M_{*}/(h^{-1}M_{\odot})) \lesssim 11$.
	This suggests a decreasing influence of the long time-scale star formation driving mechanisms on high-mass galaxies, implying that short time-scale process like feedbacks and gas recycling become more active in high-mass galaxies.
	Both at the high and low mass ends, the median $FWHM$ experiences an increase, primarily due to the decline in the number of most narrow peaks in the SFHs.
	The SFH of massive galaxies involves a large number of physical processes,
	resulting in a more complex SFH shape and more decomposable components.
	As a result, our method tends to exclude short time-scale components when there are too many pronounced components to be extracted.
	On the other hand, the evolution process in low-mass galaxies is less complex with fewer fluctuations, leading to fewer instances of fitting with narrow peaks.

	Excluding the highest and lowest mass galaxies, the peak separation of SFH decreases slightly with increasing galaxy mass.
	Based on the context of galaxy SFR rejuvenation \citep{Fang2012}, it can be inferred that larger the galaxy mass results in a shorter time required to return to a state of star formation.
	This is attributed to high-mass galaxies having stronger gas accretion rates and a higher chance of merging,
	facilitating the acquisition of necessary fuel for the star formation process.
	When distinguishing central/isolated galaxies from satellite galaxies, the correlation between peak separation and galaxy mass is weaker for satellite galaxies.
	We propose two reasons for this.
	On one hand, when satellite galaxies are influenced by central galaxies and experience gas stripping, the rise in their star formation rate is halted earlier, causing an earlier occurrence of the peak position in the star formation process.
	This suppressing effect is more pronounced in low-mass galaxies.
	Therefore, the larger peak separation of SFH at low mass is reduced more, resulting in a flatter relationship curve.
	On the other hand, the frequency of disturbance and stripping events for satellite galaxies is primarily dependent on the external environment rather than their own properties.
	This introduces more randomness and disturbance to their star formation process, thereby weakening the relationship between their SFH characteristics and their own stellar mass.

	\subsection{Color dependency}

\begin{figure*}
		\includegraphics[width=1\linewidth]{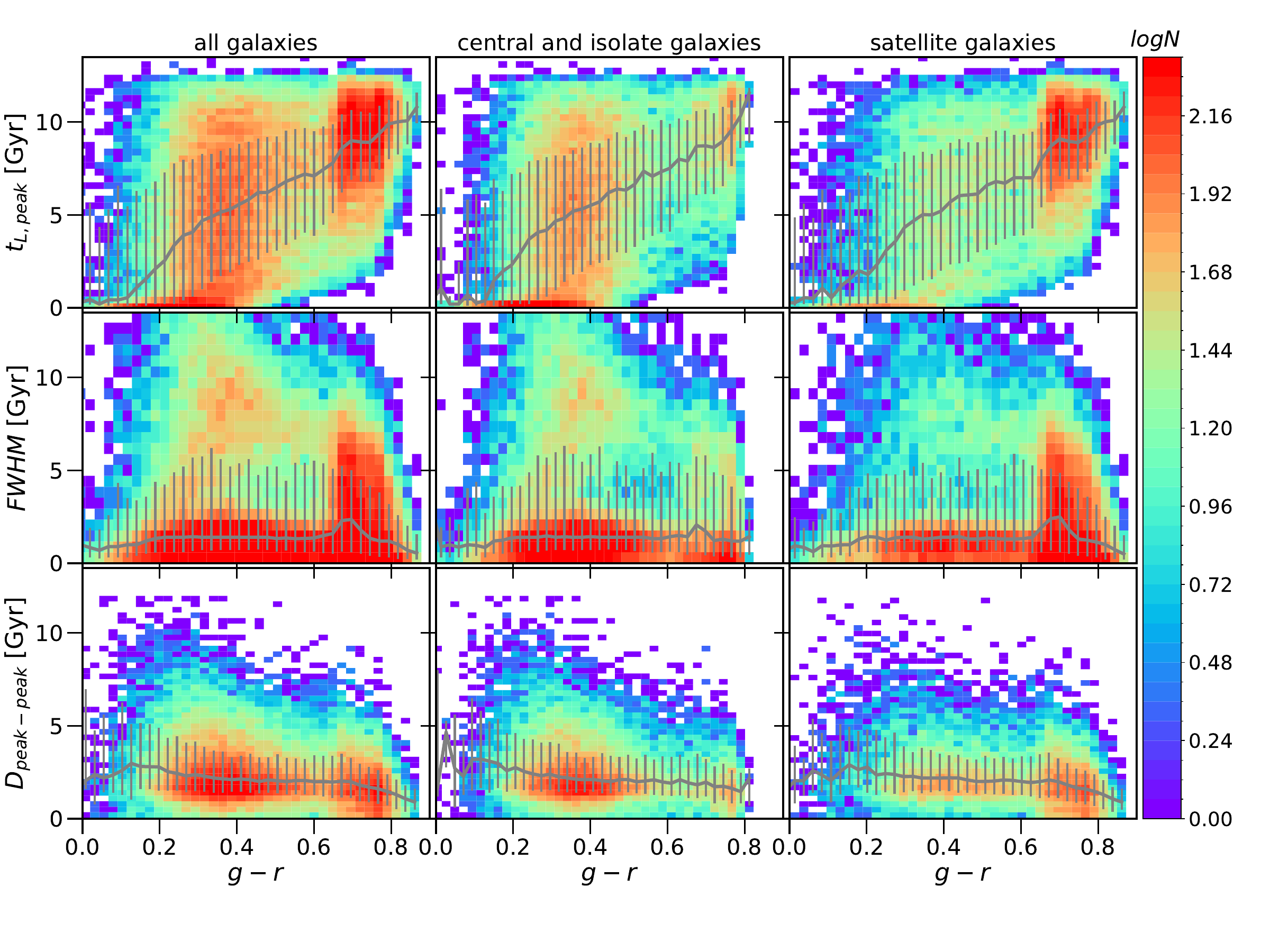}
		\caption{
			The dependence of SFH peak position (top row), FWHM (middle row) and neighboring peak separation on galaxy color $g-r$ at $z=0$.
			The colors on plots correspond to logarithmic galaxy count in the corresponding SFH-properties-color bins.
			The median values of the peak position, FWHM, or peak separation in different color bins are displayed by the gray lines.
			The error bars denote the range of the same value from $20\%$ to $80\%$ within the bin.
			The statistics for all galaxies are presented in the left row, those for central and isolated galaxies are in the middle row, while satellite galaxies are on the right row.
		}
		\label{FigColorDep}
	\end{figure*}

	Figure \ref{FigColorDep} illustrates the relationship between the galaxy color at $z=0$ and the SFH features.
	For each SFH peak, we record its position, $FWHM$ and the $g-r$ color of the galaxy it belongs to.
	Also, we select all neighboring peak pairs from all SFHs and record their separation and the $g-r$ color of the galaxy they belong to.
	Then we plot the 2D histograms of these SFH features(peak position, $FWHM$ and peaks separation) and the color of corresponding galaxies.

	According to Figure 14, there exists a clear linear correlation between the color of galaxies and the position of the SFH peak.
	It indicates that galaxies with a redder color tend to have earlier peaks,  which is naturally attributed to the fact that galaxies formed in earlier tend to have older stellar ages, thus displaying a redder color.

	The width of the SFH peaks exhibits a tri-modal distribution with the colors of galaxies.
	At the blue end with $g-r<0.2$, the peaks have smaller FWHM.
	This is because the most blue galaxies are still in the early stage of star formation, with their SFR rapidly increasing and their peak positions mainly located at a redshift of $0$.
	Therefore, the fitted peaks have narrower widths.
	When the $g-r$ value is within the range of $0.2$  to  $0.6$, the width of the peaks has no correlation with the colors of the galaxies.
	When the galaxy color exceeds $0.6$, a large number of wide peaks appear in satellite galaxies.
	These are SFH components in satellite galaxies whose star formation processes are regulated by self quenching.
	This result is consistent with previous work which summarized the quenching time scale in other simulations.
	\cite{Sales2015} found that the quenching time scale of satellite galaxies is is about $2$ to $5$ Gyr in Illustris.
	\cite{Wright2019} found that the quenching time scale is about $2.5\sim 3.3$ Gyr, and extending out to $10$ Gyr, in EAGLE.

	The peaks separation in the SFH also has a clear relationship with galaxy color.
	In redder galaxies, the stellar formation process is concentrated in earlier periods, resulting in a more concentrated peak position of star formation.

	\section{Conclusion}
	\label{sec:con}
	To summarize, this study has undertaken the following objectives and obtained statistical findings concerning the SFHs of galaxies from the \tng simulation:

	\begin{itemize}
		\item We employed a function combined by up to $6$ components, described by Gaussian or Gamma distributions, to construct fittings  for the SFHs in the \tng simulation.
		      Our fitting method can well recover the trends of the SFH, though the goodness of fit can depend on galaxy mass, color, history length, and the type of galaxy.

		\item We analyzed the SFHs  that are well-fitted by the Gaussian and Gamma distribution combinations.
		      Our finding indicates that low-mass galaxies, red galaxies, and satellite galaxies tend to have fewer components and to be majored by Gamma distribution, while high-mass galaxies, blue galaxies and central galaxies tend to have more components and to be majored by Gaussian distribution.

		\item We calculated the distribution of the position, width, and separation of SFH peaks.
		      The peak position distribution is concentrated at $t_L = 0\ Gyr$ and $t_L = 10\ Gyr$.
		      The width distribution of peaks is mainly concentrated at $FWHM \sim 0\ Gyr - 2\ Gyr$.
		      The joint distribution of peak position and width is concentrated in three regions, corresponding to different types of star formation modes.
		      In addition, we found that the separation between adjacent peaks in the TNG simulation is concentrated around $2\ Gyr$, much smaller than previously reconstructed SFHs from observations.

		\item We analyzed the relationships between three temporal attributes of the SFH, namely peak position, peak width, and peak spacing, and two properties of galaxies, namely stellar mass and galaxy color.
		      Our findings indicate that the peak position and peaks separation are related with stellar mass and galaxy color.
		      The peak width shows limited relationship with stellar mass and galaxy color. This is reasonable because the peak width reflects the time scale of basic physical processes which drives star formation.

	\end{itemize}

	Our model can accurately reproduce the trend of galaxies' stellar formation evolution.
	The method of multi-component decomposition can greatly help us understand the different physical driving factors of star formation in galaxies.
	Although more than half of the SFHs have low goodness of fit, this is mainly due to the excessive fluctuations in the SFHs, which makes it extremely difficult to fit them for all kinds of functional templates.
	Of course, the method we used is based on simulation data, the gaps between our method and SFH reconstruction methods suit for observational galaxies can not be neglected.
	Since it is difficult to give an absolute resolution criterion for the shape types and number of fitted components, our model will require better algorithms and more advanced computing power to determine these factors when applied to observational data, which will be further developed and improved in future work.

	\section*{Acknowledgements}
	\addcontentsline{toc}{section}{Acknowledgements}

	This work is supported by ``The Major Key Project of PCL''.
	YW thanks the support from Kunlun server in SYSU.
	The authors thanks the Illustris and Illustirs-TNG projects for providing the data.
	The authors thanks the referee for the valuable comments.

	\bibliographystyle{aasjournal}
	\bibliography{Lib}

\begin{thebibliography}{}
\expandafter\ifx\csname natexlab\endcsname\relax\def\natexlab#1{#1}\fi
\providecommand{\url}[1]{\href{#1}{#1}}
\providecommand{\dodoi}[1]{doi:~\href{http://doi.org/#1}{\nolinkurl{#1}}}
\providecommand{\doeprint}[1]{\href{http://ascl.net/#1}{\nolinkurl{http://ascl.net/#1}}}
\providecommand{\doarXiv}[1]{\href{https://arxiv.org/abs/#1}{\nolinkurl{https://arxiv.org/abs/#1}}}

\bibitem[{Abramson {et~al.}(2015)Abramson, Gladders, Dressler, Oemler, Poggianti, \& Vulcani}]{Abramson2015}
Abramson, L.~E., Gladders, M.~D., Dressler, A., {et~al.} 2015, \apjl, 801, L12, \dodoi{10.1088/2041-8205/801/1/L12}

\bibitem[{Behroozi {et~al.}(2010)Behroozi, Conroy, \& Wechsler}]{Behroozi2010}
Behroozi, P.~S., Conroy, C., \& Wechsler, R.~H. 2010, \apj, 717, 379, \dodoi{10.1088/0004-637X/717/1/379}

\bibitem[{Behroozi {et~al.}(2013)Behroozi, Wechsler, \& Conroy}]{Behroozi2013}
Behroozi, P.~S., Wechsler, R.~H., \& Conroy, C. 2013, \apj, 770, 57, \dodoi{10.1088/0004-637X/770/1/57}

\bibitem[{Bellstedt {et~al.}(2020)Bellstedt, Robotham, Driver, Thorne, Davies, Lagos, Stevens, Taylor, Baldry, Moffett, Hopkins, \& Phillipps}]{Bellstedt2020}
Bellstedt, S., Robotham, A. S.~G., Driver, S.~P., {et~al.} 2020, \mnras, 498, 5581, \dodoi{10.1093/mnras/staa2620}

\bibitem[{Ben{\'\i}tez-Llambay {et~al.}(2015)Ben{\'\i}tez-Llambay, Navarro, Abadi, Gottl{\"o}ber, Yepes, Hoffman, \& Steinmetz}]{BenitezLlambay2015}
Ben{\'\i}tez-Llambay, A., Navarro, J.~F., Abadi, M.~G., {et~al.} 2015, \mnras, 450, 4207, \dodoi{10.1093/mnras/stv925}

\bibitem[{Boylan-Kolchin {et~al.}(2008)Boylan-Kolchin, Ma, \& Quataert}]{BoylanKolchin2008}
Boylan-Kolchin, M., Ma, C.-P., \& Quataert, E. 2008, \mnras, 383, 93, \dodoi{10.1111/j.1365-2966.2007.12530.x}

\bibitem[{Buat {et~al.}(2008)Buat, Boissier, Burgarella, Takeuchi, Le~Floc'h, Marcillac, Huang, Nagashima, \& Enoki}]{Buat2008}
Buat, V., Boissier, S., Burgarella, D., {et~al.} 2008, \aap, 483, 107, \dodoi{10.1051/0004-6361:20078263}

\bibitem[{Carnall {et~al.}(2019)Carnall, Leja, Johnson, McLure, Dunlop, \& Conroy}]{Carnall2019}
Carnall, A.~C., Leja, J., Johnson, B.~D., {et~al.} 2019, \apj, 873, 44, \dodoi{10.3847/1538-4357/ab04a2}

\bibitem[{Carnall {et~al.}(2018)Carnall, McLure, Dunlop, \& Dav{\'e}}]{Carnall2018}
Carnall, A.~C., McLure, R.~J., Dunlop, J.~S., \& Dav{\'e}, R. 2018, \mnras, 480, 4379, \dodoi{10.1093/mnras/sty2169}

\bibitem[{Chauke {et~al.}(2018)Chauke, van~der Wel, Pacifici, Bezanson, Wu, Gallazzi, Noeske, Straatman, Mu{\~n}os-Mateos, Franx, Bari{\v{s}}i{\'c}, Bell, Brammer, Calhau, van Houdt, Labb{\'e}, Maseda, Muzzin, Rix, \& Sobral}]{Chauke2018}
Chauke, P., van~der Wel, A., Pacifici, C., {et~al.} 2018, \apj, 861, 13, \dodoi{10.3847/1538-4357/aac324}

\bibitem[{Choi {et~al.}(2014)Choi, Naab, Ostriker, Johansson, \& Moster}]{Choi2014}
Choi, E., Naab, T., Ostriker, J.~P., Johansson, P.~H., \& Moster, B.~P. 2014, \mnras, 442, 440, \dodoi{10.1093/mnras/stu874}

\bibitem[{Ciesla {et~al.}(2016)Ciesla, Boselli, Elbaz, Boissier, Buat, Charmandaris, Schreiber, B{\'e}thermin, Baes, Boquien, De~Looze, Fern{\'a}ndez-Ontiveros, Pappalardo, Spinoglio, \& Viaene}]{Ciesla2016}
Ciesla, L., Boselli, A., Elbaz, D., {et~al.} 2016, \aap, 585, A43, \dodoi{10.1051/0004-6361/201527107}

\bibitem[{Collaboration {et~al.}(2016)Collaboration, Ade, Aghanim, Arnaud, Ashdown, Aumont, Baccigalupi, Banday, Barreiro, Bartlett, Bartolo, Battaner, Battye, Benabed, Beno{\^i}t, Benoit-L{\'e}vy, Bernard, Bersanelli, Bielewicz, Bock, Bonaldi, Bonavera, Bond, Borrill, Bouchet, Boulanger, Bucher, Burigana, Butler, Calabrese, Cardoso, Catalano, Challinor, Chamballu, Chary, Chiang, Chluba, Christensen, Church, Clements, Colombi, Colombo, Combet, Coulais, Crill, Curto, Cuttaia, Danese, Davies, Davis, de~Bernardis, de~Rosa, de~Zotti, Delabrouille, D{\'e}sert, Di~Valentino, Dickinson, Diego, Dolag, Dole, Donzelli, Dor{\'e}, Douspis, Ducout, Dunkley, Dupac, Efstathiou, Elsner, En{\ss}lin, Eriksen, Farhang, Fergusson, Finelli, Forni, Frailis, Fraisse, Franceschi, Frejsel, Galeotta, Galli, Ganga, Gauthier, Gerbino, Ghosh, Giard, Giraud-H{\'e}raud, Giusarma, Gjerl{\o}w, Gonz{\'a}lez-Nuevo, G{\'o}rski, Gratton, Gregorio, Gruppuso, Gudmundsson, Hamann, Hansen, Hanson, Harrison, Helou, Henrot-Versill{\'e}, Hern{\'a}ndez-Monteagudo, Herranz, Hildebrandt, Hivon, Hobson, Holmes, Hornstrup, Hovest, Huang, Huffenberger, Hurier, Jaffe, Jaffe, Jones, Juvela, Keih{\"a}nen, Keskitalo, Kisner, Kneissl, Knoche, Knox, Kunz, Kurki-Suonio, Lagache, L{\"a}hteenm{\"a}ki, Lamarre, Lasenby, Lattanzi, Lawrence, Leahy, Leonardi, Lesgourgues, Levrier, Lewis, Liguori, Lilje, Linden-V{\o}rnle, L{\'o}pez-Caniego, Lubin, Mac{\'{\i}}as-P{\'e}rez, Maggio, Maino, Mandolesi, Mangilli, Marchini, Maris, Martin, Martinelli, Mart{\'{\i}}nez-Gonz{\'a}lez, Masi, Matarrese, McGehee, Meinhold, Melchiorri, Melin, Mendes, Mennella, Migliaccio, Millea, Mitra, Miville-Desch{\^e}nes, Moneti, Montier, Morgante, Mortlock, Moss, Munshi, Murphy, Naselsky, Nati, Natoli, Netterfield, N{\o}rgaard-Nielsen, Noviello, Novikov, Novikov, Oxborrow, Paci, Pagano, Pajot, Paladini, Paoletti, Partridge, Pasian, Patanchon, Pearson, Perdereau, Perotto, Perrotta, Pettorino, Piacentini, Piat, Pierpaoli, Pietrobon, Plaszczynski, Pointecouteau, Polenta, Popa, Pratt, Pr{\'e}zeau, Prunet, Puget, Rachen, Reach, Rebolo, Reinecke, Remazeilles, Renault, Renzi, Ristorcelli, Rocha, Rosset, Rossetti, Roudier, Rouill{\'e d'Orfeuil}, Rowan-Robinson, Rubi{\~n}o-Mart{\'{\i}}n, Rusholme, Said, Salvatelli, Salvati, Sandri, Santos, Savelainen, Savini, Scott, Seiffert, Serra, Shellard, Spencer, Spinelli, Stolyarov, Stompor, Sudiwala, Sunyaev, Sutton, Suur-Uski, Sygnet, Tauber, Terenzi, Toffolatti, Tomasi, Tristram, Trombetti, Tucci, Tuovinen, T{\"u}rler, Umana, Valenziano, Valiviita, Van~Tent, Vielva, Villa, Wade, Wandelt, Wehus, White, White, Wilkinson, Yvon, Zacchei, \& Zonca}]{Collaboration2016}
Collaboration, P., Ade, P. A.~R., Aghanim, N., {et~al.} 2016, \aap, 594, A13, \dodoi{10.1051/0004-6361/201525830}

\bibitem[{Conroy(2013)}]{Conroy2013}
Conroy, C. 2013, \araa, 51, 393, \dodoi{10.1146/annurev-astro-082812-141017}

\bibitem[{Conroy {et~al.}(2014)Conroy, Graves, \& van Dokkum}]{Conroy2014}
Conroy, C., Graves, G.~J., \& van Dokkum, P.~G. 2014, \apj, 780, 33, \dodoi{10.1088/0004-637X/780/1/33}

\bibitem[{Crain {et~al.}(2015)Crain, Schaye, Bower, Furlong, Schaller, Theuns, Dalla~Vecchia, Frenk, McCarthy, Helly, Jenkins, Rosas-Guevara, White, \& Trayford}]{Crain2015}
Crain, R.~A., Schaye, J., Bower, R.~G., {et~al.} 2015, \mnras, 450, 1937, \dodoi{10.1093/mnras/stv725}

\bibitem[{Dav{\'e} {et~al.}(2017)Dav{\'e}, Rafieferantsoa, Thompson, \& Hopkins}]{Dave2017}
Dav{\'e}, R., Rafieferantsoa, M.~H., Thompson, R.~J., \& Hopkins, P.~F. 2017, \mnras, 467, 115, \dodoi{10.1093/mnras/stx108}

\bibitem[{Diemer {et~al.}(2017)Diemer, Sparre, Abramson, \& Torrey}]{Diemer2017}
Diemer, B., Sparre, M., Abramson, L.~E., \& Torrey, P. 2017, \apj, 839, 26, \dodoi{10.3847/1538-4357/aa68e5}

\bibitem[{Digby {et~al.}(2019)Digby, Navarro, Fattahi, Simpson, Oman, Gomez, Frenk, Grand, \& Pakmor}]{Digby2019}
Digby, R., Navarro, J.~F., Fattahi, A., {et~al.} 2019, \mnras, 485, 5423, \dodoi{10.1093/mnras/stz745}

\bibitem[{Donnari {et~al.}(2019)Donnari, Pillepich, Nelson, Vogelsberger, Genel, Weinberger, Marinacci, Springel, \& Hernquist}]{Donnari2019}
Donnari, M., Pillepich, A., Nelson, D., {et~al.} 2019, \mnras, 485, 4817, \dodoi{10.1093/mnras/stz712}

\bibitem[{Fang {et~al.}(2012)Fang, Faber, Salim, Graves, \& Rich}]{Fang2012}
Fang, J.~J., Faber, S.~M., Salim, S., Graves, G.~J., \& Rich, R.~M. 2012, \apj, 761, 23, \dodoi{10.1088/0004-637X/761/1/23}

\bibitem[{Finlator {et~al.}(2007)Finlator, Dav{\'e}, \& Oppenheimer}]{Finlator2007}
Finlator, K., Dav{\'e}, R., \& Oppenheimer, B.~D. 2007, \mnras, 376, 1861, \dodoi{10.1111/j.1365-2966.2007.11578.x}

\bibitem[{Gallazzi {et~al.}(2005)Gallazzi, Charlot, Brinchmann, White, \& Tremonti}]{Gallazzi2005}
Gallazzi, A., Charlot, S., Brinchmann, J., White, S. D.~M., \& Tremonti, C.~A. 2005, \mnras, 362, 41, \dodoi{10.1111/j.1365-2966.2005.09321.x}

\bibitem[{Gavazzi {et~al.}(2002)Gavazzi, Bonfanti, Sanvito, Boselli, \& Scodeggio}]{Gavazzi2002}
Gavazzi, G., Bonfanti, C., Sanvito, G., Boselli, A., \& Scodeggio, M. 2002, \apj, 576, 135, \dodoi{10.1086/341730}

\bibitem[{Gladders {et~al.}(2013)Gladders, Oemler, Dressler, Poggianti, Vulcani, \& Abramson}]{Gladders2013}
Gladders, M.~D., Oemler, A., Dressler, A., {et~al.} 2013, \apj, 770, 64, \dodoi{10.1088/0004-637X/770/1/64}

\bibitem[{Hahn {et~al.}(2019)Hahn, Starkenburg, Choi, Dav{\'e}, Dickey, Geha, Genel, Hayward, Maller, Mandyam, Pandya, Popping, Rafieferantsoa, Somerville, \& Tinker}]{Hahn2019}
Hahn, C., Starkenburg, T.~K., Choi, E., {et~al.} 2019, \apj, 872, 160, \dodoi{10.3847/1538-4357/aafedd}

\bibitem[{Hani {et~al.}(2020)Hani, Gosain, Ellison, Patton, \& Torrey}]{Hani2020}
Hani, M.~H., Gosain, H., Ellison, S.~L., Patton, D.~R., \& Torrey, P. 2020, \mnras, 493, 3716, \dodoi{10.1093/mnras/staa459}

\bibitem[{Hopkins {et~al.}(2018)Hopkins, Wetzel, Kere{\v{s}}, Faucher-Gigu{\`e}re, Quataert, Boylan-Kolchin, Murray, Hayward, Garrison-Kimmel, Hummels, Feldmann, Torrey, Ma, Angl{\'e}s-Alc{\'a}zar, Su, Orr, Schmitz, Escala, Sanderson, Grudi{\'c}, Hafen, Kim, Fitts, Bullock, Wheeler, Chan, Elbert, \& Narayanan}]{Hopkins2018}
Hopkins, P.~F., Wetzel, A., Kere{\v{s}}, D., {et~al.} 2018, \mnras, 480, 800, \dodoi{10.1093/mnras/sty1690}

\bibitem[{Iyer \& Gawiser(2017)}]{Iyer2017}
Iyer, K., \& Gawiser, E. 2017, \apj, 838, 127, \dodoi{10.3847/1538-4357/aa63f0}

\bibitem[{{Iyer} {et~al.}(2019){Iyer}, {Gawiser}, {Faber}, {Ferguson}, {Kartaltepe}, {Koekemoer}, {Pacifici}, \& {Somerville}}]{Iyer2019}
{Iyer}, K.~G., {Gawiser}, E., {Faber}, S.~M., {et~al.} 2019, \apj, 879, 116, \dodoi{10.3847/1538-4357/ab2052}

\bibitem[{Iyer {et~al.}(2020)Iyer, Tacchella, Genel, Hayward, Hernquist, Brooks, Caplar, Dav{\'e}, Diemer, Forbes, Gawiser, Somerville, \& Starkenburg}]{Iyer2020}
Iyer, K.~G., Tacchella, S., Genel, S., {et~al.} 2020, \mnras, 498, 430, \dodoi{10.1093/mnras/staa2150}

\bibitem[{Janowiecki {et~al.}(2017)Janowiecki, Salzer, van Zee, Rosenberg, \& Skillman}]{Janowiecki2017}
Janowiecki, S., Salzer, J.~J., van Zee, L., Rosenberg, J.~L., \& Skillman, E. 2017, \apj, 836, 128, \dodoi{10.3847/1538-4357/836/1/128}

\bibitem[{Jiang {et~al.}(2008)Jiang, Jing, Faltenbacher, Lin, \& Li}]{Jiang2008}
Jiang, C.~Y., Jing, Y.~P., Faltenbacher, A., Lin, W.~P., \& Li, C. 2008, \apj, 675, 1095, \dodoi{10.1086/526412}

\bibitem[{Jim{\'e}nez-L{\'o}pez {et~al.}(2022)Jim{\'e}nez-L{\'o}pez, Corcho-Caballero, Zamora, \& Ascasibar}]{JimenezLopez2022}
Jim{\'e}nez-L{\'o}pez, D., Corcho-Caballero, P., Zamora, S., \& Ascasibar, Y. 2022, \aap, 662, A1, \dodoi{10.1051/0004-6361/202141338}

\bibitem[{Johnson {et~al.}(2013)Johnson, Weisz, Dalcanton, Johnson, Dale, Dolphin, Gil~de Paz, Kennicutt, Lee, Skillman, Boquien, \& Williams}]{Johnson2013}
Johnson, B.~D., Weisz, D.~R., Dalcanton, J.~J., {et~al.} 2013, \apj, 772, 8, \dodoi{10.1088/0004-637X/772/1/8}

\bibitem[{Joshi {et~al.}(2021)Joshi, Pillepich, Nelson, Zinger, Marinacci, Springel, Vogelsberger, \& Hernquist}]{Joshi2021}
Joshi, G.~D., Pillepich, A., Nelson, D., {et~al.} 2021, \mnras, 508, arXiv:2101.12226, \dodoi{10.1093/mnras/stab2573}

\bibitem[{Katsianis {et~al.}(2020)Katsianis, Xu, Yang, Luo, Cui, Dav{\'e}, Lagos, Zheng, \& Zhao}]{Katsianis2020}
Katsianis, A., Xu, H., Yang, X., {et~al.} 2020, arXiv e-prints, arXiv:2010.08173.
\newblock \doarXiv{2010.08173}

\bibitem[{Kauffmann(2014)}]{Kauffmann2014}
Kauffmann, G. 2014, \mnras, 441, 2717, \dodoi{10.1093/mnras/stu752}

\bibitem[{Kelson(2014)}]{Kelson2014}
Kelson, D.~D. 2014, arXiv e-prints, arXiv:1406.5191.
\newblock \doarXiv{1406.5191}

\bibitem[{Lee {et~al.}(2010)Lee, Ferguson, Somerville, Wiklind, \& Giavalisco}]{Lee2010}
Lee, S.-K., Ferguson, H.~C., Somerville, R.~S., Wiklind, T., \& Giavalisco, M. 2010, \apj, 725, 1644, \dodoi{10.1088/0004-637X/725/2/1644}

\bibitem[{Leja {et~al.}(2019)Leja, Carnall, Johnson, Conroy, \& Speagle}]{Leja2019}
Leja, J., Carnall, A.~C., Johnson, B.~D., Conroy, C., \& Speagle, J.~S. 2019, \apj, 876, 3, \dodoi{10.3847/1538-4357/ab133c}

\bibitem[{Leja {et~al.}(2017)Leja, Johnson, Conroy, van Dokkum, \& Byler}]{Leja2017}
Leja, J., Johnson, B.~D., Conroy, C., van Dokkum, P.~G., \& Byler, N. 2017, \apj, 837, 170, \dodoi{10.3847/1538-4357/aa5ffe}

\bibitem[{Lilly {et~al.}(2013)Lilly, Carollo, Pipino, Renzini, \& Peng}]{Lilly2013}
Lilly, S.~J., Carollo, C.~M., Pipino, A., Renzini, A., \& Peng, Y. 2013, \apj, 772, 119, \dodoi{10.1088/0004-637X/772/2/119}

\bibitem[{Lower {et~al.}(2020)Lower, Narayanan, Leja, Johnson, Conroy, \& Dav{\'e}}]{Lower2020}
Lower, S., Narayanan, D., Leja, J., {et~al.} 2020, \apj, 904, 33, \dodoi{10.3847/1538-4357/abbfa7}

\bibitem[{Lu {et~al.}(2016)Lu, Benson, Mao, Tonnesen, Peter, Wetzel, Boylan-Kolchin, \& Wechsler}]{Lu2016}
Lu, Y., Benson, A., Mao, Y.-Y., {et~al.} 2016, \apj, 830, 59, \dodoi{10.3847/0004-637X/830/2/59}

\bibitem[{Lu {et~al.}(2015)Lu, Mo, Lu, Katz, Weinberg, van~den Bosch, \& Yang}]{Lu2015}
Lu, Z., Mo, H.~J., Lu, Y., {et~al.} 2015, \mnras, 450, 1604, \dodoi{10.1093/mnras/stv667}

\bibitem[{Maraston {et~al.}(2010)Maraston, Pforr, Renzini, Daddi, Dickinson, Cimatti, \& Tonini}]{Maraston2010}
Maraston, C., Pforr, J., Renzini, A., {et~al.} 2010, \mnras, 407, 830, \dodoi{10.1111/j.1365-2966.2010.16973.x}

\bibitem[{Marinacci {et~al.}(2018)Marinacci, Vogelsberger, Pakmor, Torrey, Springel, Hernquist, Nelson, Weinberger, Pillepich, Naiman, \& Genel}]{Marinacci2018}
Marinacci, F., Vogelsberger, M., Pakmor, R., {et~al.} 2018, \mnras, 480, 5113, \dodoi{10.1093/mnras/sty2206}

\bibitem[{Matthee \& Schaye(2019)}]{Matthee2019}
Matthee, J., \& Schaye, J. 2019, \mnras, 484, 915, \dodoi{10.1093/mnras/stz030}

\bibitem[{Mo {et~al.}(2010)Mo, van~den Bosch, \& White}]{Mo2010}
Mo, H., van~den Bosch, F.~C., \& White, S. 2010, Galaxy Formation and Evolution (Cambridge University Press).
\newblock \url{https://ui.adsabs.harvard.edu/abs/2010gfe..book.....M}

\bibitem[{Naiman {et~al.}(2018)Naiman, Pillepich, Springel, Ramirez-Ruiz, Torrey, Vogelsberger, Pakmor, Nelson, Marinacci, Hernquist, Weinberger, \& Genel}]{Naiman2018}
Naiman, J.~P., Pillepich, A., Springel, V., {et~al.} 2018, \mnras, 477, 1206, \dodoi{10.1093/mnras/sty618}

\bibitem[{Nelson {et~al.}(2018{\natexlab{a}})Nelson, Pillepich, Springel, Weinberger, Hernquist, Pakmor, Genel, Torrey, Vogelsberger, Kauffmann, Marinacci, \& Naiman}]{Nelson2018}
Nelson, D., Pillepich, A., Springel, V., {et~al.} 2018{\natexlab{a}}, \mnras, 475, 624, \dodoi{10.1093/mnras/stx3040}

\bibitem[{Nelson {et~al.}(2018{\natexlab{b}})Nelson, Kauffmann, Pillepich, Genel, Springel, Pakmor, Hernquist, Weinberger, Torrey, Vogelsberger, \& Marinacci}]{Nelson2018a}
Nelson, D., Kauffmann, G., Pillepich, A., {et~al.} 2018{\natexlab{b}}, \mnras, 477, 450, \dodoi{10.1093/mnras/sty656}

\bibitem[{Nelson {et~al.}(2019)Nelson, Springel, Pillepich, Rodriguez-Gomez, Torrey, Genel, Vogelsberger, Pakmor, Marinacci, Weinberger, Kelley, Lovell, Diemer, \& Hernquist}]{Nelson2019}
Nelson, D., Springel, V., Pillepich, A., {et~al.} 2019, Computational Astrophysics and Cosmology, 6, 2, \dodoi{10.1186/s40668-019-0028-x}

\bibitem[{Ocvirk {et~al.}(2006)Ocvirk, Pichon, Lan{\c{c}}on, \& Thi{\'e}baut}]{Ocvirk2006}
Ocvirk, P., Pichon, C., Lan{\c{c}}on, A., \& Thi{\'e}baut, E. 2006, \mnras, 365, 46, \dodoi{10.1111/j.1365-2966.2005.09182.x}

\bibitem[{Pacifici {et~al.}(2012)Pacifici, Charlot, Blaizot, \& Brinchmann}]{Pacifici2012}
Pacifici, C., Charlot, S., Blaizot, J., \& Brinchmann, J. 2012, \mnras, 421, 2002, \dodoi{10.1111/j.1365-2966.2012.20431.x}

\bibitem[{Pacifici {et~al.}(2016)Pacifici, Kassin, Weiner, Holden, Gardner, Faber, Ferguson, Koo, Primack, Bell, Dekel, Gawiser, Giavalisco, Rafelski, Simons, Barro, Croton, Dav{\'e}, Fontana, Grogin, Koekemoer, Lee, Salmon, Somerville, \& Behroozi}]{Pacifici2016}
Pacifici, C., Kassin, S.~A., Weiner, B.~J., {et~al.} 2016, \apj, 832, 79, \dodoi{10.3847/0004-637X/832/1/79}

\bibitem[{Panter {et~al.}(2007)Panter, Jimenez, Heavens, \& Charlot}]{Panter2007}
Panter, B., Jimenez, R., Heavens, A.~F., \& Charlot, S. 2007, \mnras, 378, 1550, \dodoi{10.1111/j.1365-2966.2007.11909.x}

\bibitem[{Papovich {et~al.}(2011)Papovich, Finkelstein, Ferguson, Lotz, \& Giavalisco}]{Papovich2011}
Papovich, C., Finkelstein, S.~L., Ferguson, H.~C., Lotz, J.~M., \& Giavalisco, M. 2011, \mnras, 412, 1123, \dodoi{10.1111/j.1365-2966.2010.17965.x}

\bibitem[{Pillepich {et~al.}(2018{\natexlab{a}})Pillepich, Nelson, Hernquist, Springel, Pakmor, Torrey, Weinberger, Genel, Naiman, Marinacci, \& Vogelsberger}]{Pillepich2018}
Pillepich, A., Nelson, D., Hernquist, L., {et~al.} 2018{\natexlab{a}}, \mnras, 475, 648, \dodoi{10.1093/mnras/stx3112}

\bibitem[{Pillepich {et~al.}(2018{\natexlab{b}})Pillepich, Springel, Nelson, Genel, Naiman, Pakmor, Hernquist, Torrey, Vogelsberger, Weinberger, \& Marinacci}]{Pillepich2018a}
Pillepich, A., Springel, V., Nelson, D., {et~al.} 2018{\natexlab{b}}, \mnras, 473, 4077, \dodoi{10.1093/mnras/stx2656}

\bibitem[{Reddy {et~al.}(2012)Reddy, Pettini, Steidel, Shapley, Erb, \& Law}]{Reddy2012}
Reddy, N.~A., Pettini, M., Steidel, C.~C., {et~al.} 2012, \apj, 754, 25, \dodoi{10.1088/0004-637X/754/1/25}

\bibitem[{Robertson {et~al.}(2006)Robertson, Bullock, Cox, Di~Matteo, Hernquist, Springel, \& Yoshida}]{Robertson2006}
Robertson, B., Bullock, J.~S., Cox, T.~J., {et~al.} 2006, \apj, 645, 986, \dodoi{10.1086/504412}

\bibitem[{{Rodriguez-Gomez} {et~al.}(2015){Rodriguez-Gomez}, {Genel}, {Vogelsberger}, {Sijacki}, {Pillepich}, {Sales}, {Torrey}, {Snyder}, {Nelson}, {Springel}, {Ma}, \& {Hernquist}}]{Rodriguez-Gomez2015}
{Rodriguez-Gomez}, V., {Genel}, S., {Vogelsberger}, M., {et~al.} 2015, \mnras, 449, 49, \dodoi{10.1093/mnras/stv264}

\bibitem[{Rodr{\'\i}guez~Montero {et~al.}(2019)Rodr{\'\i}guez~Montero, Dav{\'e}, Wild, Angl{\'e}s-Alc{\'a}zar, \& Narayanan}]{RodriguezMontero2019}
Rodr{\'\i}guez~Montero, F., Dav{\'e}, R., Wild, V., Angl{\'e}s-Alc{\'a}zar, D., \& Narayanan, D. 2019, \mnras, 490, 2139, \dodoi{10.1093/mnras/stz2580}

\bibitem[{Sales {et~al.}(2015)Sales, Vogelsberger, Genel, Torrey, Nelson, Rodriguez-Gomez, Wang, Pillepich, Sijacki, Springel, \& Hernquist}]{Sales2015}
Sales, L.~V., Vogelsberger, M., Genel, S., {et~al.} 2015, \mnras, 447, L6, \dodoi{10.1093/mnrasl/slu173}

\bibitem[{Schaye {et~al.}(2015)}]{Schaye2015}
Schaye, J., {et~al.} 2015, \mnras, 446, 521, \dodoi{10.1093/mnras/stu2058}

\bibitem[{Schreiber {et~al.}(2018)Schreiber, Glazebrook, Nanayakkara, Kacprzak, Labb{\'e}, Oesch, Yuan, Tran, Papovich, Spitler, \& Straatman}]{Schreiber2018}
Schreiber, C., Glazebrook, K., Nanayakkara, T., {et~al.} 2018, \aap, 618, A85, \dodoi{10.1051/0004-6361/201833070}

\bibitem[{Simha {et~al.}(2014)Simha, Weinberg, Conroy, Dave, Fardal, Katz, \& Oppenheimer}]{Simha2014}
Simha, V., Weinberg, D.~H., Conroy, C., {et~al.} 2014, arXiv e-prints, arXiv:1404.0402, \dodoi{10.48550/arXiv.1404.0402}

\bibitem[{Somerville \& Dav{\'e}(2015)}]{Somerville2015}
Somerville, R.~S., \& Dav{\'e}, R. 2015, \araa, 53, 51, \dodoi{10.1146/annurev-astro-082812-140951}

\bibitem[{Springel(2010)}]{springel2010}
Springel, V. 2010, \mnras, 401, 791, \dodoi{10.1111/j.1365-2966.2009.15715.x}

\bibitem[{Springel {et~al.}(2001)Springel, Yoshida, \& White}]{Springel2001a}
Springel, V., Yoshida, N., \& White, S. 2001, \na, 6, 79, \dodoi{10.1016/S1384-1076(01)00042-2}

\bibitem[{Springel {et~al.}(2018)Springel, Pakmor, Pillepich, Weinberger, Nelson, Hernquist, Vogelsberger, Genel, Torrey, Marinacci, \& Naiman}]{Springel2018}
Springel, V., Pakmor, R., Pillepich, A., {et~al.} 2018, \mnras, 475, 676, \dodoi{10.1093/mnras/stx3304}

\bibitem[{Tacchella {et~al.}(2020)Tacchella, Forbes, \& Caplar}]{Tacchella2020}
Tacchella, S., Forbes, J.~C., \& Caplar, N. 2020, \mnras, 497, 698, \dodoi{10.1093/mnras/staa1838}

\bibitem[{Telles \& Melnick(2018)}]{Telles2018}
Telles, E., \& Melnick, J. 2018, \aap, 615, A55, \dodoi{10.1051/0004-6361/201732275}

\bibitem[{Thomas {et~al.}(2005)Thomas, Maraston, Bender, \& Mendes~de Oliveira}]{Thomas2005}
Thomas, D., Maraston, C., Bender, R., \& Mendes~de Oliveira, C. 2005, \apj, 621, 673, \dodoi{10.1086/426932}

\bibitem[{Tojeiro {et~al.}(2007)Tojeiro, Heavens, Jimenez, \& Panter}]{Tojeiro2007}
Tojeiro, R., Heavens, A.~F., Jimenez, R., \& Panter, B. 2007, \mnras, 381, 1252, \dodoi{10.1111/j.1365-2966.2007.12323.x}

\bibitem[{Torrey {et~al.}(2018)Torrey, Vogelsberger, Hernquist, McKinnon, Marinacci, Simcoe, Springel, Pillepich, Naiman, Pakmor, Weinberger, Nelson, \& Genel}]{Torrey2018}
Torrey, P., Vogelsberger, M., Hernquist, L., {et~al.} 2018, \mnras, 477, L16, \dodoi{10.1093/mnrasl/sly031}

\bibitem[{Vogelsberger {et~al.}(2014)Vogelsberger, Genel, Springel, Torrey, Sijacki, Xu, Snyder, Nelson, \& Hernquist}]{Vogelsberger2014}
Vogelsberger, M., Genel, S., Springel, V., {et~al.} 2014, \mnras, 444, 1518, \dodoi{10.1093/mnras/stu1536}

\bibitem[{Walcher {et~al.}(2011)Walcher, Groves, Budav{\'a}ri, \& Dale}]{Walcher2011}
Walcher, J., Groves, B., Budav{\'a}ri, T., \& Dale, D. 2011, \apss, 331, 1, \dodoi{10.1007/s10509-010-0458-z}

\bibitem[{Weinberger {et~al.}(2017)Weinberger, Springel, Hernquist, Pillepich, Marinacci, Pakmor, Nelson, Genel, Vogelsberger, Naiman, \& Torrey}]{Weinberger2017}
Weinberger, R., Springel, V., Hernquist, L., {et~al.} 2017, \mnras, 465, 3291, \dodoi{10.1093/mnras/stw2944}

\bibitem[{Wetzel {et~al.}(2016)Wetzel, Hopkins, Kim, Faucher-Gigu{\`e}re, Kere{\v{s}}, \& Quataert}]{Wetzel2016}
Wetzel, A.~R., Hopkins, P.~F., Kim, J.-h., {et~al.} 2016, \apjl, 827, L23, \dodoi{10.3847/2041-8205/827/2/L23}

\bibitem[{Wright {et~al.}(2019)Wright, Lagos, Davies, Power, Trayford, \& Wong}]{Wright2019}
Wright, R.~J., Lagos, C. d.~P., Davies, L. J.~M., {et~al.} 2019, \mnras, 487, 3740, \dodoi{10.1093/mnras/stz1410}

\bibitem[{Wuyts {et~al.}(2011)Wuyts, F{\"o}rster~Schreiber, van~der Wel, Magnelli, Guo, Genzel, Lutz, Aussel, Barro, Berta, Cava, Graci{\'a}-Carpio, Hathi, Huang, Kocevski, Koekemoer, Lee, Le~Floc'h, McGrath, Nordon, Popesso, Pozzi, Riguccini, Rodighiero, Saintonge, \& Tacconi}]{Wuyts2011}
Wuyts, S., F{\"o}rster~Schreiber, N.~M., van~der Wel, A., {et~al.} 2011, \apj, 742, 96, \dodoi{10.1088/0004-637X/742/2/96}

\bibitem[{Zhou {et~al.}(2020{\natexlab{a}})Zhou, Mo, Li, Boquien, \& Rossi}]{Zhou2020a}
Zhou, S., Mo, H.~J., Li, C., Boquien, M., \& Rossi, G. 2020{\natexlab{a}}, \mnras, 497, 4753, \dodoi{10.1093/mnras/staa2337}

\bibitem[{Zhou {et~al.}(2020{\natexlab{b}})Zhou, Zhu, Wang, \& Feng}]{Zhou2020}
Zhou, Z.-B., Zhu, W., Wang, Y., \& Feng, L.-L. 2020{\natexlab{b}}, \apj, 895, 92, \dodoi{10.3847/1538-4357/ab8d32}

\end{thebibliography}

	\appendix

	\section{Time bin resolution in SFH reconstruction}
	\label{app:timebin}

	\begin{figure*}
		\includegraphics[width=0.33\linewidth]{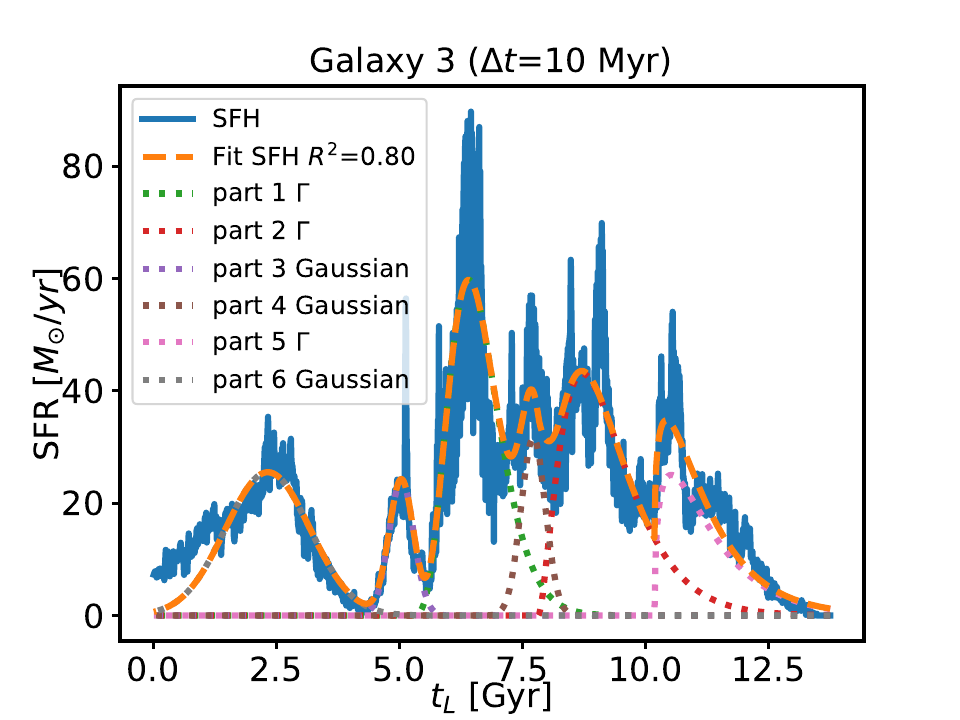}
		\includegraphics[width=0.33\linewidth]{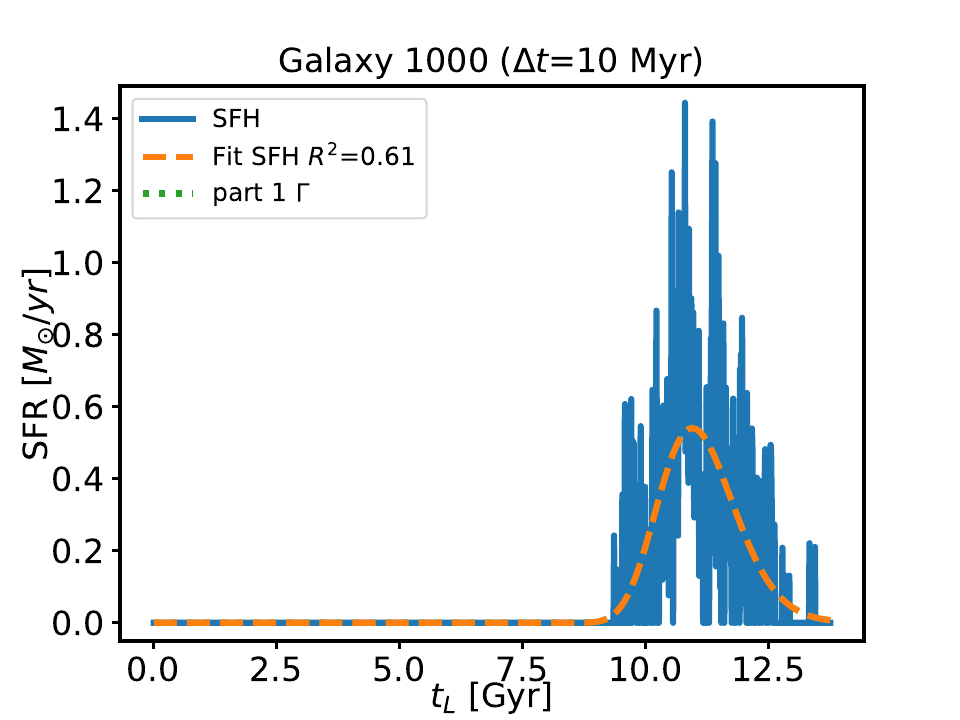}
		\includegraphics[width=0.33\linewidth]{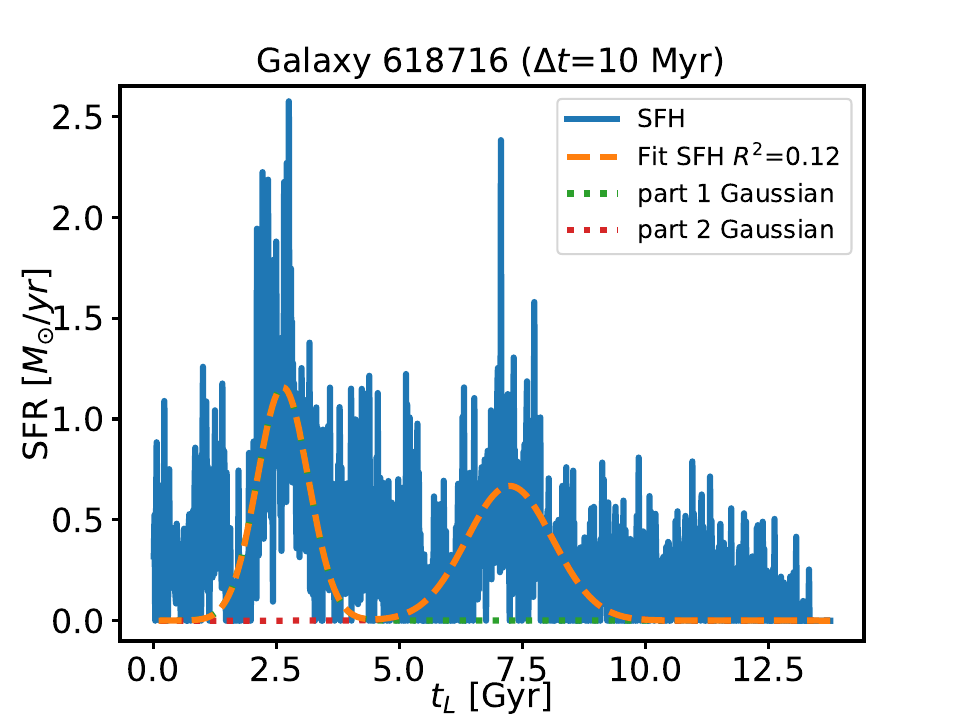}\\
		\includegraphics[width=0.33\linewidth]{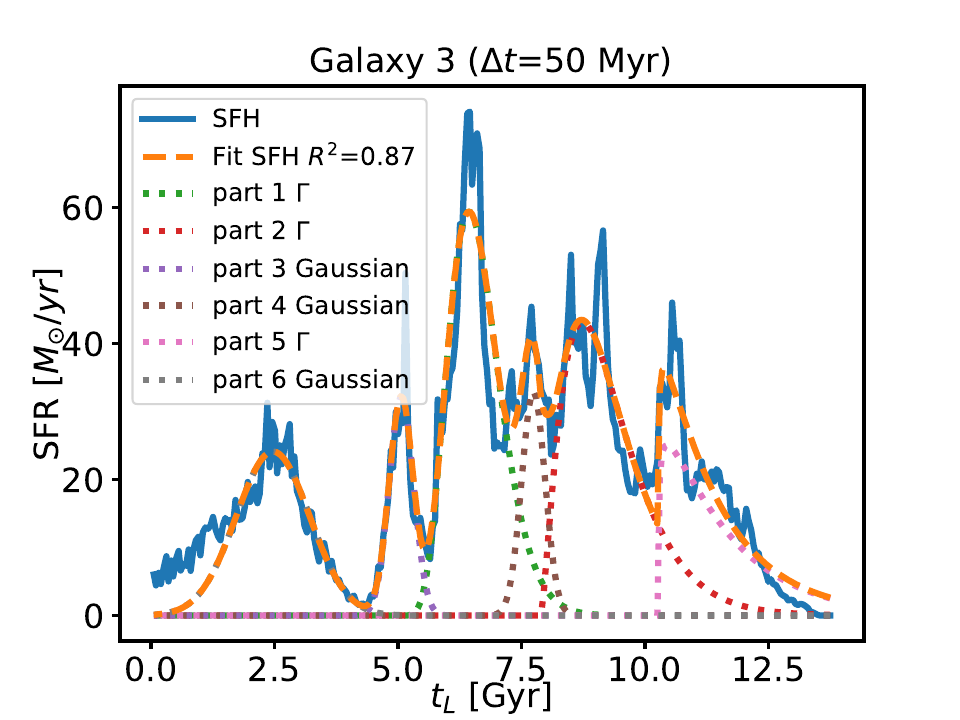}
		\includegraphics[width=0.33\linewidth]{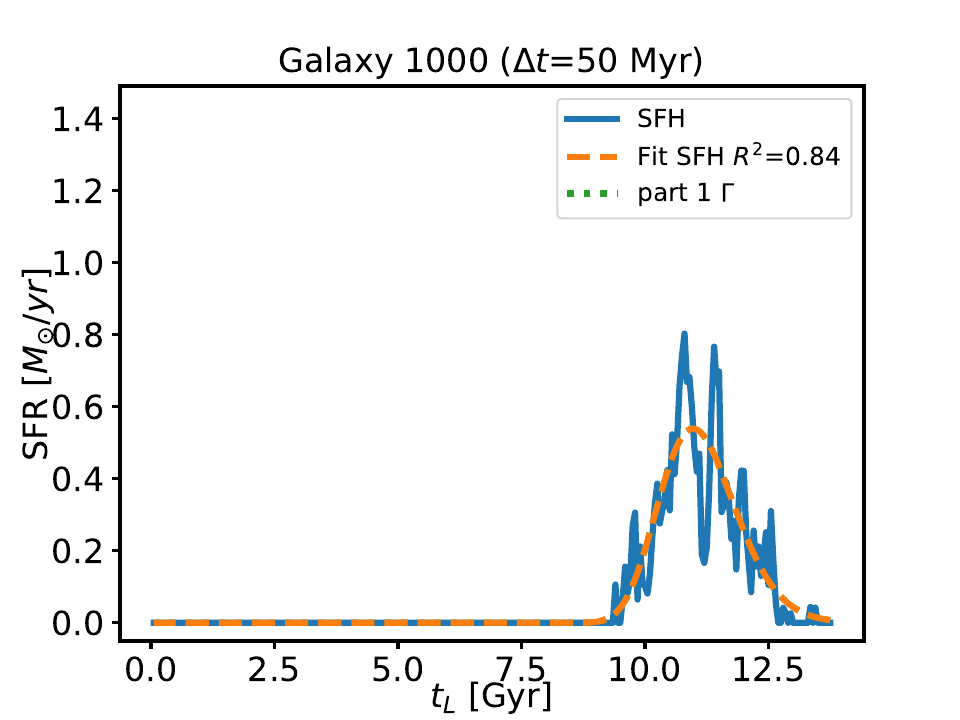}
		\includegraphics[width=0.33\linewidth]{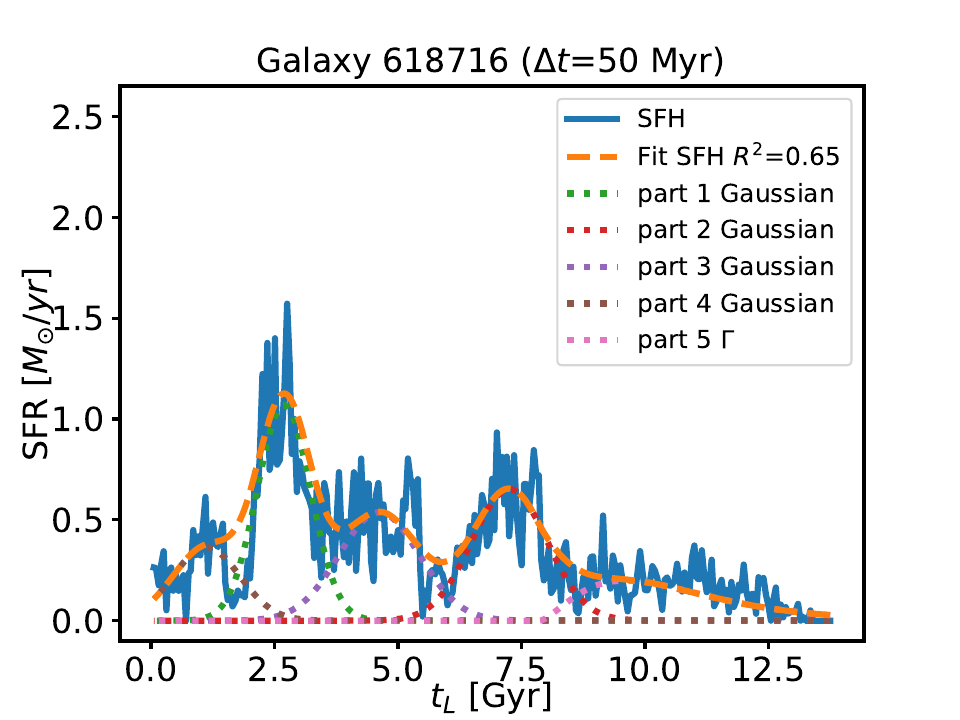}\\
		\includegraphics[width=0.33\linewidth]{Group3dt100.pdf}
		\includegraphics[width=0.33\linewidth]{Group1000dt100.pdf}
		\includegraphics[width=0.33\linewidth]{Group618716dt100.pdf}\\
		\includegraphics[width=0.33\linewidth]{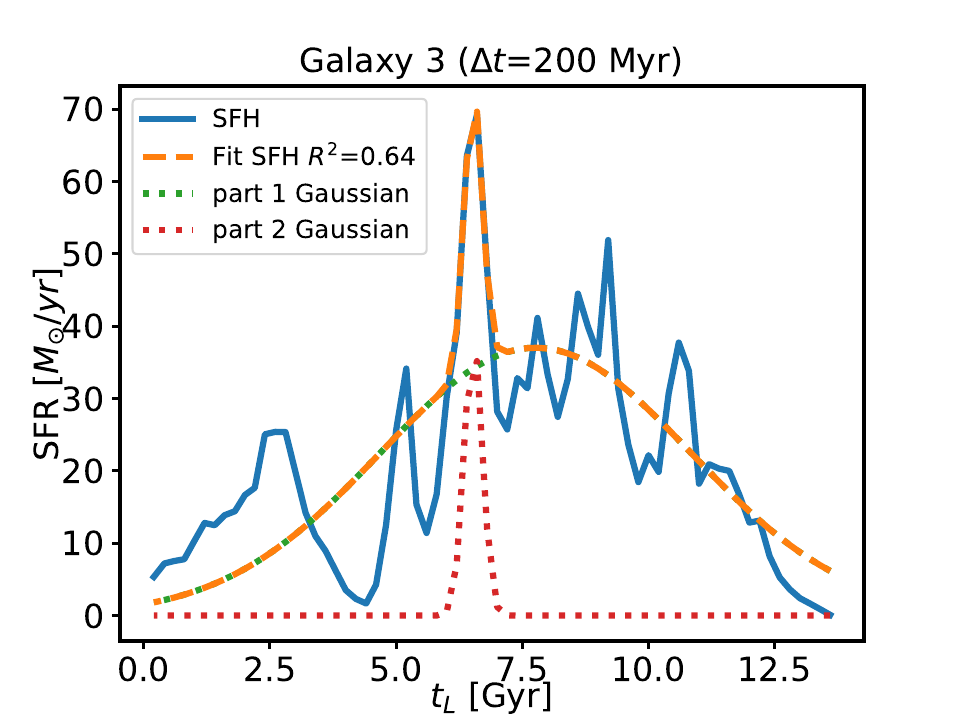}
		\includegraphics[width=0.33\linewidth]{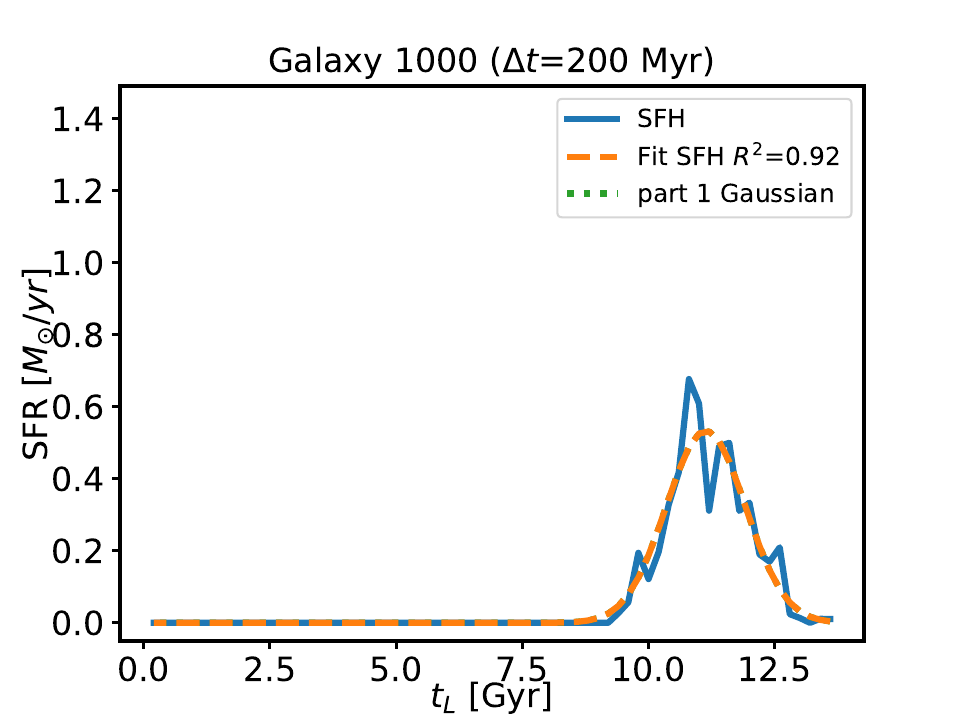}
		\includegraphics[width=0.33\linewidth]{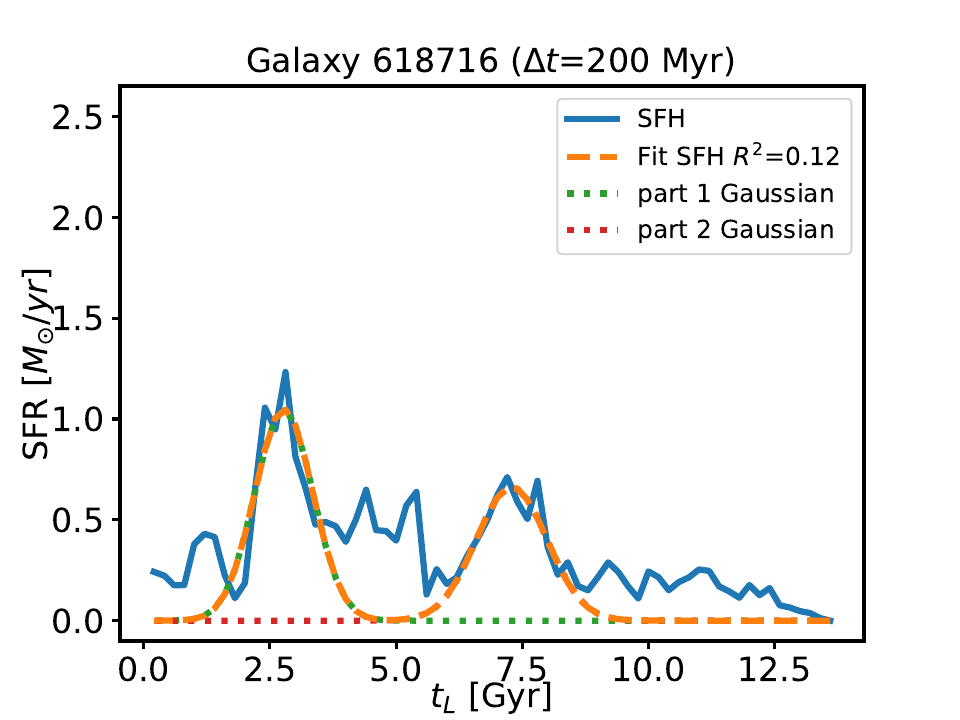}\\
		\includegraphics[width=0.33\linewidth]{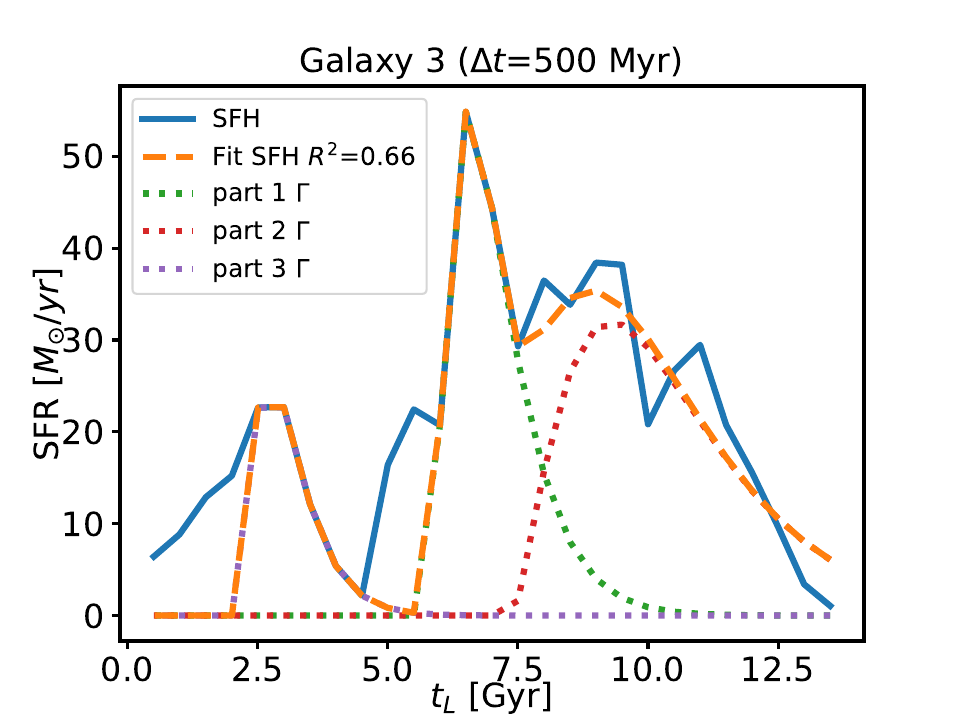}
		\includegraphics[width=0.33\linewidth]{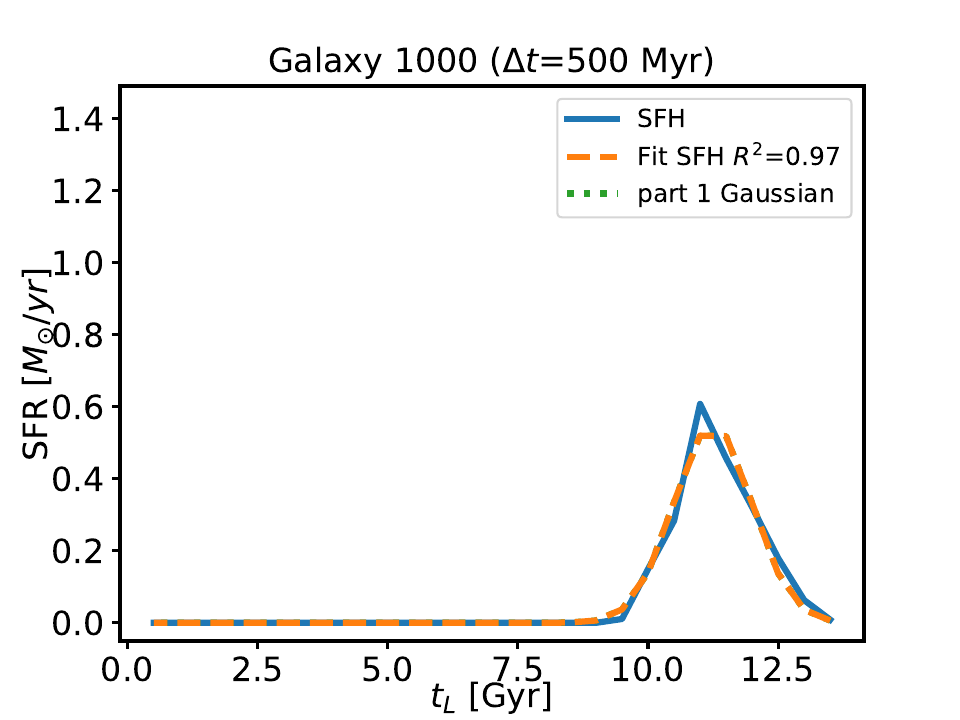}
		\includegraphics[width=0.33\linewidth]{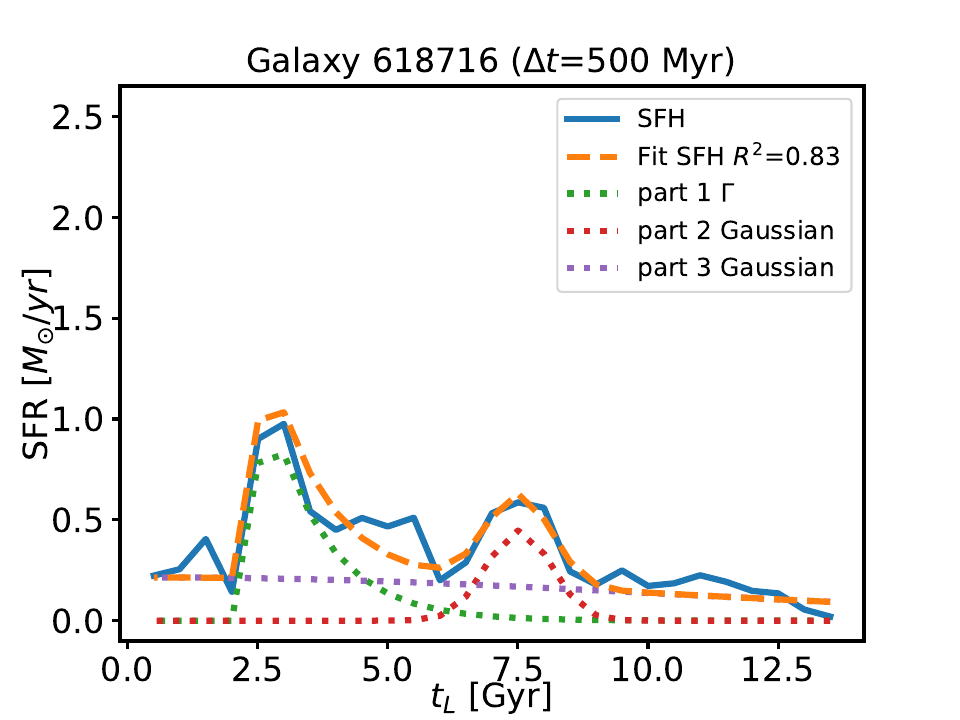}\\
		\caption{The SFHs of three galaxies obtained at different time bin resolutions (blue solid lines), along with the corresponding fitted curve (orange dashed lines) and the decomposition of components (colored dotted lines).
			Each row contains the results for one time bin.
			Each column contains the results for one galaxy.}
		\label{FigFitTimebin}
	\end{figure*}

	\begin{figure}
		\begin{center}
		\includegraphics[width=0.5\linewidth]{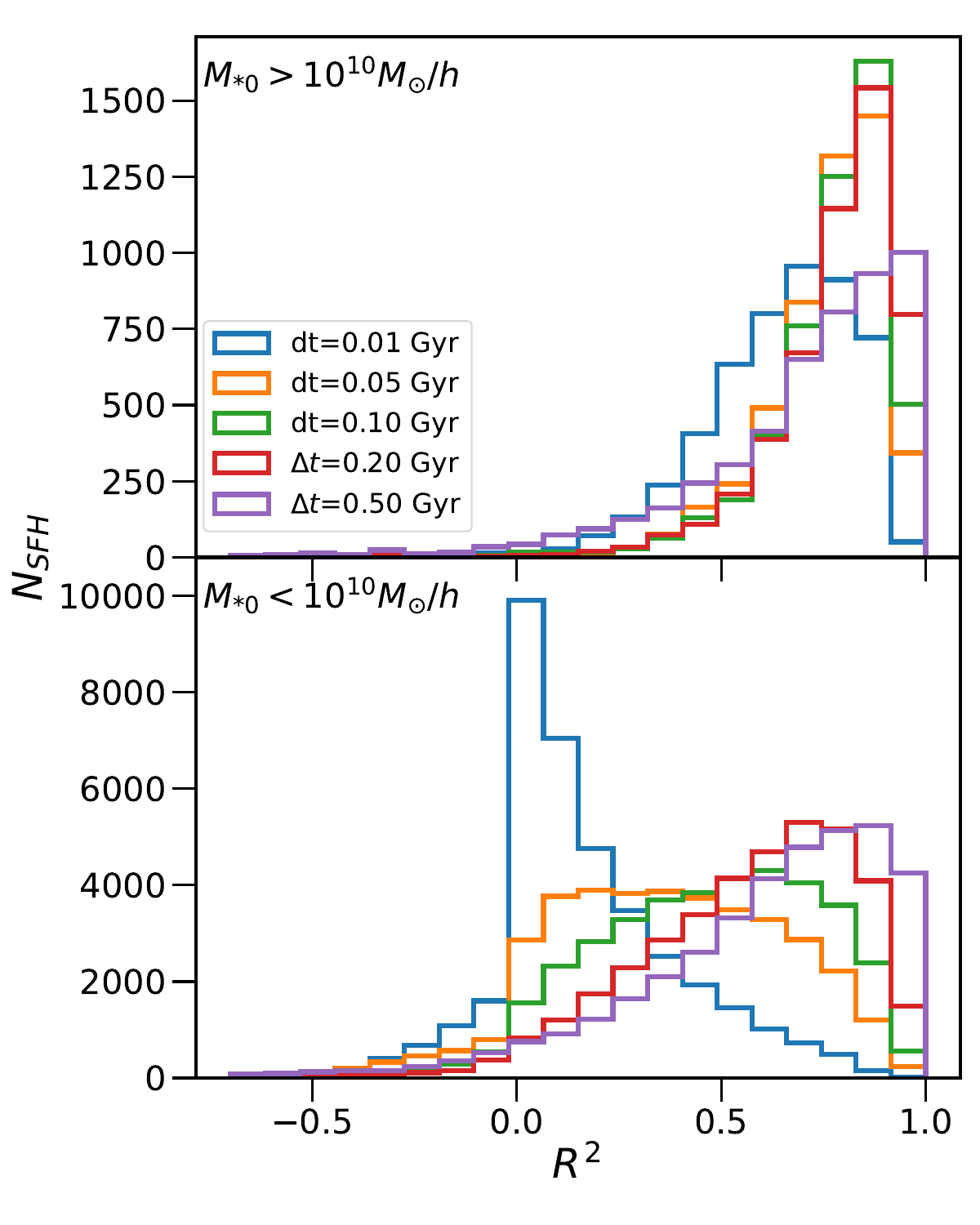}
		\caption{The histograms of fitting goodness $R^2$.
			The histograms of different colors represent the results of the SFHs obtained at different time bin resolutions.
			The upper panel shows the histograms for high mass galaxies,
			while the lower panel shows the histograms for low mass galaxies.}
			
		\end{center}
		\label{FigR2_timebin}
	\end{figure}

	\begin{figure}
		\begin{center}
		\includegraphics[width=0.5\linewidth]{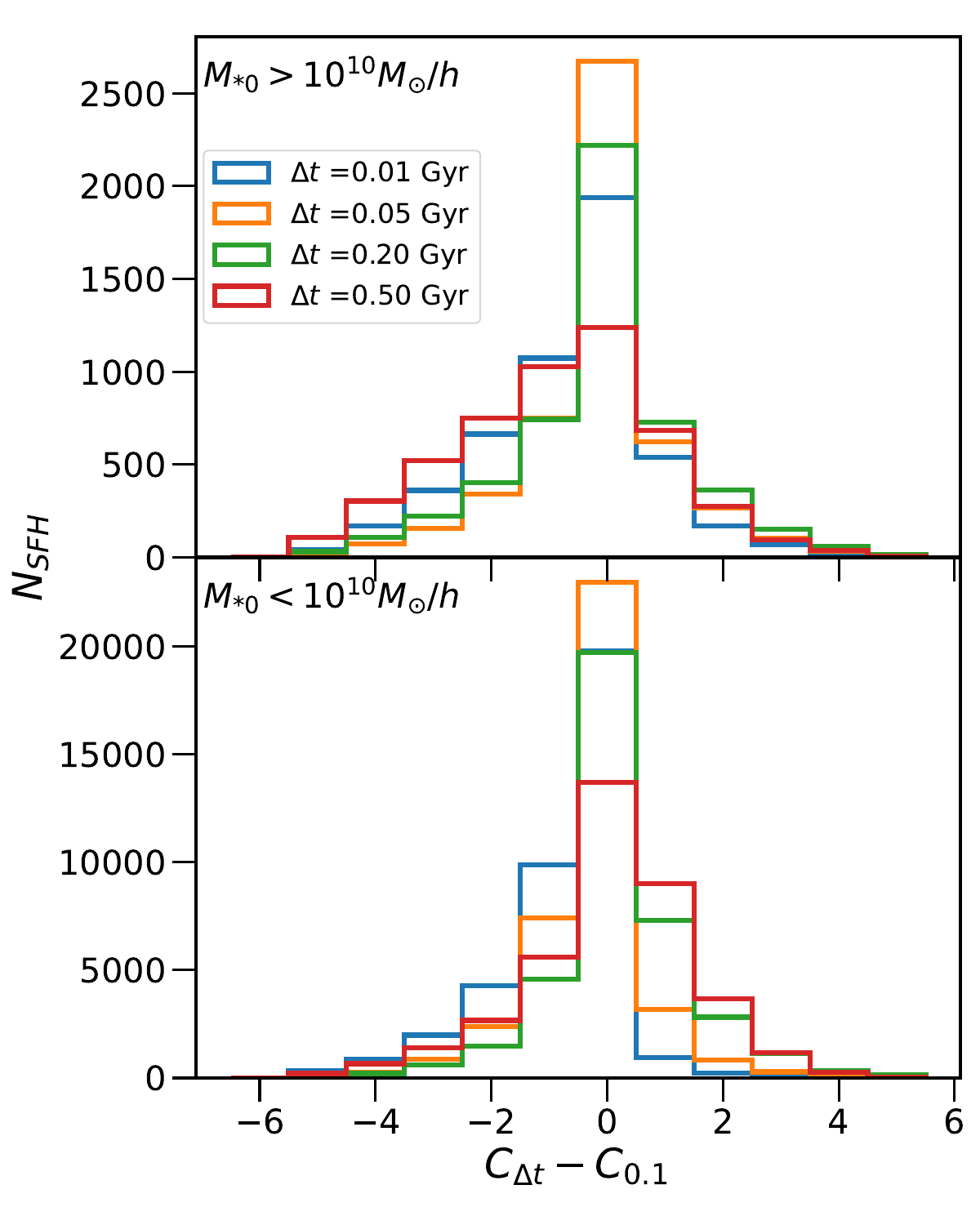}
		\caption{The histogram of the difference in the number of components in the SFH obtained at different time bin resolutions.
			$C_{dt}$ represents the number of components in an SFH.
			The x-axis shows the difference between the number of components at four other time bin resolutions and the number of components at a resolution of 100 Myr.
			The y-axis represents the corresponding number of galaxies.
			The upper panel shows the results for high-mass galaxies,
			while the lower panel shows the results for low-mass galaxies.}
		\end{center}
		\label{FigCChange_timebin}
	\end{figure}

	The resolution of the time bin significantly influences the extraction of a galaxy's star formation history and the subsequent fitting work.
	\Fig{FigFitTimebin} illustrates the SFHs (represented by the blue line) and the corresponding fitting curves (represented by the orange line) obtained using different time bin resolutions.
	The figure reveals that a larger time bin results in a smoother SFH curve, albeit at the cost of losing finer details.

	However, a smaller time bin is not always the optimal choice.
	Given the finite number of particles in a galaxy during the simulation, an excessive number of time bins would lead to fewer particles in each bin, thereby causing significant numerical fluctuations.
	This effect consequently reduces the goodness of fit for the SFHs.
	As depicted in \Fig{FigFitTimebin}, when the width of the time bin is $10$ Myr, the noise in the SFH curve itself is considerably high.
	Despite this, our fitting method can still identify the main components in the curve, although the goodness of fit is significantly reduced numerically.

	Conversely, a larger time bin results in a fitted curve that aligns better with the SFH.
	This is because using a larger time bin is akin to performing a smoothing process on the curve, which effectively eliminates noise.

	In \Fig{FigR2_timebin}, we present the statistical goodness of fit of our method for SFHs with different time bins.
	It can be observed that $R^2$ decreases as the time bin decreases.
	This relationship is particularly pronounced in low-mass galaxies, primarily due to the numerical noise caused by the insufficient number of particles in these galaxies.

	Another point of interest is whether our fitting results would be affected when different time bins are used.
	According to our inspection results (partially shown in \Fig{FigFitTimebin}), although the number of components and the type of each component (Gaussian or Gamma) may vary, the major peaks can be well restored and maintain stable positions and widths.
	The main differences lie in the handling of some small components, where different time bins can lead to different results.

	We use the fitting results of a time bin of $100$ Myr as a benchmark and compare the number of components obtained by fitting the SFHs of other time bins with it.
	As shown in \Fig{FigCChange_timebin}, most galaxies' SFH fitting does not change the number of components relative to when the time bin equals $100$ Myr ($C_{\Delta t}-C_{0.1}=0$), where $C_{\Delta t}$ is the number of fitting components of a single SFH when the time bin is $\Delta t$.

	For high-mass galaxies, whether the time bin increases or decreases, the change in the number of components is mainly a decrease, i.e., the side where $C_{\Delta t}-C_{0.1}<0$ is higher than the other side.
	As mentioned above, this is because the numerical noise brought about by the decrease in time bin can blur some small peaks, while the smoothing effect brought about by the increase in time bin may also erase these peaks.
	This suggests that $100$ Myr is a good choice of time bin for high-mass galaxies in the TNG simulation.

	For low-mass galaxies, we can see that the number of components tends to increase when the time bin increases. This suggests that the time bin of $100$ Myr may still be too small for these galaxies, and a slightly larger time bin matches their particle number better.

	At last, readers should remember that the influence of time bin resolution mainly comes from the mass resolution limit in simulation.
	In other theoretical galaxy models, i.e., SAMs, or observational data, the influence of mass resolution can be alleviated.

	\section{Stellar mass loss}
	\label{app:mass}

	\begin{figure}
		\begin{center}
			
		\includegraphics[width=0.5\linewidth]{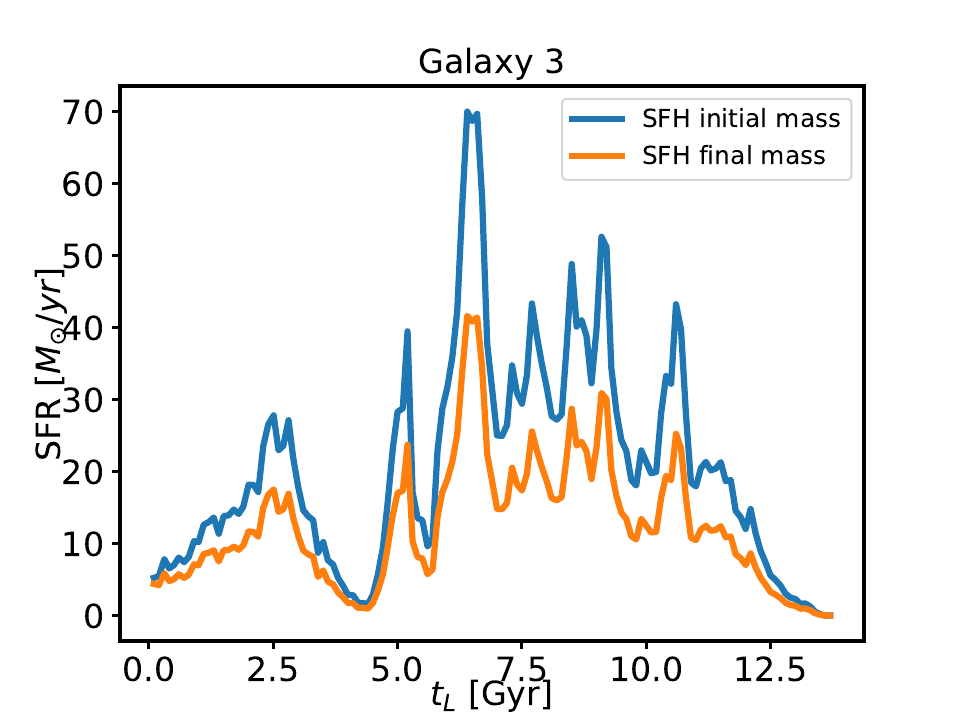}
		\caption{The SFH obtained from the initial stellar mass (blue line) and the final stellar mass (orange line) of one galaxy.}
		\end{center}
		\label{FigSFHcompare}
	\end{figure}

	Due to the loss of mass in stellar particles in simulation, the Star Formation Histories (SFHs) constructed based on the mass of particles at z=0 and those based on the initial mass of particles will differ.
	\Fig{FigSFHcompare} illustrates the differences between these two types of SFHs.
	As the age of each particle remains unchanged, with only the value of its stellar mass varying, the differences between the two SFHs are solely vertical shifts.
	Consequently, the differences in the fitting results based on the two types of SFHs primarily lie in the amplitude of the components, with no changes in the position and width of each component.

	\section{Factors affecting intrinsic SFH peak counts }
	\label{app:peak}

	\begin{table*}
		\begin{threeparttable}
			\caption{This table represents the factors that influence the intrinsic peak counts of SFH.
				The last row provides the reference for the multi-episode fraction of reconstructed SFHs in this work.
				Note that the multi-component method we proposed in this work does not require prominence parameter to define the peaks.
				Note that the prominence parameter in this work is the difference of $SFR$ rather than $logSFR$, while the latter one  is used in \cite{Iyer2019}.  }
			\begin{tabular}{cccccccc}
				\toprule
				articles                  & dataset                            & z                    & $M_*$                                & $N_{sample}$             & smooth method                                           & peak finder prominence         & $frac(N_{peak}>1$) \\
				\midrule
				\multirow{3}{*}{Iyer2017} & MUFASA                             & 1                    & \multirow{3}{*}{$\ge 10^9M_{\odot}$} & 1200                     & \multirow{3}{*}{not mentioned}                          & \multirow{3}{*}{not mentioned} & $12\% \sim 17 \%$  \\
				                          & SAMs                               & 1                    &                                      & 1200                     &                                                         &                                & $15\% \sim 20 \%$  \\
				                          & stochastic                         & 1                    &                                      & 1200                     &                                                         &                                & $\sim 15 \%$       \\
				\hline
				\multirow{3}{*}{Iyer2019} & \multirow{3}{*}{\shortstack{MUFASA                                                                                                                                                                                                          \\ \& SAMs}}         & $0.5$    & \multirow{3}{*}{$\ge 10^8M_{\odot}$} & \multirow[vpos]{3}{*}{\shortstack{$10000$  in \\ $0.5<z<3$}} & \multirow{3}{*}{\shortstack{Gaussian \\ Processes}} & \multirow{3}{*}{\shortstack{$log \frac{SFR_{peak}}{SFR_{min,local}} $ \\$> 1.5 + \frac{1.5}{4}log \frac{M_*}{10^8 M_{\odot}}$}}               & $\sim 20\%$                \\
				                          &                                    & $1$                  &                                      &                          &                                                         &                                & $10\% \sim 20\%$   \\
				                          &                                    & $2$                  &                                      &                          &                                                         &                                & $10\% \sim 15\%$   \\
				\hline
				\multirow{3}{*}{Test}     & \multirow{3}{*}{TNG}               & \multirow{3}{*}{$0$} & \multirow{3}{*}{$\ge 10^8M_{\odot}$} & \multirow{3}{*}{$23229$} & \multirow{3}{*}{\shortstack{savgol\_filter }} \tnote{a} & Iyer2019 \tnote{b}             & $16.6\%$           \\
				                          &                                    &                      &                                      &                          &                                                         & $SFR_{max}/2$  \tnote{c}       & $19.9\%$           \\
				                          &                                    &                      &                                      &                          &                                                         & $SFR_{max}/4 $                 & $66.5\%$           \\
				\hline
				This work                 & TNG                                & 0                    & $\ge 10^8M_{\odot}$                  & $23229$                  & --                                                      & --                             & $64.9\%$           \\

				\bottomrule
			\end{tabular}
			\begin{tablenotes}
				\footnotesize
				\item[a] We use the savgol\_filter function from scipy.signal module.
				\item[b] This prominence is the same as that in the line of Iyer et al. 2019.
				\item[c] $SFR_{max}$ is the maximum value of each smoothed SFH.
			\end{tablenotes}
			\label{TabPeakFrac}
		\end{threeparttable}
	\end{table*}

	\begin{figure}
		\includegraphics[width=0.5\linewidth]{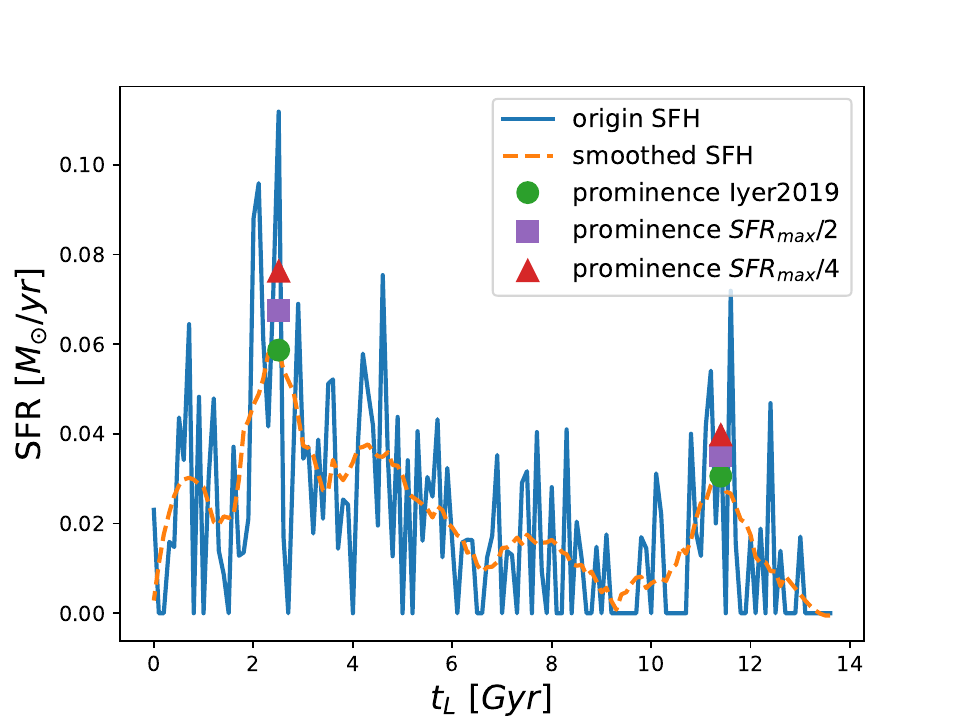}
		\includegraphics[width=0.5\linewidth]{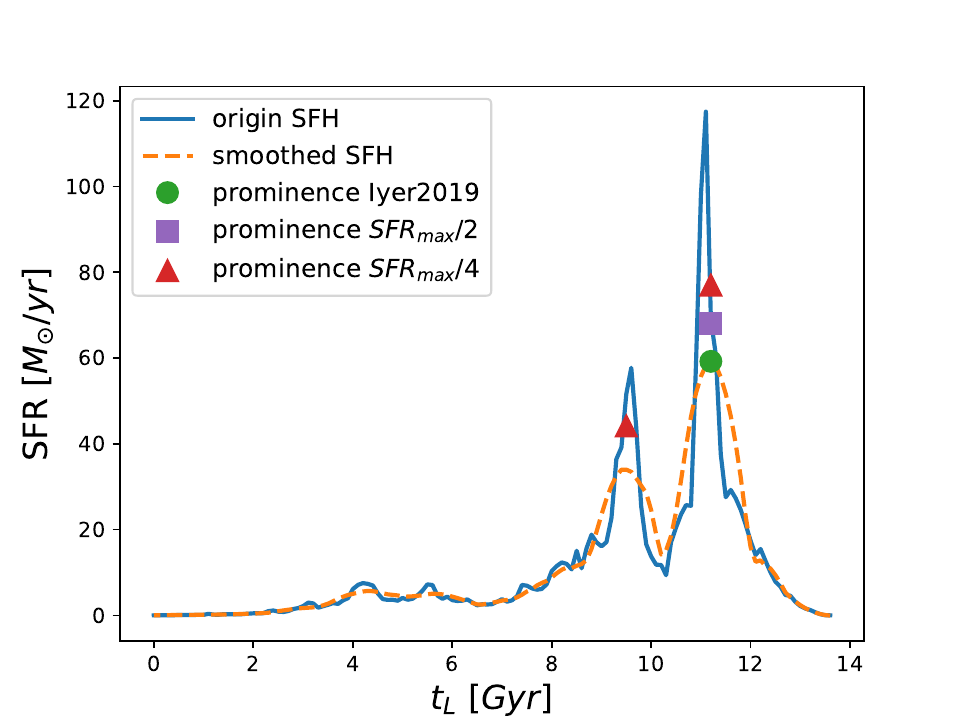}
		\caption{ Two examples of the effects of prominence. In each figure, blue solid line represents a origin SFH from the TNG simulation.
			The orange dashed line represents the smoothed curve using the scipy.signal.savgol\_filter function with a window of $20$ points and polyorder $3$.
			The peaks found with a prominence equivalent to that in \cite{Iyer2019} are indicated by green circles.
			Purple squares indicate the peaks found with a prominence of $SFR_{max}/2$,
			while red triangles indicate the peaks found with a prominence of $SFR_{max}/4$.
			To ensure the clarity of the markers, they have been slightly shifted upwards.
			In the left figure, three different sets of markers can identify the same peaks in the smoothed SFH. However, in the right figure, only the smallest prominence is able to distinguish the smaller peak.}
		\label{FigPromTest}
	\end{figure}

	Section \ref{sec:episodeunm} highlights a discrepancy between the ''true fraction" of multiple episodes SFHs in this study and those reported in \cite{Iyer2017,Iyer2019}.
	To investigate this difference, we emulate the method of identifying intrinsic peaks as described in \cite{Iyer2017} and \cite{Iyer2019}, and examine various sets of prominence for peak finding algorithm.
	As demonstrated in \Tbl{TabPeakFrac}, a higher prominence in peak finding algorithm results in a smaller fraction of multi-episode SFHs.
	Based on the value of multi-episode fraction, the SFHs reconstructed by our multi-component method closely aligns with the intrinsic SFHs defined with with a prominence of $SFR_{peak}-SFR_{min,local}>SFR_{max}/4$, where $SFR_{max}$ is the max SFR value of one smoothed SFH.
	Although the smoothing method and prominence in our test can not be aligned with \cite{Iyer2017} and \cite{Iyer2019} exactly, we can confirm that the prominence accepted in \cite{Iyer2019} is larger than $SFR_{max}/4$ for the majority of SFHs in the TNG simulation.
	We apply the prominence from \cite{Iyer2019} to our data and find that the peaks defined by this prominence are more akin to the performance of the prominence of $SFR_{max}/2$ (refer to \Fig{FigPromTest}).

\end{CJK*}
\end{document}